\documentclass[11pt]{article}

\usepackage[preprint]{acl}

\usepackage{times}
\usepackage{latexsym}

\usepackage[T1]{fontenc}
\usepackage[utf8]{inputenc}

\usepackage{microtype}

\usepackage{inconsolata}

\usepackage{graphicx}

\usepackage[table]{xcolor}   %
\usepackage{colortbl}        %

\usepackage{array}           %
\usepackage{tabularx}        %
\usepackage{booktabs}        %
\usepackage{multirow}
\usepackage{makecell}
\usepackage{arydshln}        %

\usepackage{graphicx}
\usepackage{adjustbox}
\usepackage{caption}
\usepackage{subcaption}
\usepackage{pifont}

\usepackage{siunitx}

\usepackage{colortbl}
\usepackage{tikz}

\usepackage{soul}                 %

\usepackage{listings}
\usepackage{amssymb}
\definecolor{green1}{HTML}{02C9AF}
\definecolor{LightGreen}{HTML}{DFF6DD}
\definecolor{LightYellow}{HTML}{FFF3CD} %
\definecolor{LightRed}{HTML}{F8D7DA}    %

\definecolor{DeltaPos}{RGB}{0,128,0}      %
\definecolor{DeltaNeg}{RGB}{200,0,0}      %
\definecolor{DeltaNeu}{gray}{0.45}        %

\definecolor{lightblue}{rgb}{0.8,0.85,1}
\definecolor{lightred}{rgb}{1, 0.8, 0.8}

\newcommand{\deltaup}[1]{\textsuperscript{\scriptsize\textcolor{DeltaPos}{#1}}}
\newcommand{\deltadn}[1]{\textsuperscript{\scriptsize\textcolor{DeltaNeg}{#1}}}
\newcommand{\deltane}[1]{\textsuperscript{\scriptsize\textcolor{DeltaNeu}{#1}}}

\newcommand{\deltalegend}{{\scriptsize%
\textcolor{DeltaPos}{(>\,5\%)}\;%
\textcolor{DeltaNeg}{(<\,{-}5\%)}\;%
\textcolor{DeltaNeu}{(|\,$\le$\,5\%\,|)}}}

\usepackage{dashrule} %

\usepackage{tcolorbox}
\tcbuselibrary{skins,breakable,listings}
\usepackage{listings}

\lstdefinestyle{case}{
  basicstyle=\ttfamily\scriptsize,
  breaklines=true,
  breakautoindent=false,   %
  breakindent=0pt,         %
  postbreak=\mbox{},       %
  columns=fullflexible,
  keepspaces=true,
  showstringspaces=false,
  xleftmargin=0pt,
  framexleftmargin=0pt,
  aboveskip=0pt,
  belowskip=0pt,            %
  escapeinside={(*@}{@*)}, %
}

\tcbset{
  mybox/.style={
    enhanced,
    breakable,
    arc=2pt,                        %
    colframe=black!12,
    colback=black!2,
    coltitle=black!70,
    colbacktitle=black!4,
    fonttitle=\bfseries\footnotesize,
    lefttitle=6pt, righttitle=6pt,
    toptitle=3pt,  bottomtitle=3pt,
    left=6pt, right=6pt, top=3pt, bottom=3pt,  %
    boxsep=0pt,
  },
}

\title{SonicBench: Dissecting the Physical Perception Bottleneck in \\ Large Audio Language Models}

\author{
\textbf{Yirong Sun\textsuperscript{1}\thanks{Equal contribution.}},
\textbf{Yanjun Chen\textsuperscript{1}\footnotemark[1]},
\textbf{Xin Qiu\textsuperscript{1}\footnotemark[1]},
\textbf{Gang Zhang\textsuperscript{1}},
\textbf{Hongyu Chen\textsuperscript{1}},\\
\textbf{Daokuan Wu\textsuperscript{1}},
\textbf{Chengming Li\textsuperscript{4}},
\textbf{Min Yang\textsuperscript{2,3}},
\textbf{Dawei Zhu\textsuperscript{5}},
\textbf{Wei Zhang\textsuperscript{1}},
\textbf{Xiaoyu Shen\textsuperscript{1}\thanks{Corresponding author.}}
\\
\textsuperscript{1}Ningbo Key Laboratory of Spatial Intelligence and Digital Derivative, Institute of Digital Twin, EIT\\
\textsuperscript{2}Shenzhen Institutes of Advanced Technology, Chinese Academy of Sciences
\\
\textsuperscript{3}Shenzhen University of Advanced Technology
\textsuperscript{4}Shenzhen MSU-BIT University
\textsuperscript{5}Amazon AGI\\
\small{
win1282467298@gmail.com, qiuxinzju@zju.edu.cn, xyshen@eitech.edu.cn
}
\\
\textbf{Github:} \href{https://github.com/EIT-NLP/SonicBench}{https://github.com/EIT-NLP/SonicBench}\quad
\textbf{Huggingface:} \href{https://huggingface.co/datasets/YirongSun/SonicBench}{SonicBench}
}

\begin{document}
\maketitle
\begin{abstract}
Large Audio Language Models (LALMs) excel at semantic and paralinguistic tasks, yet their ability to perceive the fundamental physical attributes of audio such as pitch, loudness, and spatial location remains under-explored. To bridge this gap, we introduce \textbf{SonicBench}, a psychophysically grounded benchmark that systematically evaluates 12 core physical attributes across five perceptual dimensions. Unlike previous datasets, SonicBench uses a controllable generation toolbox to construct stimuli for two complementary paradigms: recognition (absolute judgment) and comparison (relative judgment). This design allows us to probe not only sensory precision but also relational reasoning capabilities, a domain where humans typically exhibit greater proficiency. Our evaluation reveals a substantial deficiency in LALMs’ foundational auditory understanding; most models perform near random guessing and, contrary to human patterns, fail to show the expected advantage on comparison tasks. Furthermore, explicit reasoning yields minimal gains. However, our linear probing analysis demonstrates crucially that frozen audio encoders \emph{do} successfully capture these physical cues (accuracy $\ge 60\%$), suggesting that the primary bottleneck lies in the alignment and decoding stages, where models fail to leverage the sensory signals they have already captured.
\end{abstract}

\section{Introduction}

Large Audio Language Models (LALMs) have recently emerged as a unified interface for a wide range of auditory tasks~\cite{qwen2audio, kimiaudio}. By aligning pre-trained audio encoders with the input space of Large Language Models (LLMs), these systems inherit the strong reasoning and instruction-following capabilities of LLMs, enabling diverse audio understanding tasks within a single framework~\cite{midashenglm, liu2025voxtral, audioflamingo3}. Despite this rapid progress, existing evaluations predominantly emphasize semantic~\cite{wang2024audiobench, yang2024airbench, Sakshi2024mmau} and paralinguistic capabilities~\cite{ma2025mmar, superb, huang2025dynamicsuperbphase2}. In contrast, systematic evaluation of \emph{physical perception}, the ability to interpret intrinsic properties of audio signals, remains limited~\cite{peng2025surveyspeechllm}.

Physical perception underpins robust auditory intelligence. It encompasses fundamental attributes that anchor every audio signal, such as pitch, loudness, duration, spatial location, and timbre, and forms the basis for higher-level reasoning about acoustic events, environments, and scenes~\cite{audioset, esc50, realman}. Analogous to how visual intelligence grounds complex scene understanding in intrinsic attributes like color and geometry~\cite{mapelli1997role, gegenfurtner2000sensory}, reliable audio reasoning depends on accurate physical grounding. In real-world and embodied settings, for example, an agent must infer urgency or danger from physical cues such as pitch, tempo, and direction, even in the absence of semantic content. Without such grounding, strong performance on high-level tasks may reflect dataset shortcuts rather than genuine auditory understanding~\cite{geirhos2020shortcut}. Evaluating physical perception is therefore essential for assessing the robustness and reliability of LALMs.

To address this gap, we introduce \textbf{SonicBench}, a psychophysically grounded benchmark for evaluating LALMs’ physical perception. SonicBench covers twelve physical attributes through two complementary paradigms: \emph{recognition} (absolute judgment) and \emph{comparison} (relative judgment). The comparison paradigm reflects human psychophysics~\cite{miller1956magical, stewart2005absolute} and probes models’ relational reasoning beyond memorization. Tasks span five dimensions: Spectral \& Amplitude, Temporal, Spatial \& Environment, Timbre, and Scene-Level, ranging from low-level signal properties to high-level scene structures.

To ensure rigorous and interpretable evaluation, we develop a \textbf{SonicBench Toolbox} for controlled stimulus generation. It enforces that target cues (e.g., pitch or tempo differences) are well above human perceptual thresholds~\cite{ten2004multiple, rammsayer2012greater, sun2017pitch}, ensuring that model errors reflect representational or reasoning failures rather than sensory ambiguity.

Our evaluation reveals three consistent limitations. First, models perform poorly on basic physical perception, often approaching chance accuracy despite strong semantic performance. Second, unlike humans, they show no systematic advantage on comparison tasks, indicating weak relational reasoning over physical attributes. Third, inference-time scaling yields only marginal improvements, suggesting that increased reasoning capacity cannot compensate for missing foundational competence. Notably, linear probing shows that frozen audio encoders already capture these physical cues (accuracy $\ge 60\%$), even when end-to-end (E2E) models fail, pointing to alignment and decoding, rather than perception, as the primary bottleneck.

In summary, this work contributes: \emph{(i)} \textbf{SonicBench}, a psychophysically grounded benchmark for evaluating twelve physical audio attributes using recognition and comparison paradigms, supported by a controllable stimulus-generation toolbox; \emph{(ii)} a systematic empirical analysis demonstrating that current LALMs lack robust physical grounding and relational reasoning despite strong semantic abilities; and \emph{(iii)} evidence that performance limitations stem mainly from alignment and decoding stages, highlighting a key direction for future model development.

\section{Related Work}
\paragraph{Large Audio Language Models.}
Recent advances in multimodality have yielded a variety of models that accept audio inputs and support diverse downstream tasks. 
These prior works can be categorized into three groups: \textit{(i)} Large Audio Language Models, which pair an audio encoder with a LLM to enable joint audio-text understanding; \textit{(ii)} Large Audio Reasoning Models (LARMs), which inherit the base model like LALM equipping with explicit reasoning ability; \textit{(iii)} Omni Language Models (OLMs), which unify multiple modalities, including audio, within a shared backbone.
LALMs exhibit considerable design diversity. Some models adopt a single encoder, either preserving continuous representations~\cite{liu2025voxtral}, discretizing encoder outputs~\cite{li2025baichuanaudio}, or mapping audio into fully discrete tokens~\cite{glm4voice}; whereas others employ multiple encoders~\cite{tang2024salmonn} or introduce projection modules~\cite{ghosh2024gama} to better handle diverse audio domains.  
Upon these foundations and to handle more complex tasks, modern LARMs enhance their reasoning capabilities through specialized data or targeted post training such as fine tuning on Chain‐of‐Thought (CoT) datasets or Group Relative Policy Optimization (GRPO)~\cite{audioreasoner,li2025r1_aqa}.
In parallel, at the omni-modal frontier, modern OLMs aim for general-purpose multimodality by employing a shared backbone capable of processing text, images, audio, and sometimes video. 
While not explicitly tailored for audio, the systems have nonetheless shown strong audio understanding and generation through architecture innovations~\cite{gemini15,qwen25omni,xu2025qwen3omni}.

\begin{table*}[t]
\vspace{-1mm}
\begin{center}
\begin{adjustbox}{max width=0.95\textwidth}
    \begin{tabular}{ccccccccccccc}
        \toprule
        \multirow{2.5}{*}{\textbf{Benchmark}} & \multicolumn{4}{c}{\textbf{Spectral \& Amplitude}} &  \multicolumn{2}{c}{\textbf{Temporal}} & \multicolumn{3}{c}{\textbf{Spatial\& Environment}} &  \multicolumn{2}{c}{\textbf{Timbre}} & {\textbf{Scene Level}} \\ 
        \cmidrule(lr){2-5} \cmidrule(lr){6-7} \cmidrule(lr){8-10} \cmidrule(lr){11-12}  \cmidrule(lr){13-13} 
         & Pitch & Brightness & Loudness & Velocity & Duration & Tempo & Direction & Distance & Reverberation & Timbre & Texture & Counting\\
        \midrule
        NSynth~\cite{nsynth} & {\color{green}{\ding{52}}} & {\color{green}{\ding{52}}} & {\color{red}{\ding{56}}} & {\color{green}{\ding{52}}} & {\color{red}{\ding{56}}} & {\color{red}{\ding{56}}} & {\color{red}{\ding{56}}} & {\color{red}{\ding{56}}} & {\color{red}{\ding{56}}} & {\color{green}{\ding{52}}} & {\color{red}{\ding{56}}} & {\color{red}{\ding{56}}}\\

        SpatialSoundQA & {\color{red}{\ding{56}}} & {\color{red}{\ding{56}}} & {\color{red}{\ding{56}}} & {\color{red}{\ding{56}}} & {\color{red}{\ding{56}}} & {\color{red}{\ding{56}}} & {\color{green}{\ding{52}}} & {\color{green}{\ding{52}}} & {\color{red}{\ding{56}}} & {\color{red}{\ding{56}}} & {\color{red}{\ding{56}}} & {\color{red}{\ding{56}}}\\

        AirBench & {\color{green}{\ding{52}}} & {\color{red}{\ding{56}}} & {\color{red}{\ding{56}}} & {\color{green}{\ding{52}}} & {\color{red}{\ding{56}}} & {\color{red}{\ding{56}}} & {\color{red}{\ding{56}}} & {\color{red}{\ding{56}}} & {\color{red}{\ding{56}}} & {\color{green}{\ding{52}}} & {\color{red}{\ding{56}}} & {\color{red}{\ding{56}}}\\
        
        MMAU & {\color{red}{\ding{56}}} & {\color{red}{\ding{56}}} & {\color{red}{\ding{56}}} & {\color{red}{\ding{56}}} & {\color{green}{\ding{52}}} & {\color{green}{\ding{52}}} & {\color{red}{\ding{56}}} & {\color{red}{\ding{56}}} & {\color{red}{\ding{56}}} & {\color{green}{\ding{52}}} & {\color{red}{\ding{56}}} & {\color{green}{\ding{52}}}\\
        
        MMAR & {\color{green}{\ding{52}}} & {\color{red}{\ding{56}}} & {\color{red}{\ding{56}}} & {\color{red}{\ding{56}}} & {\color{green}{\ding{52}}} & {\color{green}{\ding{52}}} & {\color{green}{\ding{52}}} & {\color{green}{\ding{52}}} & {\color{red}{\ding{56}}} & {\color{green}{\ding{52}}} & {\color{red}{\ding{56}}} & {\color{green}{\ding{52}}}\\

        WoW-Bench~\cite{wowbench} & {\color{green}{\ding{52}}} & {\color{red}{\ding{56}}} & {\color{red}{\ding{56}}} & {\color{red}{\ding{56}}} & {\color{green}{\ding{52}}} & {\color{red}{\ding{56}}} & {\color{red}{\ding{56}}} & {\color{red}{\ding{56}}} & {\color{red}{\ding{56}}} & {\color{green}{\ding{52}}} & {\color{red}{\ding{56}}} & {\color{red}{\ding{56}}}\\
        
        MMAU-Pro & {\color{green}{\ding{52}}} & {\color{red}{\ding{56}}} & {\color{red}{\ding{56}}} & {\color{red}{\ding{56}}} & {\color{green}{\ding{52}}} & {\color{red}{\ding{56}}} & {\color{green}{\ding{52}}} & {\color{green}{\ding{52}}} & {\color{red}{\ding{56}}} & {\color{green}{\ding{52}}} & {\color{red}{\ding{56}}} & {\color{green}{\ding{52}}}\\

        Dynamic-SUPERB & {\color{red}{\ding{56}}} & {\color{red}{\ding{56}}} & {\color{red}{\ding{56}}} & {\color{red}{\ding{56}}} & {\color{red}{\ding{56}}} & {\color{red}{\ding{56}}} & {\color{red}{\ding{56}}} & {\color{green}{\ding{52}}} & {\color{green}{\ding{52}}} & {\color{green}{\ding{52}}} & {\color{green}{\ding{52}}} & {\color{red}{\ding{56}}}\\

        Dynamic-SUPERB Phase-2 & {\color{green}{\ding{52}}} & {\color{red}{\ding{56}}} & {\color{red}{\ding{56}}} & {\color{red}{\ding{56}}} & {\color{green}{\ding{52}}} & {\color{red}{\ding{56}}} & {\color{red}{\ding{56}}} & {\color{green}{\ding{52}}} & {\color{green}{\ding{52}}} & {\color{green}{\ding{52}}} & {\color{green}{\ding{52}}} & {\color{red}{\ding{56}}}\\

        STAR-Bench~\cite{liu2025starbench} & {\color{green}{\ding{52}}} & {\color{red}{\ding{56}}} & {\color{green}{\ding{52}}} & {\color{red}{\ding{56}}} & {\color{green}{\ding{52}}} & {\color{red}{\ding{56}}} & {\color{green}{\ding{52}}} & {\color{green}{\ding{52}}} & {\color{red}{\ding{56}}} & {\color{red}{\ding{56}}} & {\color{red}{\ding{56}}} & {\color{red}{\ding{56}}}\\
        
        \textbf{Ours} & {\color{green}{\ding{52}}} & {\color{green}{\ding{52}}} & {\color{green}{\ding{52}}} & {\color{green}{\ding{52}}} & {\color{green}{\ding{52}}} & {\color{green}{\ding{52}}} & {\color{green}{\ding{52}}} & {\color{green}{\ding{52}}} & {\color{green}{\ding{52}}} & {\color{green}{\ding{52}}} & {\color{green}{\ding{52}}} & {\color{green}{\ding{52}}}\\
        \bottomrule
    \end{tabular}
\end{adjustbox}
\caption{\small Benchmark coverage comparison of existing audio benchmarks and ours across perceptual attributes.}
\label{tab:data_comparison}
\end{center}
\end{table*}

\paragraph{Audio Perception Benchmarks.}
With the progress of LALMs, a variety of benchmarks have emerged to assess audio understanding. 
These benchmarks vary in their focus, as audio signals convey a rich set of cues that can be organized along an information dimension into linguistic, paralinguistic, and non-linguistic categories~\cite{peng2025surveyspeechllm}.  
Early efforts mainly target linguistic understanding, such as automatic speech recognition (ASR) and audio captioning, and have become relatively mature~\cite{du2018aishell2transformingmandarinasr,zhang2022wenetspeech10000hours,bai2024audiosetcap,xu2025fireredasr}.
More recent efforts extend to paralinguistic aspects, emphasizing cues beyond words such as speaker identity, emotion~\cite{Sakshi2024mmau,wang2025mmsu,huang2025dynamicsuperbphase2,sun2025llaso}, and to a lesser extent, to non-linguistic properties tied to the signal itself \cite{weck2024muchomusic,kumar2025mmaupro,ma2025mmar}, even though these tasks are crucial for downstream applications like interactive robotic systems.  
Furthermore, progress in the non-linguistic space remains fragmented and partial, where most studies probe individual phenomena~\cite{7953152,engel2017neuralaudiosynthesis,bogdanov2019mtg,shimada2023starss23,zheng2025bat}.  
\begin{table}[h]
    \centering
    \footnotesize
    \setlength{\tabcolsep}{8pt} %
    \renewcommand{\arraystretch}{1.15} %
    \begin{adjustbox}{width=0.95\columnwidth}
    \begin{tabular}{lc}
        \toprule
        \textbf{Statistics} & \textbf{Value} \\
        \midrule
        Total Samples & 2,400 \\
        Per-task Samples & 1,200\,/\,1,200 \\
        Perceptual Dimensions & 5 \\
        Physical Attributes & 12 \\
        \midrule
        Avg.\ Text Instruction Length & 25.5 words \\
        Avg.\ Text Answer Length & 1 word (A or B)\\
        Recognition Clip Length & 4.0 seconds \\
        Comparison Clip Length & 8.5 seconds \\
        \bottomrule
    \end{tabular}
    \end{adjustbox}
    \caption{\small \textbf{SonicBench Statistics.} Overview of the dataset scale, dimensions, and signal specifications.}
    \label{tab:benchmark_statistics}
\end{table}
Accordingly, we introduce a systematic benchmark. Rather than directly aggregating all non-linguistic tasks, this work targets perceptual attributes that directly reflect the physical characteristics of the signal, to bridge isolated non-linguistic tasks and holistic auditory perception as presented in Table~\ref{tab:data_comparison}, in analogy with the intrinsic-attribute work in vision such as color and abstract relation~\cite{colorbench,gao2025pixelspatternspoetryworld,wüst2025bongard} probing modality-specific perceptual understanding.

\section{Benchmark}
\subsection{Overview of SonicBench}
SonicBench is a comprehensive benchmark evaluating the physical perception capabilities of LALMs. It comprises 2,400 curated question-audio pairs (see Table~\ref{tab:benchmark_statistics}) targeting five core perceptual dimensions: \textit{Spectral \& Amplitude}, \textit{Temporal}, \textit{Spatial \& Environment}, \textit{Timbre}, and \textit{Scene Level}. Each perceptual dimension is further decomposed into fine-grained physical attributes, designed to assess models’ ability to perceive and reason about fundamental sound properties. For each attribute, SonicBench evaluates two complementary tasks: \textit{(i)} \textbf{Recognition}: Given a single 4-second audio clip paired with a textual question, the model must classify physical attributes into defined states (e.g., high vs. low pitch), evaluating its capability for \textbf{absolute understanding}. \textit{(ii)} \textbf{Comparison}: Given a single audio track concatenating two 4-second clips with a 0.5-second silent gap and a question, the model must distinguish the relation between segments (e.g., finding the louder clip), evaluating its sensitivity to \textbf{relative differences}.
Each attribute comprises 100 recognition and 100 comparison pairs, ensuring balanced coverage.

\subsection{Attribute Taxonomy of SonicBench}
\begin{figure*}[t]
\centering
\includegraphics[width=0.97\linewidth]{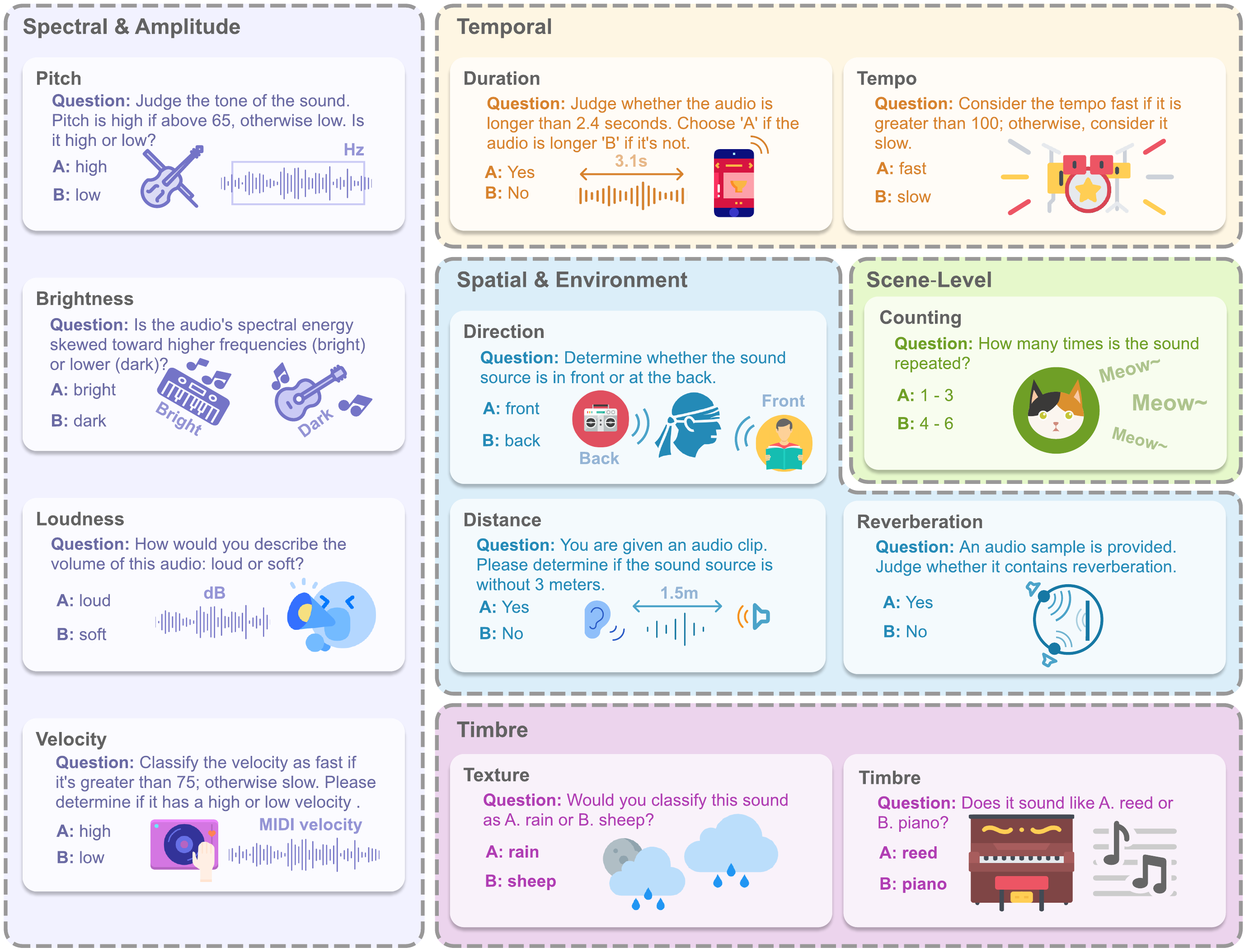}
\caption{\small \textbf{Taxonomy of SonicBench acoustic attributes.}
Overview of the twelve perceptual attributes evaluated in SonicBench, organized into five dimensions: Spectral \& Amplitude, Temporal, Spatial \& Environment, Timbre, and Scene-Level.
Each panel illustrates the auditory concept tested and the corresponding binary judgment required from models.
This taxonomy defines the perceptual scope of SonicBench, systematically covering the full range of low-level sound features to high-level scene reasoning. Examples shown here are adapted for illustration purposes; original benchmark samples are provided in Appendix~\ref{appendix:task_instruction}.}
\label{fig:benchmark_overview}
\end{figure*}

Our attribute taxonomy is designed to provide a unified and psychophysically grounded view of physical audio perception. 
While prior works have explored individual non-linguistic tasks, simply aggregating these heterogeneous datasets would introduce inconsistent distributions and confounding variables. To avoid this, we employ a unified generation pipeline to systematize physical attributes. 
As shown in Figure~\ref{fig:benchmark_overview}, we organize attributes into five perceptual dimensions, from low-level signal properties to higher-level scene understanding, including Spectral \& Amplitude, Temporal, Spatial \& Environment, Timbre, and Scene-Level, spanning twelve concrete attributes (see Appendix~\ref{appendix:attri_details} for definitions and psychophysical context). 
Across this taxonomy, we adhere to three core design principles.
\textit{(i) Systematic Coverage}. We select attributes that are not merely edge cases but are latent in virtually every real-world sound and underpin human auditory perception. 
\textit{(ii) Psychophysical Control}. Unlike uncontrolled wild audio, our stimuli are constructed with strict psychophysical margins. We ensure that attribute differences lie far above human Just-Noticeable Differences (JNDs), making the tasks trivially easy for human listeners. This guarantees that model errors reflect genuine representational deficiencies rather than sensory threshold effects.
\textit{(iii)  Controlled Comparative Paradigm}. By integrating paired comparison tasks alongside standard recognition, we establish a more controllable evaluation setting where non-target variables are rigorously held constant. This isolation allows us to test whether models possess robust relational reasoning abilities specific to the target attribute.

\subsection{Toolbox}
\label{subsec:toolbox}
To make SonicBench reusable beyond a fixed test set, we release a \emph{SonicBench Toolbox} that programmatically generates new, controllable audio samples under the same taxonomy. Constructing perception-oriented stimuli is intrinsically difficult, as one must vary a single target attribute while keeping all others as stable as possible, a process that otherwise requires substantial audio-engineering expertise and trial-and-error in a DAW\footnote{For example, when composing timbre or texture contrasts, our toolbox keeps pitch, loudness, temporal envelope, and spatial configuration fixed while only changing the instrument or spectral coloration.}. 
Inspired by prior benchmarks that pair datasets with generation utilities~\cite{cheng2025jailbreak}, the toolbox packages our rule-based signal-processing recipes e.g., spectral shaping and envelope control, so that future users can $(i)$ control attributes variables exactly, and $(ii)$ craft more samples or support more languages with minimal effort. A user only needs to provide one or two short input clips, a target attribute configuration (e.g., duration or pitch), and the desired task type (recognition or comparison); the toolbox then produces corresponding audio pairs together with task instructions and gold answers, which can be further customized if needed. Implementation details, parameter settings, and generation procedures are provided in Appendix~\ref{appendix:toolbox}.

\subsection{Data Curation Process}
\label{sec:data_curation_process}
\begin{figure*}[t]
\centering
\includegraphics[width=0.98\linewidth]{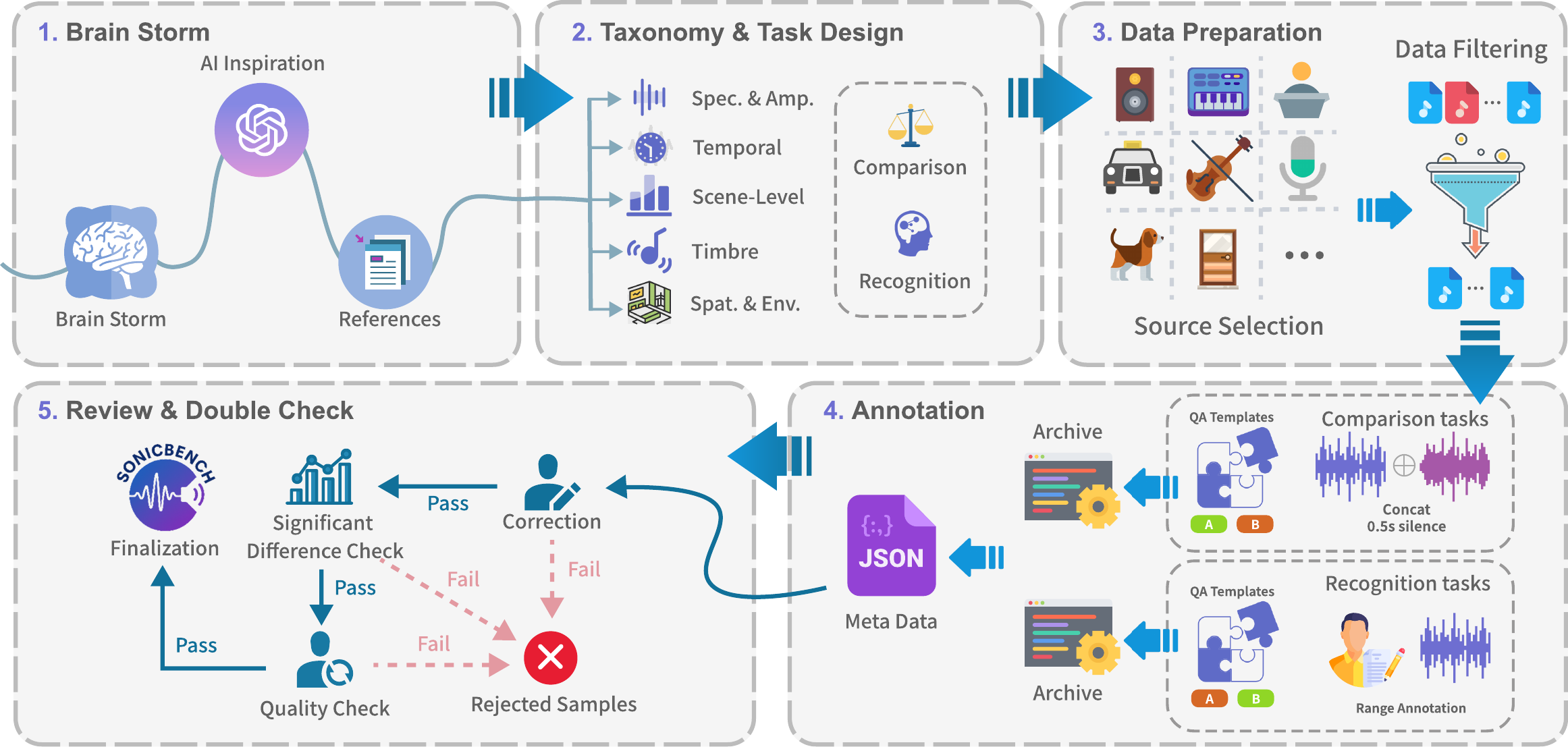}
\caption{\small \textbf{A comprehensive pipeline for constructing SonicBench.}
The process includes:
(1) Brain Storm collecting ideas through AI-assisted brainstorming and literature review;
(2) Taxonomy \& Task Design defining perceptual dimensions and formulating recognition and comparison tasks;
(3) Data Preparation selecting and filtering sound sources to ensure diversity and balance;
(4) Annotation generating QA templates and structured JSON metadata for both task types; and (5) Review \& Double Check performing multi-round manual validation including quality control, significance testing, correction, and finalization to ensure benchmark reliability.}
\label{fig:benchmark_constr_pipeline}
\end{figure*}

As shown in Figure~\ref{fig:benchmark_constr_pipeline}, we constructed our benchmark through a five-stage pipeline.
\paragraph{Brainstorming.} Given that physical attributes are foundational to both audio signals and human auditory perception, and exhibit well-studied psychophysical regularities, probing them cannot rely on ad-hoc question design. We conducted multi-round brainstorming where expert annotators, LLM-based idea generation, and targeted retrieval over psychoacoustics and audio-perception literature  interacted to iteratively expand and filter a pool of candidate attributes, task formulations, and practical annotation heuristics that subsequently grounded our taxonomy and task design (annotator qualifications are detailed in Appendix~\ref{appendix:annotator_qualifications}).  

\paragraph{Taxonomy and task design.}
We manually consolidated the candidate attributes, task formulations, and annotation heuristics from the brainstorming stage into a compact taxonomy of 12 perceptually grounded, non-semantic acoustic attributes, prioritizing coverage of five core perceptual dimensions while minimizing semantic redundancy. We then selected two core evaluation tasks, recognition and comparison, for each attribute motivated by the fact that humans typically find relative judgements easier and more reliable than absolute ones~\cite{miller1956magical,stewart2005absolute}.  
We specified primarily binary label spaces to reduce borderline cases and improve inter-annotator agreement and evaluation reliability, together with standardized question templates that define the schema for subsequent data preparation and annotation, allowing us to test whether models mirror this human advantage in relative perception.  

\paragraph{Data preparation.}
Given the taxonomy and task schema, we curated candidate audio from a mixture of public corpora to cover a broad range of everyday acoustic conditions (full source list is provided in Appendix~\ref{appendix:details_data_sources}). We then applied a series of lightweight signal- and attribute-level filters to discard clips thereby balancing attribute values (filtering details in Appendix~\ref{appendix:filtering}).

\paragraph{Annotation.}
Using the curated audio pool, we employed our Toolbox to control target attribute values and instantiate recognition samples (single 4s clips) and comparison samples (two 4s clips separated by 0.5s of silence, totalling 8.5s; generation details in Appendix~\ref{appendix:toolbox}). For each attribute \& task pair, we applied standardized QA templates, exemplified in Appendix~\ref{appendix:task_instruction} with binary label spaces to form two-option questions aligned with the audio, yielding a raw, automatically labeled JSON corpus\footnote{Avoiding subjective crowd-sourcing, labels derive from canonical parameters or physical measurements, guaranteeing objective correctness.} and corresponding audio data for subsequent human verification.

\paragraph{Quality control.}
Ensuring high data quality was central to the construction of our benchmark, so we engaged domain experts for correction, significant difference check and quality inspection to implement a three-step quality control. 
\begin{figure*}[t]
    \centering
    \includegraphics[width=1\linewidth]{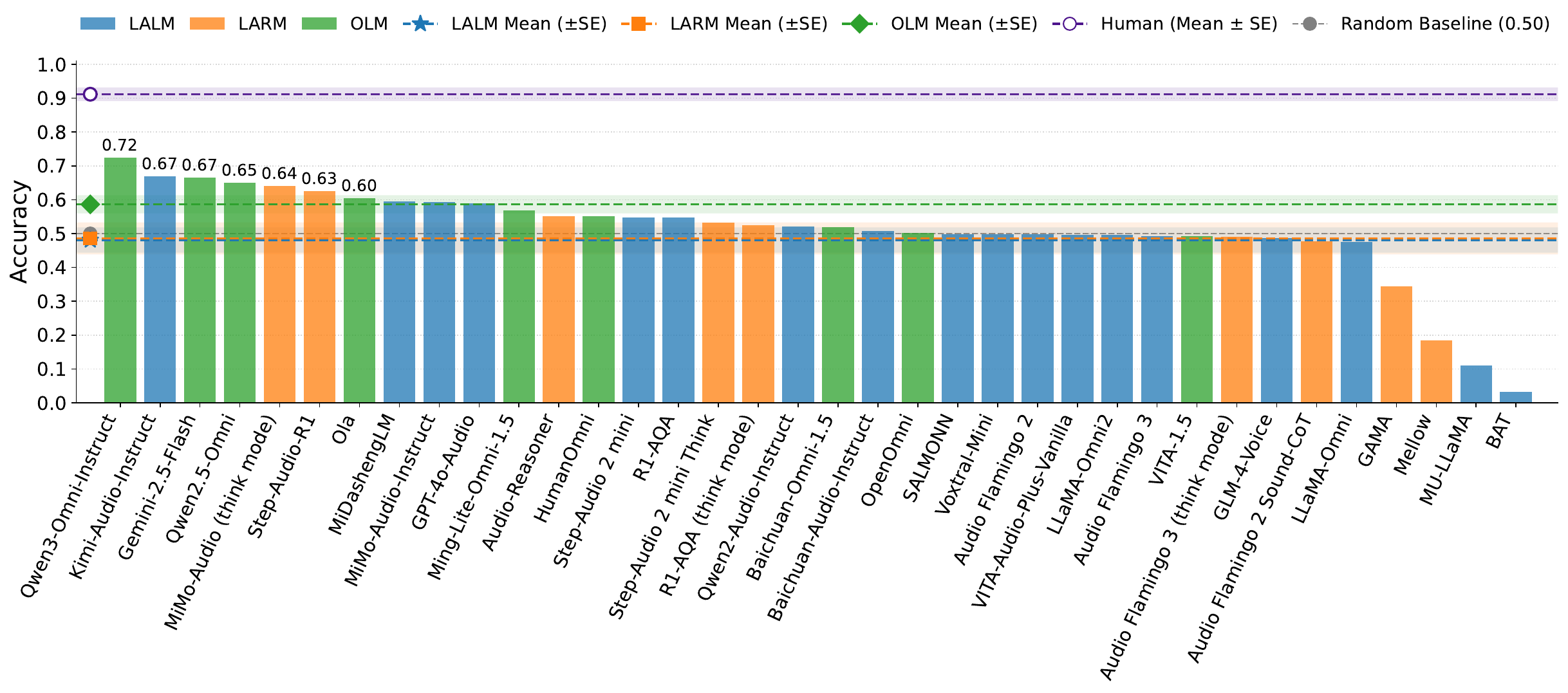}
    \caption{\small \textbf{Overall SonicBench accuracy across 36 systems.} Bars show mean accuracy over 12 attributes and 2 tasks. Colors represent model categories. Dashed lines mark each family’s mean $\pm$ SE (standard error) and the gray line marks the random-guess baseline (0.5). OLMs achieve the highest overall accuracy, while many systems cluster near chance.}
    \label{fig:overall_performance_rank}
\end{figure*}
First, we enforced role separation: each question was independently authored, corrected, and finally reviewed by different annotators. Second, expert annotators performed perceptual validation, listening to all items to confirm that the target attribute differences in both recognition and comparison trials were perceptually salient and reliably distinguishable\footnote{This step enforces strict psychophysical controls where attribute differences exceed JND thresholds, ensuring evaluation of physical perception.}. 
Third, we adopted iterative revision: any item that failed more than two rounds of inspection was revised or discarded. After this process, we retained 2,400 high-quality questions in the final version.

\begin{table*}[t]
\centering
\footnotesize
\setlength{\tabcolsep}{4pt}
\begin{adjustbox}{max width=0.98\textwidth}
\begin{tabular}{lccccccccccccc}
\toprule \toprule
\multirow{2.5}{*}{\textbf{Models}} & \multicolumn{4}{c}{\textbf{Spectral \& Amplitude}} &  \multicolumn{2}{c}{\textbf{Temporal}} & \multicolumn{3}{c}{\textbf{Spatial\& Environment}} &  \multicolumn{2}{c}{\textbf{Timbre}} & {\textbf{Scene Level}}& \multirow{2.5}{*}{\textbf{Avg.}}  \\ 
\cmidrule(lr){2-5} \cmidrule(lr){6-7} \cmidrule(lr){8-10} \cmidrule(lr){11-12}  \cmidrule(lr){13-13} 
& Pitch & Brightness & Loudness & Velocity & Duration & Tempo & Direction & Distance & Reverberation & Texture & Timbre & Counting & \\
\midrule
Random Guess & 0.50 & 0.50 & 0.50 & 0.50 & 0.50 & 0.50 & 0.50 & 0.50 & 0.50 & 0.50 & 0.50 & 0.50 & 0.50\\
\cdashline{1-14}
\noalign{\vskip 0.5mm}
Human &0.93 &0.93 &0.87 &0.83 &0.92 &0.87 &0.83 &0.80 &1.00 &1.00 &0.97 &1.00 &0.91 \\
\midrule \midrule
\multicolumn{14}{c}{\textbf{Large Audio Language Models (LALMs)}} \\ 
\midrule \midrule
BAT & \cellcolor{green1!95}0.03 & \cellcolor{green1!99}0.00 & \cellcolor{green1!98}0.02 & \cellcolor{green1!94}0.04 & \cellcolor{green1!98}0.02 & \cellcolor{green1!100}0.00 & \cellcolor{green1!100}0.00 & \cellcolor{green1!98}0.02 & \cellcolor{green1!92}0.06 & \cellcolor{green1!66}0.17 & \cellcolor{green1!86}0.06 & \cellcolor{green1!99}0.00 & \cellcolor{green1!93}0.03\\

MU-LLaMA & \cellcolor{green1!70}0.17 & \cellcolor{green1!91}0.06 & \cellcolor{green1!100}0.00 & \cellcolor{green1!63}0.19 & \cellcolor{green1!96}0.02 & \cellcolor{green1!77}0.12 & \cellcolor{green1!88}0.05 & \cellcolor{green1!86}0.09 & \cellcolor{green1!64}0.17 & \cellcolor{green1!57}0.20 & \cellcolor{green1!53}0.24 & \cellcolor{green1!92}0.04 & \cellcolor{green1!78}0.11\\

LLaMA-Omni & \cellcolor{green1!24}0.41 & \cellcolor{green1!1}0.50 & \cellcolor{green1!10}0.48 & \cellcolor{green1!6}0.49 & \cellcolor{green1!7}0.46 & \cellcolor{green1!2}0.50 & 0.50 & \cellcolor{green1!1}0.45 & \cellcolor{green1!8}0.45 & 0.52 & 0.50 & \cellcolor{green1!7}0.48 & \cellcolor{green1!5}0.48\\

Audio Flamingo 3 & 0.48 & 0.50 & 0.50 & 0.50 & 0.47 & 0.51 & 0.50 & 0.50 & 0.50 & 0.50 & 0.50 & 0.46 & 0.49\\

GLM-4-Voice & 0.50 & 0.51 & \cellcolor{green1!1}0.49 & \cellcolor{green1!1}0.52 & \cellcolor{green1!4}0.48 & 0.52 & 0.47 & \cellcolor{green1!1}0.47 & 0.48 & 0.50 & 0.50 & \cellcolor{green1!19}0.42 & \cellcolor{green1!2}0.49\\

VITA-Audio-Plus-Vanilla & 0.51 & 0.51 & 0.50 & 0.48 & 0.51 & 0.50 & 0.46 & 0.54 & 0.52 & 0.52 & 0.49 & 0.44 & 0.50\\

Audio Flamingo 2 & 0.50 & \cellcolor{green1!6}0.51 & 0.52 & 0.50 & 0.49 & 0.51 & \cellcolor{green1!7}0.48 & \cellcolor{green1!8}0.42 & \cellcolor{green1!2}0.50 & \cellcolor{green1!3}0.54 & \cellcolor{green1!2}0.53 & 0.50 & \cellcolor{green1!2}0.50\\

Voxtral-Mini & 0.53 & 0.44 & 0.51 & \cellcolor{green1!4}0.49 & \cellcolor{green1!1}0.50 & 0.52 & 0.51 & 0.47 & 0.50 & 0.54 & 0.50 & 0.51 & 0.50\\

LLaMA-Omni2 & 0.52 & 0.49 & 0.46 & \cellcolor{green1!4}0.46 & 0.49 & 0.47 & \textbf{0.53} & 0.54 & 0.53 & 0.52 & 0.49 & 0.48 & 0.50\\

SALMONN & 0.50 & 0.50 & 0.52 & 0.50 & 0.50 & 0.51 & 0.51 & 0.46 & 0.50 & 0.50 & 0.50 & 0.50 & 0.50\\

Baichuan-Audio-Instruct & 0.49 & 0.51 & 0.51 & 0.49 & 0.52 & 0.50 & 0.44 & 0.43 & 0.53 & 0.68 & 0.52 & 0.49 & 0.51\\

Qwen2-Audio-Instruct & 0.48 & \cellcolor{green1!4}0.49 & \cellcolor{green1!4}0.48 & \cellcolor{green1!1}\underline{0.55} & 0.51 & 0.53 & 0.50 & 0.53 & 0.53 & 0.59 & 0.55 & 0.53 & \cellcolor{green1!1}0.52\\

R1-AQA & 0.57 & 0.61 & 0.54 & 0.51 & 0.48 & 0.49 & 0.50 & 0.50 & 0.53 & 0.75 & 0.64 & 0.48 & 0.55\\

Step-Audio 2 mini& 0.52 & 0.59 & 0.49 & 0.51 & 0.51 & 0.50 & 0.47 & 0.48 & 0.50 & 0.95 & 0.58 & 0.51 & 0.55\\

MiMo-Audio & 0.71 	&0.70 	&0.58 &	0.53 &	0.54 &	0.49 &	0.50 &	0.51 &	0.61 	&0.89 	&0.57& 	0.50 &	0.59 \\

MiDashengLM & 0.61 & 0.70 & 0.67 & 0.54 & 0.63 & 0.55 & 0.46 & 0.51 & 0.52 & 0.85 & 0.60 & 0.52 & 0.60\\

Kimi-Audio-Instruct & \underline{0.83} & 0.81 & \cellcolor{green1!1}{\underline{0.68}} & 0.52 & \underline{0.70} & 0.57 & \cellcolor{green1!3}0.44 & \cellcolor{green1!3}0.44 & \textbf{0.77} & \underline{0.97} & 0.59 & 0.74 & \cellcolor{green1!1}\underline{0.67}\\

\cdashline{1-14}
\noalign{\vskip 0.5mm}
GPT-4o-Audio &0.71&	0.73 	&\cellcolor{green1!2}0.52 &	\cellcolor{green1!1}0.54 &	\cellcolor{green1!1}0.49 	&0.50 &	\cellcolor{green1!1}0.50 &	\cellcolor{green1!1}\underline{0.55} &	\cellcolor{green1!2}0.50 &	0.90 &	0.46 	&0.67 	&\cellcolor{green1!1}0.59 \\
\midrule \midrule
\multicolumn{14}{c}{\textbf{Large Audio Reasoning Models (LARMs)}}  \\ 
\midrule \midrule
Mellow & \cellcolor{green1!80}0.09 & \cellcolor{green1!74}0.14 & \cellcolor{green1!83}0.10 & \cellcolor{green1!72}0.14 & \cellcolor{green1!51}0.21 & \cellcolor{green1!51}0.24 & \cellcolor{green1!73}0.14 & \cellcolor{green1!61}0.18 & \cellcolor{green1!48}0.28 & \cellcolor{green1!51}0.24 & \cellcolor{green1!63}0.21 & \cellcolor{green1!43}0.27 & \cellcolor{green1!62}0.19\\

GAMA & \cellcolor{green1!42}0.30 & \cellcolor{green1!34}0.31 & \cellcolor{green1!42}0.30 & \cellcolor{green1!37}0.31 & \cellcolor{green1!37}0.27 & \cellcolor{green1!26}0.39 & \cellcolor{green1!40}0.30 & \cellcolor{green1!41}0.28 & \cellcolor{green1!33}0.33 & \cellcolor{green1!1}0.62 & \cellcolor{green1!12}0.44 & \cellcolor{green1!34}0.32 & \cellcolor{green1!31}0.34\\

Audio Flamingo 2 Sound-CoT & \cellcolor{green1!26}0.39 & \cellcolor{green1!9}0.52 & \cellcolor{green1!19}0.45 & \cellcolor{green1!26}0.36 & 0.51 & \cellcolor{green1!3}0.50 & 0.51 & \cellcolor{green1!1}0.46 & 0.50 & \cellcolor{green1!3}0.58 & \cellcolor{green1!9}0.49 & \cellcolor{green1!2}0.49 & \cellcolor{green1!8}0.48\\

Audio Flamingo 3 (think mode) & 0.51 & \cellcolor{green1!3}0.49 & \cellcolor{green1!2}0.51 & 0.50 & \cellcolor{green1!2}0.49 & \cellcolor{green1!1}0.48 & \cellcolor{green1!3}0.49 & \cellcolor{green1!4}0.48 & 0.50 & 0.50 & 0.51 & \cellcolor{green1!5}0.45 & \cellcolor{green1!2}0.49\\

R1-AQA (think mode) & 0.58 & \cellcolor{green1!14}0.48 & 0.47 & \cellcolor{green1!2}0.51 & 0.50 & 0.51 & \cellcolor{green1!1}\underline{0.52} & \cellcolor{green1!3}0.47 & 0.52 & \cellcolor{green1!3}0.77 & \cellcolor{green1!1}0.61 & \cellcolor{green1!30}0.37 & \cellcolor{green1!4}0.52\\

Step-Audio 2 mini Think& \cellcolor{green1!5}0.50 &\cellcolor{green1!5}0.53 &\cellcolor{green1!5}0.52 	&\cellcolor{green1!2}0.47 	&\cellcolor{green1!5}0.53 	&\cellcolor{green1!3}0.52 &	\cellcolor{green1!6}0.50 	&\cellcolor{green1!6}0.45 	&\cellcolor{green1!2}0.51 	&\cellcolor{green1!4}0.81 	&\cellcolor{green1!3}0.57 	&\cellcolor{green1!12}0.50 	&\cellcolor{green1!4}0.53 \\

Audio-Reasoner & \cellcolor{green1!3}0.67 & \cellcolor{green1!1}0.73 & \cellcolor{green1!1}0.56 & \cellcolor{green1!1}0.54 & \cellcolor{green1!3}0.55 & \cellcolor{green1!1}0.49 & \cellcolor{green1!4}0.48 & \cellcolor{green1!4}0.46 & \cellcolor{green1!3}0.50 & \cellcolor{green1!1}0.72 & \cellcolor{green1!2}0.54 & \cellcolor{green1!29}0.41 & \cellcolor{green1!4}0.55\\

Step-Audio-R1	&0.64	&0.67	&0.54&	0.50	&0.50	&0.57	&0.48	&\textbf{0.60}	&0.70	&0.96	&0.60	&\underline{0.77}	&0.63\\

MiMo-Audio (think mode)&0.81 & 	0.75&  	\cellcolor{green1!1}0.60 & 	\cellcolor{green1!2}0.48 	& 0.69 & 0.53 & 	0.50 & 	0.53 & 	\cellcolor{green1!1}0.67 	& 0.80 	& 0.62 & 	\cellcolor{green1!1}0.73 & 0.64 \\

\midrule \midrule

\multicolumn{14}{c}{\textbf{Omni Language Models (OLMs)}}\\ 
\midrule \midrule
VITA-1.5 & 0.51 & 0.51 & 0.52 & 0.49 & 0.51 & 0.51 & 0.49 & 0.46 & 0.55 & 0.48 & 0.44 & 0.46 & 0.49\\

OpenOmni & 0.47 & 0.48 & 0.51 & 0.51 & 0.52 & 0.50 & 0.51 & 0.47 & 0.53 & 0.51 & 0.54 & 0.51 & 0.50\\

Baichuan-Omni-1.5 & 0.57 & 0.47 & 0.53 & 0.52 & 0.48 & 0.50 & 0.50 & 0.42 & 0.49 & 0.71 & 0.51 & 0.56 & 0.52\\

HumanOmni & 0.55 & 0.61 & 0.52 & 0.49 & 0.50 & 0.50 & \underline{0.52} & 0.42 & 0.49 & 0.86 & 0.57 & 0.59 & 0.55\\

Ming-Lite-Omni-1.5	&0.57 &	0.52 &	0.53 &	0.50 &	0.53 	&0.55 &	0.51 &	0.47 &	0.53 &	0.94 &	0.57 &	0.65 	&0.57 \\

Ola & 0.74 & 0.73 & 0.64 & 0.48 & 0.56 & 0.52 & 0.49 & 0.47 & 0.58 & \underline{0.97} & 0.53 & 0.58 & 0.60\\

Qwen2.5-Omni & 0.71 & \underline{0.87} & \textbf{0.75} & 0.53 & 0.57 & 0.52 & 0.50 & 0.52 & 0.56 & \textbf{0.99} &\textbf{0.75} & 0.58 & 0.65\\

Qwen3-Omni-Instruct&	\textbf{0.87} &	\textbf{0.88} &	\textbf{0.75} &	\textbf{0.57} &	\textbf{0.77} 	&\textbf{0.65} 	&0.51 	&0.51 &	0.68 &	\textbf{0.99} &	\underline{0.73} &	\textbf{0.82} &	\textbf{0.72} \\

\cdashline{1-14}
\noalign{\vskip 0.5mm}
Gemini-2.5-Flash &0.77 	&0.79 &	\cellcolor{green1!1}0.63 	&\cellcolor{green1!3}\textbf{0.57} &	0.68 &	\underline{0.59} &	0.49 &	0.50 &	\underline{0.72} &	0.94 &	0.62 &	0.71 &	\underline{0.67} \\
\bottomrule \bottomrule
\end{tabular}
\end{adjustbox}
\vspace{2pt}
\begin{tikzpicture}[baseline]
  \fill[left color=green1!0, right color=green1!100]
       (0,0) rectangle (\dimexpr0.7\linewidth\relax, 1.8ex);
  \node[anchor=west,font=\small\bfseries] at (-2cm,0.8ex) {\textit{\%abstention}};
\end{tikzpicture}
\\[-3ex]
\caption{\small \textbf{Overall SonicBench performance Across Model Categories.} Average accuracies of all evaluated models on our benchmark, grouped into three categories, LALMs, LARMs, and OLMs. Each value represents a model’s overall accuracy aggregated across all attributes and tasks. We sorted the open-source models in each category in ascending order based on overall accuracy. The best-performing models for each attribute are \textbf{bolded}, and the second-best are \underline{underlined}. Cell colors indicate the abstention rate, with lighter shades representing lower abstention, as shown in the accompanying color bar.}
\label{tab:overall_acc}
\end{table*}

\section{Experiment}

\subsection{Models}
\label{subset:models}
We evaluate three categories of audio-capable models: \textit{(i)} Large Audio Language Models, designed for audio-text understanding; \textit{(ii)} Large Audio Reasoning Models, which enhance LALMs with explicit reasoning chains; \textit{(iii)} Omni Language Models, supporting fully multimodal input or output; Further details and model configurations are provided in Appendix~\ref{appendix:benchmarking_candidates}.

\subsection{Experiment Settings}
We tailor the prompt using prefixes and suffixes specific to each model for all open-sourced models. When it comes to reasoning models we use their official template and special tags.
Greedy decoding is performed during generation for all open-source models with setting input token length limit to 2048 and an output token length to 1024. All models are evaluated under zero-shot settings, and the instruction prompts are presented in Appendix~\ref{appendix:task_instruction}.

We additionally evaluate human performance on our task. Specifically, we sample 10\% of the data from each task-attribute cell to construct a set of 240 test examples, ensuring that the sampled set has a balanced distribution of correct answers (50\% A and 50\% B). Three human participants independently listened to each clip three times and recorded their responses. On average, each participant spent about three hours completing the evaluation.

\subsection{Evaluation}
\paragraph{Answer extraction.}
We instruct models to respond strictly with either option ``A'' or ``B''. However, some models fail to comply with this instruction and produce longer outputs. To handle such cases more flexibly, we use a regular expression to match the final option from the model output whenever possible. Refer to Appendix~\ref{app:answer_extraction} for the detailed regular expression used for answer extraction.

\paragraph{Metrics.}

We report two metrics: accuracy and abstention rate. Accuracy is measured by exact match between the answer extracted from the model output and the ground truth. If no valid option (“A” or “B”) can be extracted using our regular expression, we consider the model to have abstained. Abstentions are treated as incorrect, and the abstention rate captures how often the model fails to produce a valid choice, reflecting its robustness in adhering to task instructions.

\subsection{Main Results}
\label{subsec:main_results}

\paragraph{Clear headroom even for SOTA models.}
The overall model performance and ranking on SonicBench are presented in Figure~\ref{fig:overall_performance_rank}. Human participants achieve an average accuracy of 91\%. In contrast, the best-performing model, Qwen3-Omni, attains 72\%, and about half of the models score close to the random baseline. The four lowest-ranked models exhibit particularly low accuracy due to their high abstention rates. These results suggest that, while humans solve these tasks reliably, current models still have substantial room for improvement.

Table~\ref{tab:overall_acc} provides a detailed performance breakdown across attributes (task-wise results in App.~\ref{appendix:detailed_breakdown}). Significant variance is observed in model capabilities across individual attributes. For instance, while Kimi-Audio (the best performing LALM) excels in Pitch and Brightness, its performance is at near-random levels on Velocity, Direction, and Distance. These attributes appear to be the most challenging for all evaluated models.  

\paragraph{Current models do not exploit direct comparisons as humans do.}
Table~\ref{tab:comp-vs-rec-deltas} contrasts performance on comparison and recognition tasks (details in Appendix~\ref{appendix:Tasks_comparison_vs_recognition}). For humans, we observe a clear pattern: comparison tasks are generally easier to solve than recognition tasks. This is expected, since estimating an absolute attribute such as exact loudness can be less reliable without prior calibration, whereas deciding which of two sounds is louder is immediately intuitive for humans.

In contrast, this pattern does not consistently appear in the models we evaluated. For several attributes, including pitch, duration, and reverberation, the models exhibit a noticeable drop in performance on comparison tasks. This suggests that the internal mechanisms of these models may differ substantially from those of humans. One possible explanation is that, although the models may be trained on recognition tasks for specific attributes, the knowledge they acquire does not effectively transfer to the corresponding comparison tasks.

\begin{table}[h]
\centering
\footnotesize
\setlength{\tabcolsep}{3pt}
\renewcommand{\arraystretch}{1.1}
\begin{adjustbox}{max width=0.98\columnwidth}
\begin{tabular}{lllll}
\toprule
\textbf{Tasks} & \multicolumn{4}{c}{\textbf{Comparison \,|\, Recognition}}  \\
\cmidrule(lr){1-1}\cmidrule(lr){2-5}
\textbf{Attributes} & \textbf{\quad\;\;Human} & \textbf{Qwen3-Omni} & \textbf{Kimi-Audio} & \textbf{GPT-4o-Audio} \\
\midrule
Pitch         & 1.00\,|\,0.87\deltaup{+14.9\%} & 0.83\,|\,0.90\deltadn{$-$7.8\%}  & 0.79\,|\,0.87\deltadn{$-$9.2\%}   & 0.71\,|\,0.71\deltane{0.0\%} \\
Brightness    & 0.93\,|\,0.93\deltane{+0.0\%} & 0.92\,|\,0.83\deltaup{+10.8\%}  & 0.81\,|\,0.81\deltane{0.0\%}      & 0.68\,|\,0.78\deltadn{$-$12.8\%} \\
Loudness      & 0.97\,|\,0.77\deltaup{+26.0\%} & 0.80\,|\,0.70\deltaup{+14.3\%}  & 0.66\,|\,0.70\deltadn{$-$5.7\%}   & 0.55\,|\,0.49\deltaup{+12.2\%} \\
Velocity      & 0.83\,|\,0.83\deltane{+0.0\%} & 0.64\,|\,0.49\deltaup{+30.6\%}  & 0.53\,|\,0.50\deltaup{+6.0\%}     & 0.55\,|\,0.52\deltaup{+5.8\%} \\
Duration      & 1.00\,|\,0.83\deltaup{+20.5\%} & 0.73\,|\,0.80\deltadn{$-$8.8\%}  & 0.64\,|\,0.75\deltadn{$-$14.7\%}  & 0.44\,|\,0.54\deltadn{$-$18.5\%} \\
Tempo         & 0.97\,|\,0.77\deltaup{+26.0\%} & 0.66\,|\,0.64\deltane{+3.1\%}   & 0.58\,|\,0.55\deltaup{+5.5\%}     & 0.49\,|\,0.51\deltane{$-$3.9\%} \\
Direction     & 0.83\,|\,0.83\deltane{+0.0\%} & 0.52\,|\,0.49\deltaup{+6.1\%}   & 0.43\,|\,0.44\deltane{$-$2.3\%}   & 0.49\,|\,0.51\deltane{$-$3.9\%} \\
Distance      & 0.87\,|\,0.73\deltaup{+19.2\%} & 0.60\,|\,0.42\deltaup{+42.9\%}  & 0.39\,|\,0.48\deltadn{$-$18.8\%}  & 0.56\,|\,0.54\deltane{+3.7\%} \\
Reverberation & 1.00\,|\,1.00\deltane{+0.0\%} & 0.52\,|\,0.83\deltadn{$-$37.3\%} & 0.58\,|\,0.95\deltadn{$-$38.9\%}  & 0.50\,|\,0.50\deltane{0.0\%} \\
Texture       & 1.00\,|\,1.00\deltane{+0.0\%} & 0.99\,|\,0.98\deltane{+1.0\%}    & 0.95\,|\,0.99\deltane{$-$4.0\%}   & 0.88\,|\,0.92\deltane{$-$4.3\%} \\
Timbre        & 1.00\,|\,0.93\deltaup{+7.5\%} & 0.70\,|\,0.75\deltadn{$-$6.7\%}  & 0.49\,|\,0.68\deltadn{$-$27.9\%}  & 0.43\,|\,0.49\deltadn{$-$12.2\%} \\
Counting      & 1.00\,|\,1.00\deltane{+0.0\%} & 0.65\,|\,0.99\deltadn{$-$34.3\%} & 0.79\,|\,0.69\deltaup{+14.5\%}    & 0.66\,|\,0.68\deltane{$-$2.9\%} \\
\bottomrule
\end{tabular}
\end{adjustbox}
\caption{\small \textbf{Comparison vs. Recognition accuracy.} Cells display mean accuracy (Comparison $|$ Recognition) and relative change ($\Delta\%$). Colors denote deltas beyond $\pm 5\%$ \deltalegend. Unlike humans, models show no consistent advantage in comparison tasks. Full human results are in Table~\ref{tab:human-per-subject}.}
\label{tab:comp-vs-rec-deltas}
\end{table}

\paragraph{Inference-time scaling does not always improve performance.}
\label{finding:3}

Our evaluation included three models that offer a dedicated reasoning (think) mode: Audio Flamingo 3, MiMo-Audio, and R1-AQA. Although these models report improved performance on their downstream tasks when reasoning is enabled, this trend does not consistently extend to SonicBench. As shown in Table~\ref{tab:overall_acc}, only MiMo-Audio improves noticeably with reasoning enabled, while the other two models show marginal gains. We hypothesize that explicit reasoning is beneficial only when the model already possesses non-trivial task competence; for most SonicBench tasks, these models may lack sufficient baseline capability to benefit substantially.\footnote{An exception is MiMo-Audio on the counting task with reasoning disabled, where the model shows a systematic preference for Option~A (>90\% of cases), even when the option order is reversed. Reasoning mitigates this behavior.}

\begin{figure}[h]
    \centering
    \includegraphics[width=1.0\linewidth]{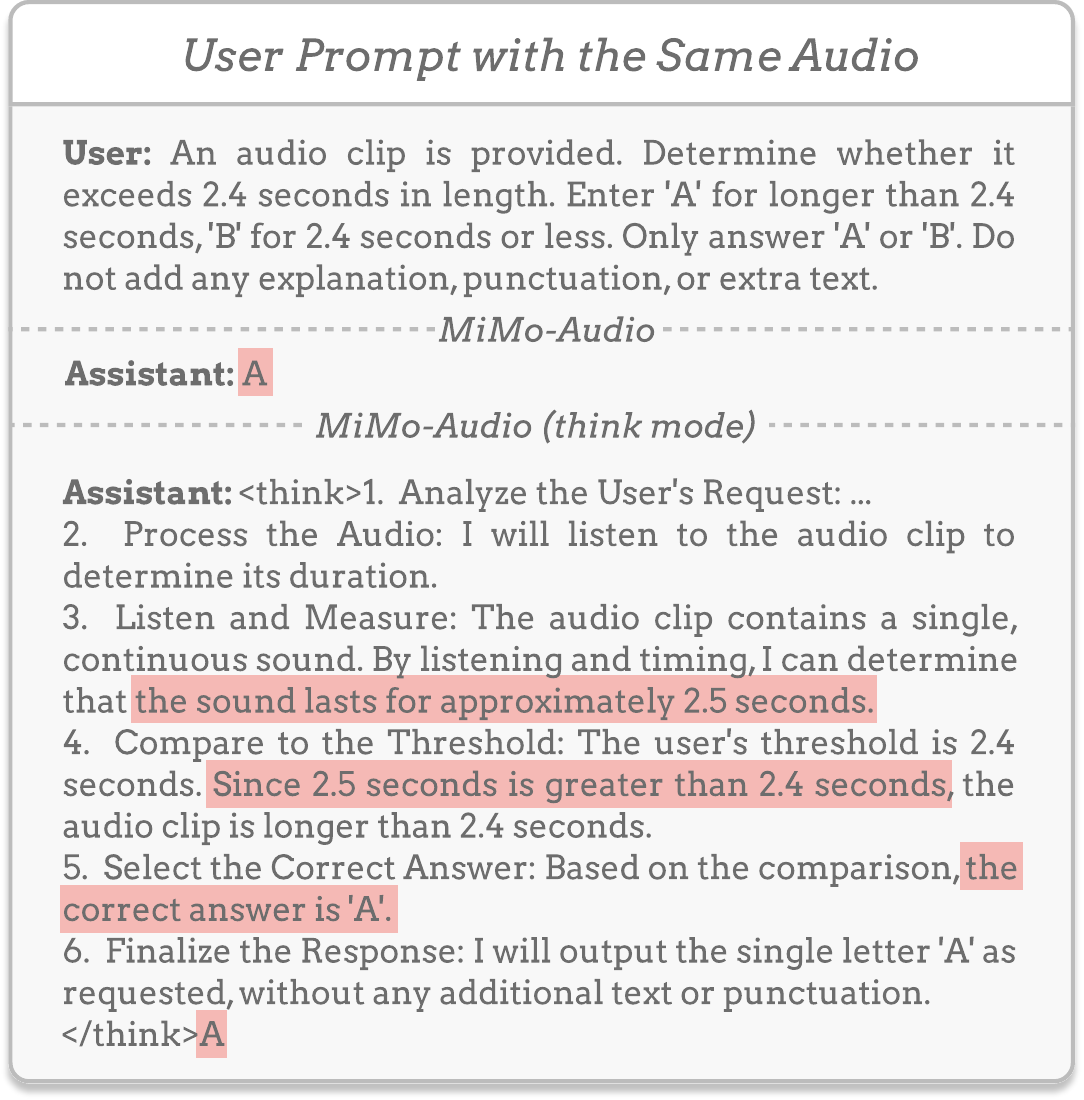}
    \caption{\small \textbf{A Case of reasoning cannot correcting perceptual failures.} Both receive the same duration recognition prompt. The model in base mode directly answers wrong, while in think mode produces a logically coherent reasoning but fails to rectify upstream perceptual errors. A case exemplifies reasoning-induced perceptual errors in Figure~\ref{fig:case_study}.}
    \label{fig:case_study2}
\end{figure}

In Figure~\ref{fig:case_study2}, we provide an example comparing MiMo-Audio with and without reasoning. Although the generated CoT is internally coherent, it is grounded in an incorrect acoustic perception, which ultimately leads to an incorrect answer. This example shows that when perceptual errors occur upstream, explicit reasoning may amplify rather than correct them, offering limited benefit.

\subsection{Attribute Difficulties}
\label{subsec:attribute_difficulties}
\begin{figure}[h]
    \centering
    \includegraphics[width=0.98\linewidth]{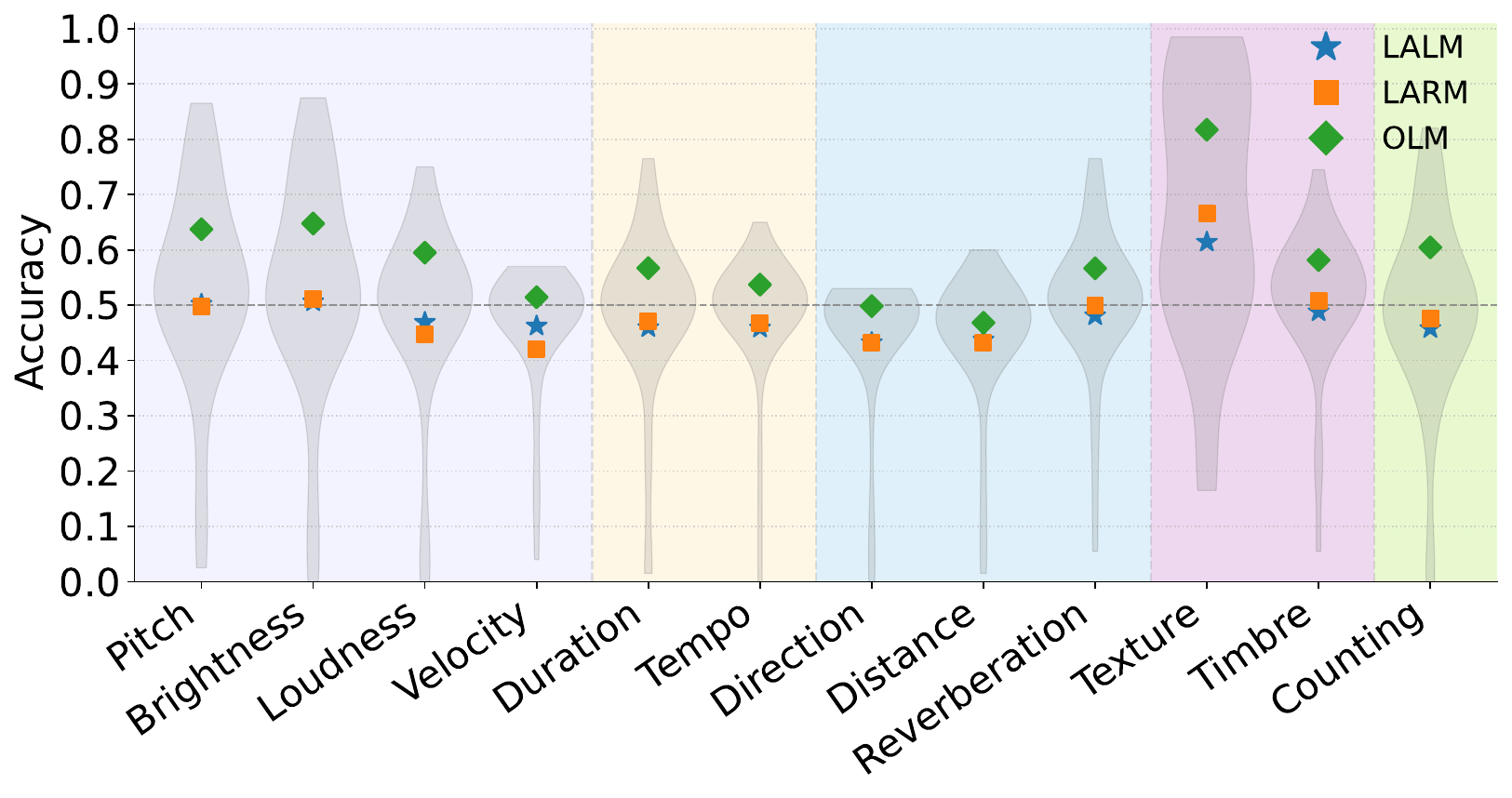}
    \caption{\small Attribute-level overall accuracy distribution across model families. Violin plots show per-attribute accuracy for all models, with background colors denoting attribute dimensions and markers indicating family means. Velocity, Direction, and Distance consistently exhibit near-chance performance (no model exceeds 0.6 accuracy), indicating persistent difficulties in capturing spatial and motion-related cues.}
    \label{fig:violin_attri}
\end{figure}

Figure~\ref{fig:violin_attri} illustrates the distribution of model performance across attributes. As shown, contemporary models exhibit substantial performance variability across all attributes. For most attributes, the best-performing systems achieve accuracies exceeding 70\%. In contrast, Velocity, Direction, and Distance remain the most challenging, with no model surpassing an accuracy of 60\%. Performance on Tempo is also notably weak. These results highlight persistent difficulties in effectively modeling temporal and especially spatial acoustic cues.
\begin{figure}[h]
    \centering
    \includegraphics[width=1.05\linewidth]{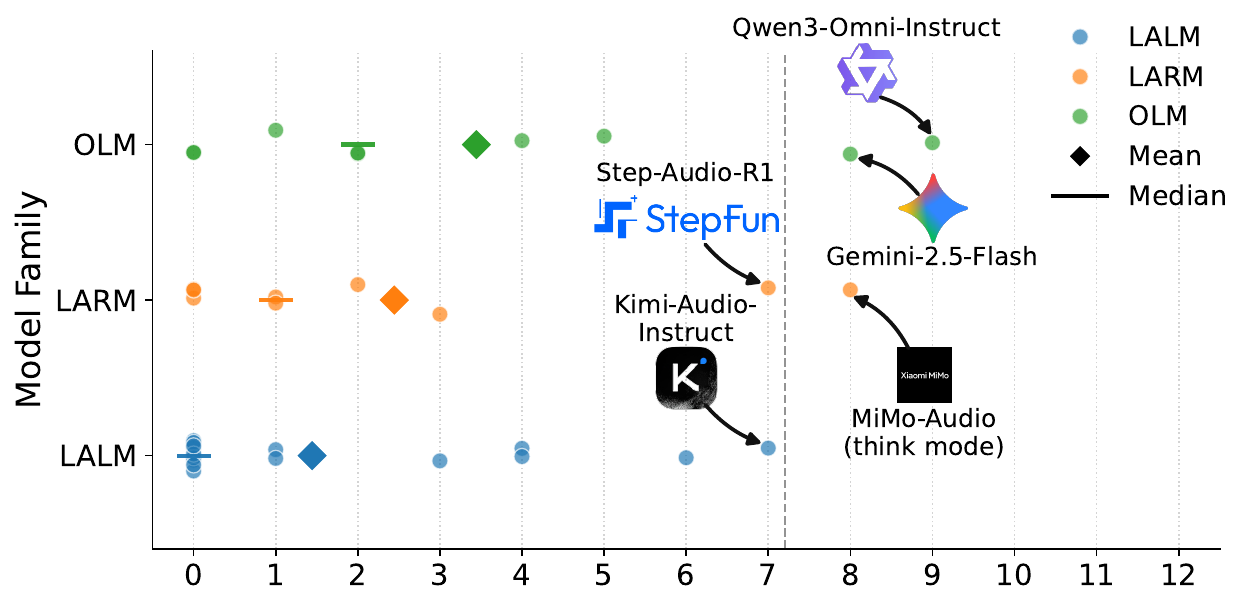}
    \caption{\small \textbf{Attribute coverage ($\geq$0.60 accuracy) across models.} Each point represents the number of attributes where a model achieves $\geq$0.60 accuracy. Colors denote families; diamonds and bars indicate family means and medians. The vertical dashed line at 7.2 attributes (60\% coverage in 12 attributes) marks the threshold for sufficient perceptual competence.}
    \label{fig:family_beeswarm_count_both_acc}
\end{figure}

We further summarize each model’s attribute coverage at $\geq$ 0.60 accuracy and present it in Figure~\ref{fig:family_beeswarm_count_both_acc}.
Within each family, the well-performing OLMs demonstrate the broadest coverage, followed by LARMs, while LALMs handle the fewest attributes.
Across all models, Qwen3-Omni performs best, successfully handling 9 attributes. It is followed by Gemini-2.5-Flash and MiMo-Audio (think mode), each covering 8 attributes. 
However, none of LALMs reaches the accuracy 7.2-attribute threshold, corresponding to 60\% coverage of all 12 attributes, a
reference point for decent perceptual competence. Even Kimi-Audio that can handle the most attributes is still below the threshold.  

\subsection{Where is the Bottleneck? A Probing Study}
\label{subsec:probe}
In this section, we aim to identify the key bottlenecks limiting model performance on SonicBench. Specifically, we examine whether the near-chance performance observed for certain attributes arises from deficiencies in perception, meaning the model fails to encode the relevant information, or from limitations in the decoding stage, where it cannot effectively use the encoded representations.
\begin{figure*}[h]
    \centering
    \includegraphics[width=0.95\linewidth]{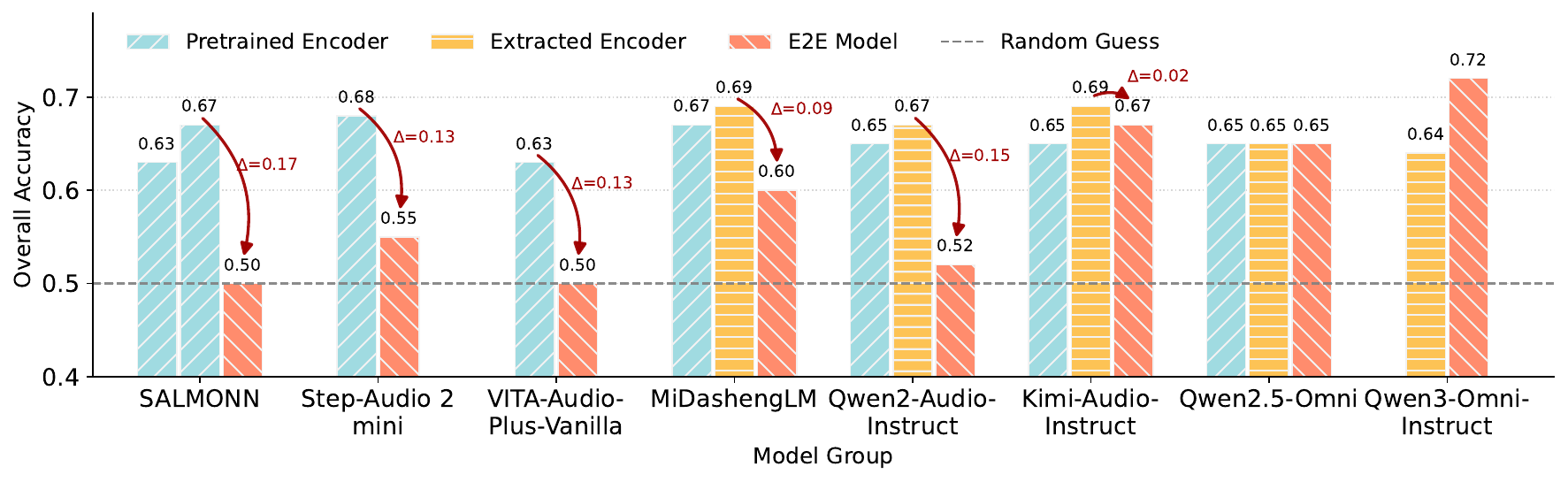}
    \caption{\small \textbf{Probe vs. E2E accuracy.} Linear probes on encoders (pretrained/extracted) consistently outperform full E2E models, with all encoders achieving $\geq 0.60$. Significant drops $\Delta$ indicate the bottleneck lies in alignment or decoding rather than perceptual encoding.}
    \label{fig:encoder_vs_model_overall_acc}
\end{figure*}

\paragraph{Setup.} Most contemporary LALMs comprise three components: an audio encoder, a projection module, and a LLM. Audio inputs are first processed by the encoder to produce intermediate representations. To assess the information captured at this perceptual stage, we apply linear probing to these representations. Specifically, we select eight E2E models from different model families, including SALMONN, Step-Audio2, VITA-Audio, MiDashengLM, Qwen2-Audio, Kimi-Audio, Qwen2.5-Omni and Qwen3-Omni, and extract their audio encoders. We attach a two-layer lightweight linear probe to the final encoder output and train only this probe, while keeping the encoder parameters frozen. Probes are trained independently for each attribute and each task (recognition vs.\ comparison) using a 50\%/50\% train-evaluation split of each JSON file. This setup results in 24 experiments in total, corresponding to 12 attributes across two task types.
We provide experiment details including model modules, encoder freeze/unfreeze status, probe architectures, and hyperparameters in Appendix~\ref{app:probe_details}.

\paragraph{Results.}
Across all eight systems, the frozen pre-trained encoder probes achieve overall accuracy $\geq 0.60$ and \emph{consistently outperform} their corresponding E2E models, which mainly cluster around $0.50$. Moreover, we observe that E2E training with an unfrozen audio encoder can yield small gains, but these improvements are modest relative to the probe-model gap. Notable exceptions are Qwen2.5-Omni and Qwen3-Omni, whose E2E model reaches and surpasses its probe, likely due to their training strategy that prevents the encoder from compensating for a frozen LLM, thereby avoiding perceptual degradation. All are presented in Figure~\ref{fig:encoder_vs_model_overall_acc} and Table~\ref{tab:model_group_accuracies}. 
These demonstrate that even though primarily pretrained in linguistic and paralinguistic tasks, task-relevant, signal-level cues are present in encoder outputs; weak E2E performance thus points to the latter modules.

On closer inspection, we find attributes including Pitch, Brightness, Loudness, Velocity, Duration, Direction, and Reverberation are typically $\geq 0.60$ at the encoder level with some approaching $0.9$, yet still degrade toward $\approx 0.5$ in E2E models.
For the other attributes, we find that attribute-wise patterns are coherent across encoders. A common performance emerges in pre-trained encoders across attributes including Timbre, Texture, Counting, Tempo, and Distance where probe accuracies in recognition task often fall below $0.60$. 
In addition, comparing pre-trained encoders and extracted encoders from E2E models trained with unfrozen encoder modules, we can tell that 
unfreezing the encoder helps the Tempo and Distance attributes more noticeably, yet brings limited gains for Timbre, Texture, and Counting. 
Conversely, E2E models exhibit the opposite, improvements on Timbre, Texture, and Counting coincide with drops on Tempo and Distance as depicted in Table~\ref{tab:comp-rec-across-groups}(b).

Across tasks, comparison remains at or below recognition even in the probe setting. This suggests a shared relational bottleneck: lacking mechanisms to explicitly compare segments within audio, both probes and full models struggle to leverage contrastive structure as presented in Table~\ref{tab:probe_full_acc_with_dimension_allmodels}.

\section{Conclusion}
We introduce SonicBench to evaluate the foundational physical perception capabilities of LALMs. By testing twelve psychophysically grounded attributes through recognition and comparison tasks, we revealed a critical gap: current models, despite their semantic fluency, struggle to perceive basic physical properties such as pitch, loudness, and spatial direction, and fail to exhibit the relational reasoning advantages observed in human listeners. Crucially, our linear probing analysis demonstrates that this deficiency does not stem from the audio encoders, which successfully capture these physical cues, but rather from the alignment and decoding stages where these signals are lost or misinterpreted. These findings imply that optimizing alignment and decoding is critical to unlock the full potential of current encoders. Ultimately, SonicBench aims to foster physically grounded LALMs essential for robust real-world interaction.

\clearpage

\paragraph{Limitations.}

First, regarding probing methodology, we employed lightweight linear classifiers on 50\% splits. While non-linear probes or larger training data might yield higher absolute accuracies, the substantial gap between these simple probes and full E2E models is sufficient to validate our core conclusion that the bottleneck lies in alignment and decoding, not in the encoder's raw accessibility.

Second, concerning data diversity, SonicBench prioritizes datasets with verifiable ground-truth labels, e.g., measured impulse responses over unconstrained ``in-the-wild'' recordings. While this ensures physical precision, it limits acoustic variability compared to web-scale data. However, precise physical annotation for general audio is currently cost-prohibitive and prone to noise. We posit that mastering these canonical physical properties is a necessary prerequisite for generalizability, and hope our findings can serve as a foundational step to inspire future scaling of physical attribute annotations.

Third, regarding linguistic coverage, SonicBench utilizes exclusively English text instructions. Given the known sensitivity of LALMs to textual phrasing and prompt languages, our current evaluation does not account for potential performance fluctuations in multilingual contexts. Future work should investigate whether physical perception remains robust across different languages to ensure true language-agnostic grounding.

\bibliography{custom}

\begin{thebibliography}{119}
\providecommand{\natexlab}[1]{#1}

\bibitem[{ebu(2014)}]{ebur128}
 2014.
\newblock Loudness normalisation and permitted maximum level of audio signals.

\bibitem[{itu(2015)}]{iturbs1770}
 2015.
\newblock Algorithms to measure audio programme loudness and true-peak audio level.

\bibitem[{Adavanne et~al.(2019)Adavanne, Politis, and Virtanen}]{dcase2019}
Sharath Adavanne, Archontis Politis, and Tuomas Virtanen. 2019.
\newblock A multi-room reverberant dataset for sound event localization and detection.
\newblock In \emph{Proceedings of the Detection and Classification of Acoustic Scenes and Events 2019 Workshop (DCASE2019)}, pages 10--14, New York University, NY, USA.

\bibitem[{AI et~al.(2025{\natexlab{a}})AI, :, Ma, Zou, Yan, Jin, Shen, Zheng, Wang, Xu, Yao, Zhou, Chen, Li, Sun, Liu, Zhu, Jiang, Peng, Ji, Ren, Wang, Ru, Tan, Wang, Bai, Gao, Guo, Zhang, Xu, Liu, Xiong, Zheng, Gao, Li, Liu, Chai, Xiao, Wang, Wang, Lu, Li, Dong, Yu, Yuan, Gao, Xiao, Sun, Chen, Mao, Wu, Lyu, Ma, Fang, Qiu, Huang, Yang, and He}]{ai2025mingflashomni}
Inclusion AI, :, Bowen Ma, Cheng Zou, Canxiang Yan, Chunxiang Jin, Chunjie Shen, Dandan Zheng, Fudong Wang, Furong Xu, GuangMing Yao, Jun Zhou, Jingdong Chen, Jianing Li, Jianxin Sun, Jiajia Liu, Jianjiang Zhu, Jianping Jiang, Jun Peng, and 39 others. 2025{\natexlab{a}}.
\newblock \href {https://arxiv.org/abs/2510.24821} {Ming-flash-omni: A sparse, unified architecture for multimodal perception and generation}.
\newblock \emph{Preprint}, arXiv:2510.24821.

\bibitem[{AI et~al.(2025{\natexlab{b}})AI, Gong, Zou, Zheng, Zhou, Yan, Jin, Shen, Zheng, Wang, Xu, Yao, Zhou, Chen, Sun, Liu, Zhu, Peng, Ji, Song, Ren, Wang, Ru, Xie, Tan, Xue, Wang, Bai, Gao, Chen, Guo, Zhang, Xu, Liu, Xiong, Gao, Liu, Li, Chai, Xiao, Wang, Chen, Lu, Li, Dong, Yu, Yuan, Gao, Sun, Chen, Wu, Lyu, Ma, Feng, Fang, Qiu, Huang, and He}]{ai2025mingomniunifiedmultimodalmodel}
Inclusion AI, Biao Gong, Cheng Zou, Chuanyang Zheng, Chunluan Zhou, Canxiang Yan, Chunxiang Jin, Chunjie Shen, Dandan Zheng, Fudong Wang, Furong Xu, GuangMing Yao, Jun Zhou, Jingdong Chen, Jianxin Sun, Jiajia Liu, Jianjiang Zhu, Jun Peng, Kaixiang Ji, and 39 others. 2025{\natexlab{b}}.
\newblock \href {https://arxiv.org/abs/2506.09344} {Ming-omni: A unified multimodal model for perception and generation}.
\newblock \emph{Preprint}, arXiv:2506.09344.

\bibitem[{Almeida et~al.(2017)Almeida, Schubert, Smith, and Wolfe}]{almeida2017brightness}
Andre Almeida, Emery Schubert, John Smith, and Joe Wolfe. 2017.
\newblock Brightness scaling of periodic tones.
\newblock \emph{Attention, Perception, \& Psychophysics}, 79(7):1892--1896.

\bibitem[{Amengual~Garí et~al.(2020)Amengual~Garí, Sahin, Eddy, and Kob}]{srirs}
Sebastià Amengual~Garí, Banu Sahin, Dustin Eddy, and Malte Kob. 2020.
\newblock \href {https://doi.org/10.5281/zenodo.4007387} {Open database of spatial room impulse responses at detmold university of music}.

\bibitem[{An et~al.(2024)An, Chen, Deng, Du, Gao, Gao, Gu, He, Hu, Hu, Ji, Li, Li, Lu, Luo, Lv, Ma, Ma, Ni, Song, Shi, Shi, Wang, Wang, Wang, Xiao, Yan, Yang, Zhang, Zhang, Zhang, Zhao, and Zheng}]{an2024funaudiollm}
Keyu An, Qian Chen, Chong Deng, Zhihao Du, Changfeng Gao, Zhifu Gao, Yue Gu, Ting He, Hangrui Hu, Kai Hu, Shengpeng Ji, Yabin Li, Zerui Li, Heng Lu, Haoneng Luo, Xiang Lv, Bin Ma, Ziyang Ma, Chongjia Ni, and 14 others. 2024.
\newblock \href {https://arxiv.org/abs/2407.04051} {Funaudiollm: Voice understanding and generation foundation models for natural interaction between humans and llms}.
\newblock \emph{Preprint}, arXiv:2407.04051.

\bibitem[{Anobile et~al.(2021)Anobile, Castaldi, Maldonado~Moscoso, Arrighi, and Burr}]{anobile2021groupitizing}
Giovanni Anobile, Elisa Castaldi, Paula~A Maldonado~Moscoso, Roberto Arrighi, and David Burr. 2021.
\newblock Groupitizing improves estimation of numerosity of auditory sequences.
\newblock \emph{Frontiers in Human Neuroscience}, 15:687321.

\bibitem[{Bai et~al.(2023)Bai, Bai, Chu, Cui, Dang, Deng, Fan, Ge, Han, Huang, Hui, Ji, Li, Lin, Lin, Liu, Liu, Lu, Lu, Ma, Men, Ren, Ren, Tan, Tan, Tu, Wang, Wang, Wang, Wu, Xu, Xu, Yang, Yang, Yang, Yang, Yao, Yu, Yuan, Yuan, Zhang, Zhang, Zhang, Zhang, Zhou, Zhou, Zhou, and Zhu}]{qwen}
Jinze Bai, Shuai Bai, Yunfei Chu, Zeyu Cui, Kai Dang, Xiaodong Deng, Yang Fan, Wenbin Ge, Yu~Han, Fei Huang, Binyuan Hui, Luo Ji, Mei Li, Junyang Lin, Runji Lin, Dayiheng Liu, Gao Liu, Chengqiang Lu, Keming Lu, and 29 others. 2023.
\newblock Qwen technical report.
\newblock \emph{arXiv preprint arXiv:2309.16609}.

\bibitem[{Bai et~al.(2024)Bai, Liu, Wang, Shi, Wang, Plumbley, Gan, and Chen}]{bai2024audiosetcap}
Jisheng Bai, Haohe Liu, Mou Wang, Dongyuan Shi, Wenwu Wang, Mark~D. Plumbley, Woon-Seng Gan, and Jianfeng Chen. 2024.
\newblock \href {https://arxiv.org/abs/2411.18953} {Audiosetcaps: An enriched audio-caption dataset using automated generation pipeline with large audio and language models}.
\newblock \emph{Preprint}, arXiv:2411.18953.

\bibitem[{Bianchi et~al.(2016)Bianchi, Santurette, Wendt, and Dau}]{bianchi2016pitch}
Federica Bianchi, S{\'e}bastien Santurette, Dorothea Wendt, and Torsten Dau. 2016.
\newblock Pitch discrimination in musicians and non-musicians: Effects of harmonic resolvability and processing effort.
\newblock \emph{Journal of the Association for Research in Otolaryngology}, 17(1):69--79.

\bibitem[{Blauert and Hearing(1997)}]{blauert1997psychophysics}
Jens Blauert and Spatial Hearing. 1997.
\newblock The psychophysics of human sound localization.
\newblock \emph{Spatial hearing}.

\bibitem[{Bogdanov et~al.(2019)Bogdanov, Won, Tovstogan, Porter, and Serra}]{bogdanov2019mtg}
Dmitry Bogdanov, Minz Won, Philip Tovstogan, Alastair Porter, and Xavier Serra. 2019.
\newblock \href {http://hdl.handle.net/10230/42015} {The mtg-jamendo dataset for automatic music tagging}.
\newblock In \emph{Machine Learning for Music Discovery Workshop, International Conference on Machine Learning (ICML 2019)}, Long Beach, CA, United States.

\bibitem[{Boltz(1998)}]{boltz1998tempo}
Marilyn~G Boltz. 1998.
\newblock Tempo discrimination of musical patterns: Effects due to pitch and rhythmic structure.
\newblock \emph{Perception \& Psychophysics}, 60(8):1357--1373.

\bibitem[{Bregman(1994)}]{bregman1994auditory}
Albert~S Bregman. 1994.
\newblock \emph{Auditory scene analysis: The perceptual organization of sound}.
\newblock MIT press.

\bibitem[{Chen et~al.(2022)Chen, Wu, Wang, Liu, Tompkins, Chen, and Wei}]{Chen2022beats}
Sanyuan Chen, Yu~Wu, Chengyi Wang, Shujie Liu, Daniel Tompkins, Zhuo Chen, and Furu Wei. 2022.
\newblock \href {https://arxiv.org/abs/2212.09058} {Beats: Audio pre-training with acoustic tokenizers}.

\bibitem[{Cheng et~al.(2025)Cheng, Xiao, Shao, Wang, Yang, Shen, Torr, Gu, and Xu}]{cheng2025jailbreak}
Hao Cheng, Erjia Xiao, Jing Shao, Yichi Wang, Le~Yang, Chao Shen, Philip Torr, Jindong Gu, and Renjing Xu. 2025.
\newblock Jailbreak-audiobench: In-depth evaluation and analysis of jailbreak threats for large audio language models.
\newblock \emph{arXiv preprint arXiv:2501.13772}.

\bibitem[{Chu et~al.(2024)Chu, Xu, Yang, Wei, Wei, Guo, Leng, Lv, He, Lin, Zhou, and Zhou}]{qwen2audio}
Yunfei Chu, Jin Xu, Qian Yang, Haojie Wei, Xipin Wei, Zhifang Guo, Yichong Leng, Yuanjun Lv, Jinzheng He, Junyang Lin, Chang Zhou, and Jingren Zhou. 2024.
\newblock \href {https://arxiv.org/abs/2407.10759} {Qwen2-audio technical report}.
\newblock \emph{Preprint}, arXiv:2407.10759.

\bibitem[{Comanici et~al.(2025)Comanici, Bieber, Schaekermann, Pasupat, Sachdeva, Dhillon, Blistein, Ram, Zhang, Rosen, Marris, Petulla, Gaffney, Aharoni, Lintz, Pais, Jacobsson, Szpektor, Jiang, Haridasan, Omran, Saunshi, Bahri, Mishra, Chu, Boyd, Hekman, Parisi, Zhang, Kawintiranon, Bedrax-Weiss, Wang, Xu, Purkiss, Mendlovic, Deutel, Nguyen, Langley, Korn, Rossazza, Ramé, Waghmare, Miller, Byrd, Sheshan, Hadsell, Bhardwaj, Janus, Rissa, Horgan, Abdagic, Belenki, Allingham, Singh, Guidroz, Srinivasan, Schmit, Chiafullo, Elisseeff, Jha, Kolhar, Berrada, Ding, Si, Mallick, Och, Erell, Ni, Latkar, Yang, Sirkovic, Feng, Leland, Hornung, Wu, Blundell, Alvari, Huang, Yip, Deur, Liu, Surita, Duque, Damen, Jia, Guez, Mircea, Sinha, Magni, Stradomski, Marian, Galić, Chen, Husain, Singhal, Grewe, Aubet, Song, Blanco, Rechis, Ho, Munoz, Zheng, Hamrick, Mather, Taitelbaum, Rutherford, Lei, Chen, Shukla, Moreira, Doi, Isik, Shabat, Rogozińska, Kolipaka, Chang, Vušak, Venkatachary, Noghabi, Bharti, Jun, Zaks, Green,
  Challagundla, Wong, Mohammad, Hirsch, Cheng, Naim, Proleev, Vincent, Singh, Krikun, Krishnan, Ghahramani, Atias, Aggarwal, Kirov, Vytiniotis, Koh, Chronopoulou, Dogra, Ion, Tyen, Lee, Weissenberger, Strohman, Balakrishna, Rae, Velic, de~Liedekerke, Elyada, Yuan, Liu, Shani, Kishchenko, Alessio, Li, Song, Kwei, Jankowski, Pappu, Namiki, Ma, Tripuraneni, Cherry, Ikonomidis, Ling, Ji, Westberg, Wright, Yu, Parkinson, Ramaswamy, Connor, Yeganeh, Grover, Kenwright, Litchev, Apps, Tomala, Halim, Castro-Ros, Li, Boral, Sho, Yarom, Malmi, Klinghoffer, Lin, Ansell, S, Zhao, Zuo, Santoro, Cheng, Demmessie, Liu, Brichtova, Culp, Braun, Graur, Ng, Mehta, Phillips, Sundberg, Godbole, Liu, Katariya, Rim, Seyedhosseini, Ammirati, Valfridsson, Malihi, Knight, Toor, Lampe, Ittycheriah, Chiang, Yeung, Fréchette, Rao, Wang, Srivastava, Zhang, Rhodes, Brand, Weesner, Figotin, Gimeno, Fellinger, Marcenac, Leal, Marcus, Cotruta, Cabrera, Luo, Garrette, Axelrod, Baltateanu, Barker, Chen, Toma, Ingram, Riesa, Kulkarni, Zhang,
  Liu, Wang, Polacek, Wu, Hui, Reyes, Su, Barnes, Malhi, Siddiqui, Feng, Damaschin, Pighin, Steiner, Yang, Boppana, Ivanov, Kandoor, Shah, Mujika, Huang, Choquette-Choo, Patel, Yu, Creswell, Jerry, Liu, Barros, Razeghi, Roy, Culliton, Xiong, Pan, Strohmann, Powell, Seal, DeCarlo, Shyam, Katircioglu, Wang, Hardin, Odisho, Broder, Chang, Nair, Shtefan, O'Brien, Agarwal, Potluri, Goyal, Jhindal, Thakur, Stuken, Lyon, Toutanova, Feng, Wu, Horn, Wang, Cullum, Taubman, Shrivastava, Shi, Tomlinson, Patel, Tu, Oflazer, Pongetti, Yang, Taïga, Perot, Pierse, Han, Drori, Iturrate, Chakrabarti, Yeung, Dopson, ting Chen, Kulshreshtha, Guo, Pham, Schuster, Chen, Polozov, Xing, Zhou, Kacham, Kukliansky, Miech, Yaroshenko, Chi, Douglas, Fei, Blondel, Myla, Madmoni, Wu, Keysers, Kjems, Albuquerque, Yu, D'sa, Plantan, Ionescu, Elias, Gupta, Vuyyuru, Alcober, Zhou, Ji, Hartmann, Puttagunta, Song, Amid, Stefanoiu, Lee, Pucciarelli, Wang, Raul, Petrov, Tian, Anklin, Nti, Gomes, Schumacher, Vesom, Panagopoulos, Bousmalis, Andor,
  Jacob, Zhang, Rosgen, Kecman, Tung, Belias, Goodman, Covington, Wieder, Saxena, Davoodi, Huang, Maddineni, Roulet, Campbell-Ajala, Sessa, Xintian, Wu, Lai, Collins, Haig, Sakenas, Xu, Giustina, Shafey, Charoenpanit, Garg, Ainslie, Severson, Arenas, Pathak, Rajayogam, Feng, Bakker, Li, Wichers, Rogers, Geng, Li, Jagerman, Jia, Olmert, Sharon, Mauger, Mariserla, Ma, Mohabey, Kim, Andreev, Pollom, Love, Jain, Agrawal, Schroecker, Fortin, Warmuth, Liu, Leach, Blok, Girirajan, Aharoni, Uria, Sozanschi, Goldberg, Ionita, Ribeiro, Zlocha, Birodkar, Lachgar, Yuan, Choudhury, Ginsberg, Zheng, Dibb, Graves, Lokhande, Rasskin, Muraru, Quick, Tata, Sermanet, Chawla, Karo, Wang, Zhang, Keller, Dragan, Su, Chou, Liu, Tao, Prabhakara, Wilson, Liu, Wang, Evans, Du, Castaño, Prasad, Mahdy, Gerlach, Reid, Kahn, Zait, Pillai, Ulrich, Wang, Wassenberg, Farkash, Yalasangi, Wang, Bauza, Bucher, Liu, Yan, Leung, Sindhwani, Barnes, Singh, Jurin, Chang, Bhumihar, Eiger, Citovsky, Withbroe, Li, Xue, Santo, Stoyanov, Raimond, Zheng,
  Gao, Listík, Kwasiborski, Saputro, Ozturel, Mallya, Majmundar, West, Caron, Wei, Castrejon, Vikram, Ramachandran, Dhawan, Park, Smoot, van~den Driessche, Blau, Malik, Liang, Hirsch, dos Santos, Weinstein, van~den Oord, Lall, FitzGerald, Jiang, Yang, Webster, Elqursh, Pope, Rotival, Raposo, Zhu, Dean, Alabed, Tran, Gupta, Gleicher, Austin, Rosseel, Umekar, Das, Sun, Chen, Misiunas, Zhou, Di, Loo, Newlan, Li, Ramasesh, Xu, Chen, Gandhe, Soricut, Gupta, Hu, El-Sayed, Garcia, Brusilovsky, Chen, Bolt, Huang, Gurney, Zhang, Pritzel, Wilkiewicz, Seybold, Shamanna, Fischer, Dean, Gill, Mcilroy, Bhowmick, Selier, Yang, Cheng, Magay, Tan, Varma, Walder, Kocisky, Nakashima, Natsev, Kwong, Gog, Zhang, Dieleman, Jimma, Ryabtsev, Brahma, Steiner, Du, Žužul, Žanić, Raghavachari, Gierke, Zheng, Petrova, Dauphin, Liu, Kessler, Hand, Duvarney, Kim, Lee, Hussenot, Hui, Smith, Jain, Xia, Tomar, Amiri, Phan, Fuchs, Weyand, Tomasev, Cordell, Liu, Mallinson, Joshi, Crawford, Suggala, Chien, Fernando, Sanchez-Vargas,
  Williams, Crone, Luo, Karpov, Shan, Thurk, Strudel, Voigtlaender, Patil, Dozat, Khodaei, Singla, Ambroszczyk, Wu, Chang, Roark, Hegde, Ding, Filos, Wu, Pinto, Liu, Khanna, Pandey, Mcloughlin, Li, Haves, Zhou, Buchatskaya, Leal, de~Boursac, Akazawa, Anderson, Chen, Somandepalli, Liang, Goenka, Winkler, Grushetsky, Ding, Smith, Ye, Pont-Tuset, Li, Li, Golany, Wegner, Jiang, Barak, Shangguan, Vértes, Wong, Bornschein, Tudor, Bevilacqua, Schaul, Rawat, Zhao, Axiotis, Meng, McLean, Lai, Beattie, Kushman, Liu, Kutzman, Lang, Ye, Netrapalli, Mishra, Khan, Goel, Willoughby, Tian, Zhuang, Chen, Tsai, Kementsietsidis, Khare, Keeling, Xu, Waters, Altché, Popat, Mittal, Saxton, Badawy, Mathieu, Zheng, Zhou, Ranka, Shin, Duan, Salimans, Mihailescu, Shaham, Chang, Assael, Dikkala, Izzard, Cohen-Addad, Graves, Feinberg, Chung, Strouse, Karmon, Sharifzadeh, Ashwood, Pham, Blanton, Vasiloff, Barber, Geller, Zhou, Zubach, Huang, Zhang, Gupta, Young, Proskurnia, Votel, Gabeur, Barcik, Tripathi, Yu, Yan, Changpinyo,
  Pavetić, Coyle, Fujii, Mendez, Zhou, Rajamani, Hechtman, Cao, Juan, Tan, Dalibard, Du, Clay, Yao, Jia, Vijaykumar, Zhou, Bai, Hung, Pecht, Todorov, Khadke, Gupta, Lahoti, Autef, Duddu, Lee-Thorp, Bykovsky, Misiunas, Flennerhag, Thangaraj, McGiffin, Nado, Kunesch, Noever, Hertz, Liang, Stone, Palmer, Daruki, Pramanik, Põder, Kyker, Khan, Sluzhaev, Ritter, Ruderman, Zhou, Nagpal, Vodrahalli, Necula, Barham, Pavlick, Hartford, Shafran, Zhao, Mikuła, Eccles, Shimokawa, Garg, Vilnis, Chen, Shumailov, Lee, Abdelhamed, Xie, Cohen, Hlavnova, Malkin, Sitawarin, Lottes, Coquinot, Yu, Kumar, Zhang, Mahendru, Ahmed, Martens, Chen, Boag, Peng, Devin, Klimovskiy, Phuong, Vainstein, Xie, Ramabhadran, Howard, Yu, Goswami, Cui, Shleifer, Pinto, Yeh, Yang, Javanmardi, Ethier, Lee, Orbay, Kotecha, Bromberg, Shaw, Thornton, Rosenthal, Gu, Thomas, Gemp, Ayyar, Ushio, Selvan, Wee, Liu, Majzoubi, Yu, Abernethy, Liechty, Pan, Nguyen, Qiong, Hu, Perrin, Arora, Pitler, Wang, Shivakumar, Prost, Limonchik, Wang, Gao, Cour, Buch,
  Gui, Ivanova, Neubeck, Chan, Kim, Chen, Goyal, Chung, Liu, Su, Petrushkina, Shen, Joulin, Xu, Lin, Kulizhskaya, Chelba, Vasudevan, Collins, Bashlovkina, Lu, Fritz, Park, Zhou, Su, Tanburn, Sushkov, Rasquinha, Li, Prendki, Li, LV, Sharma, Fitoussi, Huang, Dai, Dao, Burrows, Prior, Qin, Pundak, Sjoesund, Khurshudov, Zhu, Webson, Kemp, Tan, Agrawal, Sargsyan, Cheng, Stephan, Kwiatkowski, Reid, Byravan, Michaely, Heess, Zhou, Goenka, Carpenter, Levskaya, Wang, Roberts, Leblond, Chikkerur, Ginzburg, Chang, Riachi, Chuqiao, Xu, Borsos, Pliskin, Pawar, Lustman, Kirkwood, Anand, Chaudhary, Kalb, Milan, Augenstein, Goldie, Prince, Raman, Sun, Xia, Cohen, Huo, Camp, Ellis, Zilka, Torres, Patel, Arora, Chan, Adler, Ayoub, Liang, Jamil, Jiang, Baumgartner, Sun, Karov, Akulov, Zheng, Cai, Fantacci, Rubin, Acha, Wang, D'Souza, Sathyanarayana, Dai, Rowe, Simanovsky, Goldman, Kuang, Pan, Rosenberg, Rojas-Esponda, Dutta, Zeng, Jurenka, Farquhar, Bansal, Iqbal, Roelofs, Joung, Beak, Ryu, Poplin, Wu, Alayrac, Buthpitiya,
  Ronneberger, Habtegebriel, Li, Cavallaro, Wei, Bensky, Denk, Ganapathy, Stanway, Joshi, Bertolini, Lo, Ma, Charles, Sampemane, Sahni, Chen, Askham, Gaddy, Young, Tan, Eyal, Bražinskas, Zhong, Wu, Epstein, Bailey, Hard, Lee, Goldshtein, Ruiz, Badawi, Lochbrunner, Kearns, Brown, Pardo, Weber, Yang, Jiang, Akin, Fu, Wainwright, Zou, Gaba, Manzagol, Kan, Song, Zainullina, Lin, Ko, Deshmukh, Jindal, Svensson, Tyam, Zhao, Kaeser-Chen, Baird, Moradi, Hall, Guo, Tsang, Liang, Pereira, Ganesh, Korotkov, Adamek, Thiagarajan, Tran, Chen, Tar, Jain, Dasgupta, Bilal, Reitter, Zhao, Vezzani, Gehman, Mehta, Beltrone, Dotiwalla, Guadarrama, Abbas, Karp, Georgiev, Ferng, Brockschmidt, Peng, Hirnschall, Verma, Bi, Xiao, Dabush, Xu, Wallis, Parker, Wang, Xu, Safarli, Tewari, Zhang, Kim, Gesmundo, Thomas, Levi, Chowdhury, Rao, Garst, Conway-Rahman, Ran, McKinney, Xiao, Yu, Agrawal, Stjerngren, Ionescu, Chen, Sharma, Chiu, Liu, Franko, Sanford, Cai, Michel, Ganapathy, Labanowski, Garrett, Vargas, Sun, Gale, Buschmann,
  Desjardins, Ghelani, Jain, Verma, Asawaroengchai, Eisenschlos, Harlalka, Kazawa, Metzler, Howland, Jian, Ades, Shah, Gangwani, Lee, Ring, Hernandez, Reich, Sinha, Sathe, Kovac, Gill, Kannan, D'olimpio, Sevenich, Whang, Kim, Sim, Chen, Zhang, Lall, Matias, Jia, Friesen, Nasso, Thapliyal, Perozzi, Yu, Shekhawat, Huda, Grabowski, Wang, Sreevatsa, Dib, Hassen, Schuh, Milutinovic, Welty, Quinn, Shah, Wang, Barth-Maron, Frye, Axelsson, Zhu, Ma, Giannoumis, Sedghi, Ye, Luan, Aydin, Chandra, Sampathkumar, Huang, Lavrenko, Eleryan, Hong, Hansen, Carthy, Samanta, Ćevid, Wang, Li, Voznesensky, Hoffman, Terzis, Sehwag, Fidel, He, Cai, He, Feng, Nikoltchev, Phatale, Chase, Lawton, Zhang, Ouyang, Tragut, Manshadi, Narayanan, Shen, Gao, Bolukbasi, Roy, Li, Golovin, Panait, Qin, Han, Anthony, Kudugunta, Patraucean, Ray, Chen, Yang, Bhatia, Talluri, Morris, Ražnatović, Brownfield, An, Peng, Kane, Zheng, Duduta, Kessinger, Noraky, Liu, Rong, Veličković, Rush, Goldin, Wei, Garlapati, Pantofaru, Kwon, Ni, Noland, Trapani,
  Beaufays, Roy, Chow, Turker, Cideron, Mei, Clark, Dou, Bošnjak, Leith, Du, Yazdanbakhsh, Nasr, Kwak, Sheth, Kaskasoli, Anand, Lakshminarayanan, Jerome, Bieber, Chu, Senges, Shen, Sridhar, Ndebele, Beyret, Mohamed, Chen, Freitag, Guo, Liu, Roit, Chen, Yan, Stone, Co-Reyes, Cole, Scellato, Azizi, Hashemi, Jin, Iyer, Valentine, György, Ahuja, Diaz, Lee, Clement, Kong, Garmon, Watts, Bhatia, Gupta, Miecnikowski, Vallet, Taly, Loper, Joshi, Atwood, Chick, Collier, Iliopoulos, Trostle, Gunel, Leal-Cavazos, Hrafnkelsson, Guzman, Ju, Forbes, Emond, Chauhan, Caine, Xiao, Zeng, Moufarek, Murphy, Meng, Gupta, Riedel, Das, Lawal, Narayan, Sosea, Swirhun, Friso, Neyshabur, Lu, Girgin, Wunder, Yvinec, Pyne, Carbune, Rijhwani, Guo, Doshi, Briukhov, Bain, Hitron, Wang, Gupta, Chen, Du, Zhang, Shah, Akula, Dylla, Kachra, Kuo, Zou, Wang, Xu, Zhu, Snyder, Menon, Firat, Mordatch, Yuan, Ponomareva, Blevins, Moore, Wang, Chen, Scholz, Dwornik, Lin, Li, Antognini, I, Song, Miller, Kalra, Raveret, Akerlund, Wu, Nystrom, Godbole,
  Liu, DeBalsi, Zhao, Liu, Caciularu, Lax, Khandelwal, Langston, Bailey, Lattanzi, Wang, Kovelamudi, Mondal, Guruganesh, Hua, Roval, Wesołowski, Ingale, Halcrow, Sohn, Angermueller, Raad, Stickgold, Lu, Kosik, Xie, Lillicrap, Huang, Zhang, Paulus, Farabet, Wertheim, Wang, Joshi, ling Ko, Wu, Agrawal, Lin, Sheng, Sung, Breland-King, Butterfield, Gawde, Singh, Zhang, Apte, Shetty, Hutter, Li, Salesky, Lebron, Kanerva, Paganini, Nguyen, Vallu, Peter, Velury, Kao, Hoover, Bortsova, Bishop, Jakobovits, Agostini, Agarwal, Liu, Kwong, Tavakkol, Bica, Greve, GP, Marcus, Hou, Duerig, Moroshko, Lacey, Davis, Amelot, Wang, Kim, Strinopoulos, Wan, Lan, Krishnan, Tang, Humphreys, Bai, Shtacher, Machado, Pang, Burke, Liu, Aravamudhan, Song, Hirst, Singh, Jou, Bai, Piccinno, Fu, Alazard, Meiri, Winter, Chen, Zhang, Heitkaemper, Lambert, Lee, Frömmgen, Rogulenko, Nair, Niemczyk, Bulyenov, Xu, Shemtov, Zadimoghaddam, Toropov, Wirth, Dai, Gollapudi, Zheng, Kurakin, Lee, Bullard, Serrano, Balazevic, Li, Schalkwyk, Murphy,
  Zhang, Sequeira, Datta, Agrawal, Sutton, Attaluri, Chiang, Farhan, Thornton, Lin, Choma, Nguyen, Dasgupta, Robinson, Comşa, Riley, Pillai, Mustafa, Golan, Zandieh, Lespiau, Porter, Ross, Rajayogam, Agarwal, Venugopalan, Shahriari, Yan, Xu, Tobin, Dubov, Shi, Recasens, Kovsharov, Borgeaud, Dery, Vasanth, Gribovskaya, Qiu, Mahdieh, Skut, Nielsen, Zheng, Yu, Bostock, Gupta, Archer, Rawles, Davies, Svyatkovskiy, Tsai, Halpern, Reisswig, Wydrowski, Chang, Puigcerver, Taege, Li, Schnider, Li, Dena, Xu, Telang, Shi, Zen, Kastner, Ko, Subramaniam, Kumar, Blois, Dai, Wieting, Lu, Zeldes, Xie, Hauth, Ţifrea, Li, El-Husseini, Abolafia, Zhou, Ding, Ghalebikesabi, Guía, Maksai, Ágoston Weisz, Arik, Sukhanov, Świetlik, Jia, Yu, Wang, Brand, Bloxwich, Kirmani, Chen, Go, Sprechmann, Kannen, Carin, Sandhu, Edkins, Nooteboom, Gupta, Maggiore, Azizi, Pritch, Yin, Gupta, Tarlow, Smith, Ivanov, Babaeizadeh, Goel, Kambala, Chu, Kastelic, Liu, Soltau, Stone, Agrawal, Kim, Soparkar, Tadepalli, Bunyan, Soh, Kannan, Kim, Chen,
  Halumi, Roy, Wang, Sercinoglu, Gibson, Bhatnagar, Sano, von Dincklage, Ren, Mitrevski, Olšák, She, Doersch, Jilei, Wang, Liu, Tan, Yakar, Warkentin, Ramirez, Lebsack, Dillon, Mathews, Cobley, Wu, Chen, Simon, Nath, Sainath, Bendebury, Julian, Mankalale, Ćurko, Zacchello, Brown, Sodhia, Howard, Caelles, Gupta, Evans, Bulanova, Katzen, Goldenberg, Tsitsulin, Stanton, Schillings, Kovalev, Fry, Shah, Lin, Upadhyay, Li, Radpour, Maggioni, Xiong, Haas, Brennan, Kamath, Savinov, Nagrani, Yacovone, Kappedal, Andriopoulos, Lao, Li, Rozhdestvenskiy, Hashimoto, Audibert, Austin, Rodriguez, Ruoss, Honke, Karkhanis, Xiong, Wei, Huang, Leng, Premachandran, Bileschi, Evangelopoulos, Mensink, Pavagadhi, Teplyashin, Chang, Xue, Tanzer, Goldman, Patel, Li, Wiesner, Zheng, Stewart-Binks, Han, Li, Luo, Lenc, Lučić, Xue, Mullins, Guseynov, Chang, Galatzer-Levy, Zhang, Bingham, Hu, Hartman, Ma, Griffith, Irpan, Radebaugh, Yue, Fan, Ungureanu, Sorokin, Teufel, Li, Anil, Paparas, Wang, Lin, Peng, Shum, Petrovic, Brady,
  Nguyen, Macherey, Li, Singh, Yenugula, Iinuma, Chen, Kopparapu, Stern, Dave, Thekkath, Perot, Kumar, Li, Xiao, Bilotti, Bateni, Noble, Lee, Vázquez-Reina, Salazar, Yang, Wang, Gruzewska, Rao, Raghuram, Xu, Ben-David, Mei, Dalmia, Zhang, Liu, Bansal, Pankov, Schwarcz, Burns, Chan, Sanghai, Liang, Liang, He, Stuart, Narayanan, Zhu, Frank, Fatemi, Sabne, Lang, Bhattacharya, Settle, Wang, McMahan, Tacchetti, Soares, Hadian, Cabi, Chung, Putikhin, Li, Chen, Tarango, Michalewski, Kazemi, Masoom, Sheftel, Shivanna, Vadali, Comanescu, Reid, Moore, Neelakantan, Sander, Herzig, Rosenberg, Dehghani, Choi, Fink, Hayes, Ge, Weng, Ho, Karro, Krishna, Thiet, Skerry-Ryan, Eppens, Andreetto, Sarma, Bonacina, Ayan, Nawhal, Shan, Dusenberry, Thakoor, Gubbi, Nguyen, Tsarfaty, Albanie, Mitrović, Gandhi, Chen, Epasto, Stephanov, Jin, Gehman, Amini, Weber, Behbahani, Xu, Allamanis, Chen, Ott, Sha, Jastrzebski, Qi, Greene, Wu, Toki, Vlasic, Shapiro, Kotikalapudi, Shen, Saeki, Xie, Cassirer, Bharadwaj, Kiyono, Bhojanapalli,
  Rosenfeld, Ritter, Mao, Oliveira, Egyed, Bandemer, Parisotto, Kinoshita, Pluto, Maniatis, Li, Guo, Ghiasi, Tarbouriech, Chatterjee, Jin, Katrina, Xu, Palomaki, Arnold, Sewak, Piccinini, Sharma, Albrecht, Purser-haskell, Vaswani, Chen, Wisniewski, Cao, Aslanides, Phu, Sieb, Agubuzu, Zheng, Sohn, Selvi, Andreassen, Subudhi, Eruvbetine, Woodman, Mery, Krause, Ren, Ma, Luo, Chen, Fan, Griffiths, Schuler, Li, Zhang, Sarr, Luo, Patana, Watson, Naboulsi, Collins, Sidhwani, Hoogeboom, Silver, Caveness, Zhao, Rodriguez, Deines, Bai, Griffin, Tagliasacchi, Xue, Babbula, Pang, Ding, Shen, Peake, Crocker, Raghvendra, Swisher, Han, Singh, Wu, Pchelin, Munkhdalai, Alon, Bacon, Robles, Bulian, Johnson, Powell, Ferreira, Li, Benzing, Velimirović, Soyer, Kong, Tony, Nguyên, Yang, Liu, van Amersfoort, Gillick, Sun, Rauschmayr, Zhang, Zhan, Zhou, Frolov, Yang, Vnukov, Rouillard, Li, Mandhane, Fallen, Venkataraman, Hu, Brennan, Lee, Chang, Sundermeyer, Pan, Ke, Tong, Fabrikant, Bono, Gu, Foley, Mao, Delakis, Bhaswar,
  Frostig, Li, Zipori, Hope, Kozlova, Mishra, Djolonga, Schiff, Merey, Briakou, Morgan, Wan, Hassidim, Skerry-Ryan, Sengupta, Jasarevic, Kallakuri, Kunkle, Brennan, Lieber, Mansoor, Walker, Zhang, Xie, Žužić, Chukwuka, Druinsky, Cho, Yao, Naeem, Butt, Kim, Jia, Jordan, Lelkes, Kurzeja, Wang, Zhao, Over, Chakladar, Prasetya, Jha, Ganapathy, Cong, Shroff, Saroufim, Miryoosefi, Hammad, Nasir, Xi, Gao, Maeng, Hora, Cheng, Haghani, Lewenberg, Lu, Matysiak, Raisinghani, Wang, Baugher, Sukthankar, Giang, Schultz, Fiedel, Chen, Lee, Dey, Zheng, Paul, Smith, Ly, Wang, Bansal, Perz, Ricco, Blank, Keshava, Sharma, Chow, Lad, Jalan, Osindero, Swanson, Scott, Ilić, Li, Jonnalagadda, Soudagar, Xiong, Batsaikhan, Jarrett, Kumar, Shah, Lawlor, Waters, Graham, May, Ramos, Lefdal, Cankara, Cano, O'Donoghue, Borovik, Liu, Grimstad, Alnahlawi, Tsihlas, Hudson, Grigorev, Jia, Huang, Igwe, Lebedev, Tang, Krivokon, Garcia, Tan, Jia, Stys, Vashishth, Liang, Venkatraman, Gu, Kementsietsidis, Zhu, Jung, Bai, Hosseini, Ahmed,
  Gupta, Yuan, Ashraf, Nigam, Vasudevan, Awasthi, Gilady, Mariet, Eskander, Li, Hu, Garrido, Schlattner, Zhang, Saxena, Dević, Muralidharan, Murthy, Zhou, Choi, Wongpanich, Wang, Shah, Xu, Huang, Spencer, Chen, Cohan, Wang, Tompson, Wu, Haroun, Li, Huergo, Yang, Yin, Wendt, Bendersky, Chaabouni, Snaider, Ferret, Jindal, Thompson, Xue, Bishop, Phal, Sharma, Sung, Radhakrishnan, Shomrat, Ingle, Vij, Gilmer, Istin, Sobell, Lu, Nottage, Sadigh, Willcock, Zhang, Xu, Brown, Lee, Wang, Zhu, Tay, Kim, Gutierrez, Sharma, Xian, Seo, Cui, Pochernina, Baetu, Jastrzębski, Ly, Elhawaty, Suh, Sezener, Wang, Yuen, Tucker, Cai, Yang, Wang, Muzio, Qian, Yoo, Lockhart, McKee, Guo, Mehrotra, Mendonça, Mehta, Ben, Tekur, Mu, Zhu, Krakovna, Lee, Maschinot, Cevey, Choe, Bai, Srinivasan, Gasaway, Young, Siegler, Holtmann-Rice, Piratla, Baumli, Yogev, Hofer, van Hasselt, Grant, Chervonyi, Silver, Hogue, Agarwal, Wang, Singh, Flynn, Lipschultz, David, Bellot, Yang, Le, Graziano, Olszewska, Hui, Maurya, Parotsidis, Chen, Oguntebi,
  Kelley, Baddepudi, Mauerer, Shaw, Siegman, Yang, Shetty, Roy, Song, Stokowiec, Burnell, Savant, Busa-Fekete, Miao, Ghosh, MacDermed, Lippe, Dektiarev, Behrman, Mentzer, Nguyen, Wei, Verma, Knutsen, Dasari, Yan, Mitrichev, Wang, Shejwalkar, Austin, Sunkara, Potti, Virin, Wright, Liu, Riva, Pot, Kochanski, Le, Balasubramaniam, Dhar, Liao, Bloniarz, Shukla, Cole, Lee, Zhang, Kafle, Vashishtha, Mahmoudieh, Chen, Hoffmann, Srinivasan, Lago, Shalom, Wang, Elabd, Sharma, Oh, Kothawade, Le, Monteiro, Yang, Alarakyia, Geirhos, Mincu, Garnes, Kobayashi, Mariooryad, Krasowiak, Zhixin, Lai, Mourad, Wang, Bu, Aharoni, Chen, Goyal, Zubov, Bapna, Dabir, Kothari, Lamerigts, Cao, Shar, Yew, Kulkarni, Mahaarachchi, Joshi, Zhu, Lichtarge, Zhou, Muckenhirn, Selo, Vinyals, Chen, Brohan, Mehta, Cogan, Wang, Geri, Ko, Chen, Viola, Shivam, Wang, Elish, Popa, Pereira, Liu, Koster, Kim, Zhang, Ebrahimi, Talukdar, Zheng, Poklukar, Mikhalap, Johnson, Vijayakumar, Omernick, Dibb, Dubey, Hu, Suman, Aggarwal, Kornakov, Xia, Lowe,
  Kolganov, Xiao, Nikolaev, Hemingray, Li, Iljazi, Rybiński, Sandhu, Lu, Luong, Jenatton, Govindaraj, Hui, Li, Dulac-Arnold, Park, Wang, Modi, Pouget-Abadie, Greller, Gupta, Berry, Ramachandran, Xie, McCafferty, Wang, Gupta, Lim, Bratanič, Brock, Akolzin, Sproch, Karliner, Kim, Goedeckemeyer, Shazeer, Schmid, Calandriello, Bhatia, Choromanski, Montgomery, Dua, Ramalho, King, Gao, Nguyen, Lindner, Pitta, Johnson, Salama, Ardila, Han, Farnese, Odoom, Wang, Ding, Rink, Smith, Lehri, Cohen, Vats, He, Gopavarapu, Paszke, Patel, Gansbeke, Loher, Castro, Voitovich, von Glehn, George, Niklaus, Eaton-Rosen, Rakićević, Jue, Perel, Zhang, Bahat, Pouget, Xing, Huot, Shenoy, Bos, Coriou, Richter, Noy, Wang, Ontanon, Qin, Makarchuk, Hassabis, Li, Sharma, Venkatesan, Kemaev, Daniel, Huang, Shah, Ponce, Warren, Chen, Faruqui, Wu, Andačić, Payrits, McDuff, Hume, Cao, Tessler, Wang, Wang, Rendulic, Agustsson, Johnson, Lando, Howard, Padmanabhan, Daswani, Banino, Kilgore, Heek, Ji, Caceres, Li, Kassner, Vlaskin, Liu,
  Grills, Hou, Sukkerd, Cheon, Shetty, Markeeva, Stanczyk, Iyer, Gong, Gao, Gopalakrishnan, Blyth, Reynolds, Bhoopchand, Bilenko, Gharibian, Zayats, Faust, Singh, Ma, Jiao, Vijayanarasimhan, Aroyo, Yadav, Chakera, Kakarla, Meshram, Gregor, Botea, Senter, Jia, Kovacs, Sharma, Baur, Kang, He, Zhuo, Kostelac, Laish, Peng, O'Bryan, Kasenberg, Rao, Leurent, Zhang, Stevens, Salazar, Zhang, Lobov, Walker, Porter, Redshaw, Ke, Rao, Lee, Lam, Moffitt, Kim, Qiao, Koo, Dadashi, Song, Sundararajan, Xu, Kawamoto, Zhong, Barbu, Reddy, Verzetti, Li, Papamakarios, Klimczak-Plucińska, Cassin, Kavukcuoglu, Swavely, Vaucher, Zhao, Hemsley, Tschannen, Ge, Menghani, Yu, Ha, He, Wu, Song, Sterneck, Zinke, Calian, Marsden, Ruiz, Hessel, Gueta, Lee, Farris, Gupta, Li, Saleh, Misra, Xiao, Mendolicchio, Buttimore, Krayvanova, Nayakanti, Wiethoff, Pande, Mirhoseini, Lao, Liu, Hua, Chen, Malkov, Kalashnikov, Gupta, Audhkhasi, Zhai, Kopalle, Jain, Ofek, Meyer, Baatarsukh, Strejček, Qian, Freedman, Figueira, Sokolik, Bachem, Lin,
  Kharrat, Hidey, Xu, Duan, Li, Ersoy, Everett, Cen, Santamaria-Fernandez, Taubenfeld, Mackinnon, Deng, Zablotskaia, Viswanadha, Goel, Yates, Deng, Choy, Chen, Sinha, Mossin, Wang, Szlam, Hao, Rubenstein, Toksoz-Exley, Aperghis, Zhong, Ahn, Isard, Lacombe, Luisier, Anastasiou, Kalley, Prabhu, Dunleavy, Bijwadia, Mao-Jones, Chen, Pasumarthi, Wood, Dostmohamed, Hurley, Simsa, Parrish, Pajarskas, Harvey, Skopek, Kochinski, Rey, Rieser, Zhou, Lee, Acharya, Li, Jiang, Zhang, Gipson, Mahintorabi, Gelmi, Khajehnouri, Yeh, Lee, Matthey, Baker, Pham, Fu, Pak, Gupta, Vasconcelos, Sadovsky, Walker, Hsiao, Zochbauer, Marzoca, Velan, Zeng, Baechler, Driess, Jain, Huang, Tao, Maggs, Levine, Schneider, Gemzer, Petit, Han, Fisher, Zelle, Biles, Ie, Fadeeva, Liu, Franco, Collister, Zhang, Wang, Zhao, Kieliger, Shuster, Zhu, Gong, Chan, Sun, Basu, Zimmermann, Hayes, Bapna, Snoek, Yang, Datta, Abdallah, Kilgour, Li, Mah, Jun, Rivière, Karmarkar, Spalink, Huang, Gonzalez, Tran, Nowak, Palowitch, Chadwick, Talius, Mehta, Sellam,
  Fränken, Nicosia, He, Kini, Amos, Basu, Jobe, Shaw, Xu, Evans, Ikeda, Yan, Jin, Wang, Yadav, Labzovsky, Sampath, Ma, Schumann, Siddhant, Shah, Youssef, Agarwal, Dabney, Tonioni, Ambar, Li, Guyon, Li, Soergel, Fang, Karadzhov, Udrescu, Trinh, Raunak, Noury, Guo, Gupta, Finkelstein, Petek, Liang, Billock, Sun, Wood, Song, Yu, Matejovicova, Cohen, Andra, D'Ambrosio, Deng, Nallatamby, Songhori, Dangovski, Lampinen, Botadra, Hillier, Cao, Baddi, Kuncoro, Yoshino, Bhagatwala, Ranzato, Schaeffer, Liu, Ye, Sarvana, Nham, Kuang, Gao, Baek, Mittal, Wahid, Gergely, Ni, Feldman, Muir, Lamblin, Macherey, Dyer, Kilpatrick, Campos, Bhutani, Fort, Ahmad, Severyn, Chatziprimou, Ferludin, Dimarco, Kusupati, Heyward, Bahir, Villela, Millican, Marcus, Bahargam, Unlu, Roth, Wei, Gopal, Ghoshal, Lee, Lin, Lees, Lee, Hosseini, Fan, Neel, Wu, Altun, Cai, Piqueras, Woodward, Bissacco, Haykal, Bordbar, Sundaram, Hodkinson, Toyama, Polovets, Myers, Sinha, Levinboim, Krishnakumar, Chhaparia, Sholokhova, Gundavarapu, Jawahar, Qureshi,
  Hu, Momchev, Rahtz, Wu, S, Dhamdhere, Guo, Gupta, Eslami, Schain, Blokzijl, Welling, Orr, Bolelli, Perez-Nieves, Sirotenko, Prasad, Kar, Pigem, Terzi, Weisz, Ghosh, Mavalankar, Madeka, Daugaard, Adam, Shah, Berman, Tran, Baker, Andrejczuk, Chole, Raboshchuk, Mirzazadeh, Kagohara, Wu, Schallhart, Orlando, Wang, Rrustemi, Xiong, Liu, Vezer, Ramsden, yiin Chang, Mudgal, Li, Vieillard, Hoshen, Ahmad, Slone, Hua, Potikha, Rossini, Stritar, Prakash, Wang, Dong, Nazari, Nehoran, Tekelioglu, Li, Badola, Funkhouser, Li, Yerram, Ganeshan, Formoso, Langner, Shi, Li, Yamamori, Panda, Saade, Scarpati, Breaux, Carey, Zhou, Hsieh, Bridgers, Butryna, Gupta, Tulsyan, Woo, Eltyshev, Grathwohl, Parks, Benjamin, Panigrahy, Dodhia, Freitas, Sauer, Song, Alet, Tolins, Paduraru, Zhou, Albert, Zhang, Shu, Bansal, Nguyen, Globerson, Xiao, Manyika, Hennigan, Rong, Matak, Bakalov, Sharma, Sinopalnikov, Pierson, Roller, Brown, Gao, Fukuzawa, Ghafouri, Vassigh, Barr, Wang, Korsun, Jayaram, Ren, Zaman, Khan, Lunts, Deutsch, Uthus, Katz,
  Samsikova, Khalifa, Sethi, Sun, Tang, Alon, Luo, Yu, Nayyar, Petrini, Truong, Hellendoorn, Chinaev, Alberti, Wang, Hu, Mirrokni, Balashankar, Aharon, Mehta, Iscen, Kready, Manning, Mohananey, Chen, Tripathi, Wu, Petrovski, Hwang, Baeuml, Chandrakaladharan, Liu, Coaguila, Chen, Ma, Tafti, Tatineni, Spitz, Ye, Vicol, Rosca, Puigdomènech, Yahav, Ghemawat, Lin, Kirk, Nabulsi, Brin, Bohnet, Caluwaerts, Veerubhotla, Zheng, Dai, Petrov, Xu, Mehran, Xu, Zintgraf, Choi, Hombaiah, Thoppilan, Reddi, Lew, Li, Webster, Sawhney, Lamprou, Shakeri, Lunayach, Chen, Bagri, Salcianu, Chen, Donchev, Magister, Nørly, Rodrigues, Izo, Noga, Zou, Köppe, Zhou, Lee, Long, Eisenbud, Chen, Schenck, To, Zhong, Taropa, Truong, Levy, Martins, Zhang, Semturs, Zhang, Yakubovich, Moreno, McConnaughey, Lu, Redmond, Weerts, Bitton, Refice, Lacasse, Conmy, Tallec, Odell, Forbes-Pollard, Socala, Hoech, Kohli, Walton, Wang, Sazanovich, Zhu, Kapishnikov, Galt, Denton, Murdoch, Sikora, Mohamed, Wei, First, McConnell, Cobo, Qin, Avrahami, Balle,
  Watanabe, Louis, Kraft, Ariafar, Gu, Rives, Yoon, Rusu, Cobon-Kerr, Hahn, Luo, Yuvein, Zhu, Ahuja, Benenson, Kaufman, Yu, Hightower, Zhang, Ni, Hendricks, Wang, Yona, Jain, Barrio, Bhupatiraju, Velusamy, Dafoe, Riedel, Thomas, Yuan, Bellaiche, Panthaplackel, Kloboves, Jauhari, Akbulut, Davchev, Gladchenko, Madras, Chuklin, Hill, Yuan, Madhavan, Leonhard, Scandinaro, Chen, Niu, Douillard, Damoc, Onoe, Pedregosa, Bertsch, Leichner, Pagadora, Malmaud, Ponda, Twigg, Duzhyi, Shen, Wang, Garg, Chen, Evci, Lee, Liu, Kojima, Yamaguchi, Rajendran, Piergiovanni, Rajendran, Fornoni, Ibagon, Ragan, Khan, Blitzer, Bunner, Sun, Kosakai, Lundberg, Elue, Guu, Park, Park, Narayanaswamy, Wu, Mudigonda, Cohn, Mu, Kumar, Graesser, Zhang, Killam, Zhuang, Giménez, Jishi, Ley-Wild, Zhai, Osawa, Cedillo, Liu, Upadhyay, Sieniek, Sharma, Paine, Angelova, Addepalli, Parada, Majumder, Lamp, Kumar, Deng, Myaskovsky, Sabolić, Dudek, York, de~Chaumont~Quitry, Nie, Cattle, Gunjan, Piot, Khawaja, Bang, Wang, Khodadadeh, R, Rawlani,
  Powell, Lee, Griesser, Oh, Magalhaes, Li, Tokumine, Vogel, Hsu, BC, Jindal, Cohen, Yang, Yuan, de~Cesare, Bruguier, Xu, Roy, Jacovi, Belov, Arya, Meadowlark, Cohen-Ganor, Ye, Morris-Suzuki, Banzal, Song, Ponnuramu, Zhang, Scrivener, Zaiem, Rochman, Han, Ghazi, Lee, Drath, Suo, Girgis, Shenoy, Nguyen, Eck, Gupta, Yan, Carreira, Gulati, Sang, Mirylenka, Cooney, Chou, Ling, Fan, Coleman, Tubone, Kumar, Baldridge, Hernandez-Campos, Lazaridou, Besley, Yona, Bulut, Wellens, Pierigiovanni, George, Green, Han, Tao, Clark, You, Abdolmaleki, Fu, Chen, Chaugule, Chandorkar, Rahman, Thompson, Koanantakool, Bernico, Ren, Vlasov, Vassilvitskii, Kula, Liang, Kim, Huang, Ye, Lepikhin, and Helmholz}]{comanici2025gemini25}
Gheorghe Comanici, Eric Bieber, Mike Schaekermann, Ice Pasupat, Noveen Sachdeva, Inderjit Dhillon, Marcel Blistein, Ori Ram, Dan Zhang, Evan Rosen, Luke Marris, Sam Petulla, Colin Gaffney, Asaf Aharoni, Nathan Lintz, Tiago~Cardal Pais, Henrik Jacobsson, Idan Szpektor, Nan-Jiang Jiang, and 3416 others. 2025.
\newblock \href {https://arxiv.org/abs/2507.06261} {Gemini 2.5: Pushing the frontier with advanced reasoning, multimodality, long context, and next generation agentic capabilities}.
\newblock \emph{Preprint}, arXiv:2507.06261.

\bibitem[{Cowan(2001)}]{cowan2001magical}
Nelson Cowan. 2001.
\newblock The magical number 4 in short-term memory: A reconsideration of mental storage capacity.
\newblock \emph{Behavioral and brain sciences}, 24(1):87--114.

\bibitem[{Deshmukh et~al.(2025)Deshmukh, Dixit, Singh, and Raj}]{mellow}
Soham Deshmukh, Satvik Dixit, Rita Singh, and Bhiksha Raj. 2025.
\newblock \href {https://arxiv.org/abs/2503.08540} {Mellow: a small audio language model for reasoning}.
\newblock \emph{Preprint}, arXiv:2503.08540.

\bibitem[{Dinkel et~al.(2025)Dinkel, Li, Liu, Luan, Niu, Sun, Wang, Xiao, Zhang, and Zhou}]{midashenglm}
Heinrich Dinkel, Gang Li, Jizhong Liu, Jian Luan, Yadong Niu, Xingwei Sun, Tianzi Wang, Qiyang Xiao, Junbo Zhang, and Jiahao Zhou. 2025.
\newblock \href {https://arxiv.org/abs/2508.03983} {Midashenglm: Efficient audio understanding with general audio captions}.
\newblock \emph{Preprint}, arXiv:2508.03983.

\bibitem[{Dinkel et~al.(2024)Dinkel, Yan, Wang, Zhang, Wang, and Wang}]{dinkel2024scalingmaskedaudioencoder}
Heinrich Dinkel, Zhiyong Yan, Yongqing Wang, Junbo Zhang, Yujun Wang, and Bin Wang. 2024.
\newblock \href {https://arxiv.org/abs/2406.06992} {Scaling up masked audio encoder learning for general audio classification}.
\newblock \emph{Preprint}, arXiv:2406.06992.

\bibitem[{Drake and Botte(1993)}]{drake1993tempo}
Carolyn Drake and Marie-Claire Botte. 1993.
\newblock Tempo sensitivity in auditory sequences: Evidence for a multiple-look model.
\newblock \emph{Perception \& psychophysics}, 54(3):277--286.

\bibitem[{Du et~al.(2018)Du, Na, Liu, and Bu}]{du2018aishell2transformingmandarinasr}
Jiayu Du, Xingyu Na, Xuechen Liu, and Hui Bu. 2018.
\newblock \href {https://arxiv.org/abs/1808.10583} {Aishell-2: Transforming mandarin asr research into industrial scale}.
\newblock \emph{Preprint}, arXiv:1808.10583.

\bibitem[{Engel et~al.(2017{\natexlab{a}})Engel, Resnick, Roberts, Dieleman, Eck, Simonyan, and Norouzi}]{nsynth}
Jesse Engel, Cinjon Resnick, Adam Roberts, Sander Dieleman, Douglas Eck, Karen Simonyan, and Mohammad Norouzi. 2017{\natexlab{a}}.
\newblock \href {https://arxiv.org/abs/1704.01279} {Neural audio synthesis of musical notes with wavenet autoencoders}.
\newblock \emph{Preprint}, arXiv:1704.01279.

\bibitem[{Engel et~al.(2017{\natexlab{b}})Engel, Resnick, Roberts, Dieleman, Eck, Simonyan, and Norouzi}]{engel2017neuralaudiosynthesis}
Jesse Engel, Cinjon Resnick, Adam Roberts, Sander Dieleman, Douglas Eck, Karen Simonyan, and Mohammad Norouzi. 2017{\natexlab{b}}.
\newblock \href {https://arxiv.org/abs/1704.01279} {Neural audio synthesis of musical notes with wavenet autoencoders}.
\newblock \emph{Preprint}, arXiv:1704.01279.

\bibitem[{Evers et~al.(2020)Evers, Löllmann, Mellmann, Schmidt, Barfuss, Naylor, and Kellermann}]{locata}
Christine Evers, Heinrich~W. Löllmann, Heinrich Mellmann, Alexander Schmidt, Hendrik Barfuss, Patrick~A. Naylor, and Walter Kellermann. 2020.
\newblock \href {https://doi.org/10.1109/TASLP.2020.2990485} {The locata challenge: Acoustic source localization and tracking}.
\newblock \emph{IEEE/ACM Transactions on Audio, Speech, and Language Processing}, 28:1620--1643.

\bibitem[{Fang et~al.(2025{\natexlab{a}})Fang, Guo, Zhou, Ma, Zhang, and Feng}]{llamaomni}
Qingkai Fang, Shoutao Guo, Yan Zhou, Zhengrui Ma, Shaolei Zhang, and Yang Feng. 2025{\natexlab{a}}.
\newblock \href {https://arxiv.org/abs/2409.06666} {Llama-omni: Seamless speech interaction with large language models}.
\newblock \emph{Preprint}, arXiv:2409.06666.

\bibitem[{Fang et~al.(2025{\natexlab{b}})Fang, Zhou, Guo, Zhang, and Feng}]{llamaomni2}
Qingkai Fang, Yan Zhou, Shoutao Guo, Shaolei Zhang, and Yang Feng. 2025{\natexlab{b}}.
\newblock \href {https://arxiv.org/abs/2505.02625} {Llama-omni2: Llm-based real-time spoken chatbot with autoregressive streaming speech synthesis}.
\newblock \emph{Preprint}, arXiv:2505.02625.

\bibitem[{Florentine et~al.(1987)Florentine, Buus, and Mason}]{florentine1987level}
Mary Florentine, S{\o}ren Buus, and Christine~R. Mason. 1987.
\newblock Level discrimination as a function of level for tones from 0.25 to 16 khz.
\newblock \emph{The Journal of the Acoustical Society of America}, 81(5):1528--1541.

\bibitem[{Fu et~al.(2025)Fu, Lin, Long, Shen, Dai, Zhao, Zhang, Dong, Li, Wang, Cao, Yin, Ma, Zheng, Ji, Wu, He, Shan, and Sun}]{vita}
Chaoyou Fu, Haojia Lin, Zuwei Long, Yunhang Shen, Yuhang Dai, Meng Zhao, Yi-Fan Zhang, Shaoqi Dong, Yangze Li, Xiong Wang, Haoyu Cao, Di~Yin, Long Ma, Xiawu Zheng, Rongrong Ji, Yunsheng Wu, Ran He, Caifeng Shan, and Xing Sun. 2025.
\newblock \href {https://arxiv.org/abs/2408.05211} {Vita: Towards open-source interactive omni multimodal llm}.
\newblock \emph{Preprint}, arXiv:2408.05211.

\bibitem[{Gao et~al.(2025)Gao, Huang, Xu, Tang, Li, Liu, Li, Hu, Lin, Yang, Wu, Bi, Chen, and Zhang}]{gao2025pixelspatternspoetryworld}
Hongcheng Gao, Zihao Huang, Lin Xu, Jingyi Tang, Xinhao Li, Yue Liu, Haoyang Li, Taihang Hu, Minhua Lin, Xinlong Yang, Ge~Wu, Balong Bi, Hongyu Chen, and Wentao Zhang. 2025.
\newblock \href {https://arxiv.org/abs/2507.16863} {Pixels, patterns, but no poetry: To see the world like humans}.
\newblock \emph{Preprint}, arXiv:2507.16863.

\bibitem[{Gegenfurtner and Rieger(2000)}]{gegenfurtner2000sensory}
Karl~R Gegenfurtner and Jochem Rieger. 2000.
\newblock Sensory and cognitive contributions of color to the recognition of natural scenes.
\newblock \emph{Current Biology}, 10(13):805--808.

\bibitem[{Geirhos et~al.(2020)Geirhos, Jacobsen, Michaelis, Zemel, Brendel, Bethge, and Wichmann}]{geirhos2020shortcut}
Robert Geirhos, J{\"o}rn-Henrik Jacobsen, Claudio Michaelis, Richard Zemel, Wieland Brendel, Matthias Bethge, and Felix~A Wichmann. 2020.
\newblock Shortcut learning in deep neural networks.
\newblock \emph{Nature Machine Intelligence}, 2(11):665--673.

\bibitem[{{Gemini Team}(2024)}]{gemini15}
{Gemini Team}. 2024.
\newblock \href {https://arxiv.org/abs/2403.05530} {Gemini 1.5: Unlocking multimodal understanding across millions of tokens of context}.
\newblock arXiv preprint arXiv:2403.05530.

\bibitem[{Gemmeke et~al.(2017)Gemmeke, Ellis, Freedman, Jansen, Lawrence, Moore, Plakal, and Ritter}]{audioset}
Jort~F. Gemmeke, Daniel P.~W. Ellis, Dylan Freedman, Aren Jansen, Wade Lawrence, R.~Channing Moore, Manoj Plakal, and Marvin Ritter. 2017.
\newblock \href {https://doi.org/10.1109/ICASSP.2017.7952261} {Audio set: An ontology and human-labeled dataset for audio events}.
\newblock In \emph{2017 IEEE International Conference on Acoustics, Speech and Signal Processing (ICASSP)}, pages 776--780.

\bibitem[{Ghosh et~al.(2025)Ghosh, Kong, Kumar, Sakshi, Kim, Ping, Valle, Manocha, and Catanzaro}]{audioflamingo2}
Sreyan Ghosh, Zhifeng Kong, Sonal Kumar, S~Sakshi, Jaehyeon Kim, Wei Ping, Rafael Valle, Dinesh Manocha, and Bryan Catanzaro. 2025.
\newblock \href {https://arxiv.org/abs/2503.03983} {Audio flamingo 2: An audio-language model with long-audio understanding and expert reasoning abilities}.
\newblock \emph{Preprint}, arXiv:2503.03983.

\bibitem[{Ghosh et~al.(2024)Ghosh, Kumar, Seth, Evuru, Tyagi, Sakshi, Nieto, Duraiswami, and Manocha}]{ghosh2024gama}
Sreyan Ghosh, Sonal Kumar, Ashish Seth, Chandra Kiran~Reddy Evuru, Utkarsh Tyagi, S~Sakshi, Oriol Nieto, Ramani Duraiswami, and Dinesh Manocha. 2024.
\newblock \href {https://arxiv.org/abs/2406.11768} {Gama: A large audio-language model with advanced audio understanding and complex reasoning abilities}.
\newblock \emph{Preprint}, arXiv:2406.11768.

\bibitem[{Gillick et~al.(2019)Gillick, Roberts, Engel, Eck, and Bamman}]{groovemidi}
Jon Gillick, Adam Roberts, Jesse Engel, Douglas Eck, and David Bamman. 2019.
\newblock Learning to groove with inverse sequence transformations.
\newblock In \emph{International Conference on Machine Learning (ICML)}.

\bibitem[{Giordano(2005)}]{giordano2005impactsounds}
Bruno~L. Giordano. 2005.
\newblock \emph{Sound source perception in impact sounds}.
\newblock Ph.D. thesis, Università degli Studi di Padova.

\bibitem[{Giordano and McAdams(2006)}]{giordano2006materials}
Bruno~L Giordano and Stephen McAdams. 2006.
\newblock Material identification of real impact sounds: Effects of size variation in steel, glass, wood, and plexiglass plates.
\newblock \emph{The Journal of the Acoustical Society of America}, 119(2):1171--1181.

\bibitem[{Goebl and Bresin(2001)}]{goebl2001disklavier}
Werner Goebl and Roberto Bresin. 2001.
\newblock Are computer-controlled pianos a reliable tool in music performance research? recording and reproduction precision of a yamaha disklavier grand piano.
\newblock In \emph{Proceedings of the 2001 Workshop on Current Research Directions in Computer Music}, pages 45--50.

\bibitem[{Goel et~al.(2025)Goel, Ghosh, Kim, Kumar, Kong, gil Lee, Yang, Duraiswami, Manocha, Valle, and Catanzaro}]{audioflamingo3}
Arushi Goel, Sreyan Ghosh, Jaehyeon Kim, Sonal Kumar, Zhifeng Kong, Sang gil Lee, Chao-Han~Huck Yang, Ramani Duraiswami, Dinesh Manocha, Rafael Valle, and Bryan Catanzaro. 2025.
\newblock \href {https://arxiv.org/abs/2507.08128} {Audio flamingo 3: Advancing audio intelligence with fully open large audio language models}.
\newblock \emph{Preprint}, arXiv:2507.08128.

\bibitem[{Grondin(1993)}]{grondin1993duration}
Simon Grondin. 1993.
\newblock Duration discrimination of empty and filled intervals marked by auditory and visual signals.
\newblock \emph{Perception \& psychophysics}, 54(3):383--394.

\bibitem[{Hall et~al.(2008)Hall, Buss, and Grose}]{hall2008auditory}
John~W. Hall, Emily Buss, and John~H. Grose. 2008.
\newblock Auditory intensity discrimination.
\newblock In William~A. Yost, Arthur~N. Popper, and Richard~R. Fay, editors, \emph{Auditory Perception of Sound Sources}, pages 115--154. Springer.

\bibitem[{Kim et~al.(2025)Kim, Yun, Woo, Yang, and Kim}]{wowbench}
Jaeyeon Kim, Heeseung Yun, Sang~Hoon Woo, Chao-Han~Huck Yang, and Gunhee Kim. 2025.
\newblock \href {https://arxiv.org/abs/2508.20976} {Wow-bench: Evaluating fine-grained acoustic perception in audio-language models via marine mammal vocalizations}.
\newblock \emph{Preprint}, arXiv:2508.20976.

\bibitem[{KimiTeam et~al.(2025)KimiTeam, Ding, Ju, Leng, Liu, Liu, Shang, Shen, Song, Tan, Tang, Wang, Wei, Xin, Xu, Yu, Zhang, Zhou, Charles, Chen, Chen, Du, He, Hu, Lai, Li, Liu, Sun, Wang, Wang, Wu, Wu, Yang, Yang, Yang, Yang, Yin, Yuan, Zhang, and Zhou}]{kimiaudio}
KimiTeam, Ding Ding, Zeqian Ju, Yichong Leng, Songxiang Liu, Tong Liu, Zeyu Shang, Kai Shen, Wei Song, Xu~Tan, Heyi Tang, Zhengtao Wang, Chu Wei, Yifei Xin, Xinran Xu, Jianwei Yu, Yutao Zhang, Xinyu Zhou, Y.~Charles, and 21 others. 2025.
\newblock \href {https://arxiv.org/abs/2504.18425} {Kimi-audio technical report}.
\newblock \emph{Preprint}, arXiv:2504.18425.

\bibitem[{Ko et~al.(2017)Ko, Peddinti, Povey, Seltzer, and Khudanpur}]{7953152}
Tom Ko, Vijayaditya Peddinti, Daniel Povey, Michael~L. Seltzer, and Sanjeev Khudanpur. 2017.
\newblock \href {https://doi.org/10.1109/ICASSP.2017.7953152} {A study on data augmentation of reverberant speech for robust speech recognition}.
\newblock In \emph{2017 IEEE International Conference on Acoustics, Speech and Signal Processing (ICASSP)}, pages 5220--5224.

\bibitem[{Kolarik et~al.(2016)Kolarik, Moore, Zahorik, Cirstea, and Pardhan}]{kolarik2016auditory}
Andrew~J Kolarik, Brian~CJ Moore, Pavel Zahorik, Silvia Cirstea, and Shahina Pardhan. 2016.
\newblock Auditory distance perception in humans: a review of cues, development, neuronal bases, and effects of sensory loss.
\newblock \emph{Attention, Perception, \& Psychophysics}, 78(2):373--395.

\bibitem[{Kong et~al.(2025)Kong, Goel, Santos, Ghosh, Valle, Ping, and Catanzaro}]{audioflamingosoundcot}
Zhifeng Kong, Arushi Goel, Joao~Felipe Santos, Sreyan Ghosh, Rafael Valle, Wei Ping, and Bryan Catanzaro. 2025.
\newblock \href {https://arxiv.org/abs/2508.11818} {Audio flamingo sound-cot technical report: Improving chain-of-thought reasoning in sound understanding}.
\newblock \emph{Preprint}, arXiv:2508.11818.

\bibitem[{Kumar et~al.(2025)Kumar, Šimon Sedláček, Lokegaonkar, López, Yu, Anand, Ryu, Chen, Plička, Hlaváček, Ellingwood, Udupa, Hou, Ferner, Barahona, Bolaños, Rahi, Herrera-Alarcón, Dixit, Patil, Deshmukh, Koroshinadze, Liu, Perera, Zanou, Stafylakis, Chung, Harwath, Zhang, Manocha, Lozano-Diez, Kesiraju, Ghosh, and Duraiswami}]{kumar2025mmaupro}
Sonal Kumar, Šimon Sedláček, Vaibhavi Lokegaonkar, Fernando López, Wenyi Yu, Nishit Anand, Hyeonggon Ryu, Lichang Chen, Maxim Plička, Miroslav Hlaváček, William~Fineas Ellingwood, Sathvik Udupa, Siyuan Hou, Allison Ferner, Sara Barahona, Cecilia Bolaños, Satish Rahi, Laura Herrera-Alarcón, Satvik Dixit, and 15 others. 2025.
\newblock \href {https://arxiv.org/abs/2508.13992} {Mmau-pro: A challenging and comprehensive benchmark for holistic evaluation of audio general intelligence}.
\newblock \emph{Preprint}, arXiv:2508.13992.

\bibitem[{Kuttruff(2016)}]{kuttruff2016room}
Heinrich Kuttruff. 2016.
\newblock \emph{Room acoustics}.
\newblock Crc Press.

\bibitem[{Lemaitre and Heller(2012)}]{lemaitre2012action}
Guillaume Lemaitre and Laurie~M Heller. 2012.
\newblock Auditory perception of material is fragile while action is strikingly robust.
\newblock \emph{The Journal of the Acoustical Society of America}, 131(2):1337--1348.

\bibitem[{Li et~al.(2025{\natexlab{a}})Li, Liu, Dinkel, Niu, Zhang, and Luan}]{li2025r1_aqa}
Gang Li, Jizhong Liu, Heinrich Dinkel, Yadong Niu, Junbo Zhang, and Jian Luan. 2025{\natexlab{a}}.
\newblock \href {https://arxiv.org/abs/2503.11197} {Reinforcement learning outperforms supervised fine-tuning: A case study on audio question answering}.
\newblock \emph{Preprint}, arXiv:2503.11197.

\bibitem[{Li et~al.(2023)Li, Li, Savarese, and Hoi}]{li2023blip2}
Junnan Li, Dongxu Li, Silvio Savarese, and Steven Hoi. 2023.
\newblock \href {https://arxiv.org/abs/2301.12597} {Blip-2: Bootstrapping language-image pre-training with frozen image encoders and large language models}.
\newblock \emph{Preprint}, arXiv:2301.12597.

\bibitem[{Li et~al.(2025{\natexlab{b}})Li, Liu, Zhang, Fang, Pan, Wang, Liang, Li, Lin, Dong, Xu, Sun, Zhou, and Chen}]{li2025baichuanaudio}
Tianpeng Li, Jun Liu, Tao Zhang, Yuanbo Fang, Da~Pan, Mingrui Wang, Zheng Liang, Zehuan Li, Mingan Lin, Guosheng Dong, Jianhua Xu, Haoze Sun, Zenan Zhou, and Weipeng Chen. 2025{\natexlab{b}}.
\newblock \href {https://arxiv.org/abs/2502.17239} {Baichuan-audio: A unified framework for end-to-end speech interaction}.
\newblock \emph{Preprint}, arXiv:2502.17239.

\bibitem[{Li et~al.(2025{\natexlab{c}})Li, Liu, Zhang, Zhang, Chen, Li, Li, Liu, Ming, Dong, Pan, Li, Fang, Kuang, Wang, Zhu, Zhang, Guo, Zhang, Wang, Ding, Song, Li, Huo, Liang, Zhang, Wu, Zhao, Xiong, Wu, Ye, Lu, Li, Zhang, Zhou, Chen, Su, Zhang, Chen, Dong, Nie, Wu, Xiao, Li, Dang, Zhang, Sun, Wu, Yang, Lin, Ma, Wu, li, Yang, Liu, Zhang, Chen, Ai, Zhang, Chen, Huang, Li, Luo, Duan, Zhu, Xiao, Su, Pu, Wang, Jia, Zhang, Ai, Wang, Qiao, Zhang, Shen, Yang, Zhen, Zhou, Chen, Li, Zhu, Lu, Zhao, Liang, Li, Qin, Sun, Xu, Sun, Lin, Zhou, and Chen}]{li2025baichuanomni15}
Yadong Li, Jun Liu, Tao Zhang, Tao Zhang, Song Chen, Tianpeng Li, Zehuan Li, Lijun Liu, Lingfeng Ming, Guosheng Dong, Da~Pan, Chong Li, Yuanbo Fang, Dongdong Kuang, Mingrui Wang, Chenglin Zhu, Youwei Zhang, Hongyu Guo, Fengyu Zhang, and 74 others. 2025{\natexlab{c}}.
\newblock \href {https://arxiv.org/abs/2501.15368} {Baichuan-omni-1.5 technical report}.
\newblock \emph{Preprint}, arXiv:2501.15368.

\bibitem[{Liang et~al.(2025)Liang, Li, Fan, Li, Nguyen, Cobbina, Bhardwaj, Chen, Liu, and Zhou}]{colorbench}
Yijun Liang, Ming Li, Chenrui Fan, Ziyue Li, Dang Nguyen, Kwesi Cobbina, Shweta Bhardwaj, Jiuhai Chen, Fuxiao Liu, and Tianyi Zhou. 2025.
\newblock \href {https://arxiv.org/abs/2504.10514} {Colorbench: Can vlms see and understand the colorful world? a comprehensive benchmark for color perception, reasoning, and robustness}.
\newblock \emph{Preprint}, arXiv:2504.10514.

\bibitem[{Liu et~al.(2025{\natexlab{a}})Liu, Ehrenberg, Lo, Denoix, Barreau, Lample, Delignon, Chandu, von Platen, Muddireddy, Gandhi, Ghosh, Mishra, Foubert, Rastogi, Yang, Jiang, Sablayrolles, Héliou, Martin, Agarwal, Roux, Darcet, Mensch, Bout, Rozière, Monicault, Bamford, Wallenwein, Renaudin, Lanfranchi, Dabert, Chaplot, Mizelle, de~las Casas, Chane-Sane, Fugier, Hanna, Berrada, Delerce, Guinet, Novikov, Martin, Jaju, Ludziejewski, Rute, Chabran, Chudnovsky, Studnia, Barmentlo, Amar, Roberts, Denize, Saxena, Yadav, Khandelwal, Jain, Lavaud, Blier, Zhao, Martin, Saulnier, Gao, Pellat, Guillaumin, Felardos, Dinot, Darrin, Augustin, Seznec, Gupta, Raghuraman, Duchenne, Wang, Saffer, Jacob, Wambergue, Kurylowicz, Chagniot, Stock, Agrawal, Delacourt, Sauvestre, Soletskyi, Vaze, Subramanian, Garg, Dalal, Gandhi, Aithal, Antoniak, Scao, Schueller, Lavril, Robert, Wang, Lacroix, Bewley, Nemychnikova, Paltz, Richard, Li, Marshall, Zhang, Wan, and Tang}]{liu2025voxtral}
Alexander~H. Liu, Andy Ehrenberg, Andy Lo, Clément Denoix, Corentin Barreau, Guillaume Lample, Jean-Malo Delignon, Khyathi~Raghavi Chandu, Patrick von Platen, Pavankumar~Reddy Muddireddy, Sanchit Gandhi, Soham Ghosh, Srijan Mishra, Thomas Foubert, Abhinav Rastogi, Adam Yang, Albert~Q. Jiang, Alexandre Sablayrolles, Amélie Héliou, and 87 others. 2025{\natexlab{a}}.
\newblock \href {https://arxiv.org/abs/2507.13264} {Voxtral}.
\newblock \emph{Preprint}, arXiv:2507.13264.

\bibitem[{Liu et~al.(2023)Liu, Hussain, Sun, and Shan}]{mullama}
Shansong Liu, Atin~Sakkeer Hussain, Chenshuo Sun, and Ying Shan. 2023.
\newblock \href {https://arxiv.org/abs/2308.11276} {Music understanding llama: Advancing text-to-music generation with question answering and captioning}.
\newblock \emph{Preprint}, arXiv:2308.11276.

\bibitem[{Liu et~al.(2025{\natexlab{b}})Liu, Niu, Xiao, Zheng, Yuan, Zang, Cao, Dong, Liang, Chen, Sun, Lin, and Wang}]{liu2025starbench}
Zihan Liu, Zhikang Niu, Qiuyang Xiao, Zhisheng Zheng, Ruoqi Yuan, Yuhang Zang, Yuhang Cao, Xiaoyi Dong, Jianze Liang, Xie Chen, Leilei Sun, Dahua Lin, and Jiaqi Wang. 2025{\natexlab{b}}.
\newblock \href {https://arxiv.org/abs/2510.24693} {Star-bench: Probing deep spatio-temporal reasoning as audio 4d intelligence}.
\newblock \emph{Preprint}, arXiv:2510.24693.

\bibitem[{Liu et~al.(2025{\natexlab{c}})Liu, Dong, Wang, Liu, Hu, Lu, and Rao}]{ola}
Zuyan Liu, Yuhao Dong, Jiahui Wang, Ziwei Liu, Winston Hu, Jiwen Lu, and Yongming Rao. 2025{\natexlab{c}}.
\newblock \href {https://arxiv.org/abs/2502.04328} {Ola: Pushing the frontiers of omni-modal language model}.
\newblock \emph{Preprint}, arXiv:2502.04328.

\bibitem[{Long et~al.(2025)Long, Shen, Fu, Gao, Li, Chen, Zhang, Shao, Li, Peng, Cao, Li, Ji, and Sun}]{vitaaudio}
Zuwei Long, Yunhang Shen, Chaoyou Fu, Heting Gao, Lijiang Li, Peixian Chen, Mengdan Zhang, Hang Shao, Jian Li, Jinlong Peng, Haoyu Cao, Ke~Li, Rongrong Ji, and Xing Sun. 2025.
\newblock \href {https://arxiv.org/abs/2505.03739} {Vita-audio: Fast interleaved cross-modal token generation for efficient large speech-language model}.
\newblock \emph{Preprint}, arXiv:2505.03739.

\bibitem[{Luo et~al.(2025)Luo, Lin, Zhang, Wu, Liu, Yang, Li, Chen, Li, Zhang, Xia, Alinejad-Rokny, and Huang}]{openomni}
Run Luo, Ting-En Lin, Haonan Zhang, Yuchuan Wu, Xiong Liu, Min Yang, Yongbin Li, Longze Chen, Jiaming Li, Lei Zhang, Xiaobo Xia, Hamid Alinejad-Rokny, and Fei Huang. 2025.
\newblock \href {https://arxiv.org/abs/2501.04561} {Openomni: Advancing open-source omnimodal large language models with progressive multimodal alignment and real-time self-aware emotional speech synthesis}.
\newblock \emph{Preprint}, arXiv:2501.04561.

\bibitem[{Ma et~al.(2025)Ma, Ma, Zhu, Yang, Chao, Xu et~al.}]{ma2025mmar}
Ziyang Ma, Yinghao Ma, Yanqiao Zhu, Chen Yang, Yi-Wen Chao, Ruiyang Xu, and 1 others. 2025.
\newblock Mmar: A challenging benchmark for deep reasoning in speech, audio, music, and their mix.
\newblock \emph{arXiv preprint arXiv:2505.13032}.

\bibitem[{Makous and Middlebrooks(1990)}]{makous1990twodim}
James~C Makous and John~C Middlebrooks. 1990.
\newblock Two-dimensional sound localization by human listeners.
\newblock \emph{The journal of the Acoustical Society of America}, 87(5):2188--2200.

\bibitem[{Mapelli and Behrmann(1997)}]{mapelli1997role}
Daniela Mapelli and Marlene Behrmann. 1997.
\newblock The role of color in object recognition: Evidence from visual agnosia.
\newblock \emph{Neurocase}, 3(4):237--247.

\bibitem[{Mathias(2010)}]{mathias2010individual}
Samuel Mathias. 2010.
\newblock \emph{Individual Differences in Pitch Perception}.
\newblock Ph.D. thesis, University of York.

\bibitem[{McDermott and Simoncelli(2011)}]{MCDERMOTT2011926}
Josh H. McDermott and Eero P. Simoncelli. 2011.
\newblock \href {https://doi.org/10.1016/j.neuron.2011.06.032} {Sound texture perception via statistics of the auditory periphery: Evidence from sound synthesis}.
\newblock \emph{Neuron}, 71(5):926--940.

\bibitem[{Michon(1964)}]{michon1964studies}
John~A Michon. 1964.
\newblock Studies on subjective duration: I. differential sensitivity in the perception of repeated temporal intervals.
\newblock \emph{Acta Psychologica}, 22:441--450.

\bibitem[{Miller(1956)}]{miller1956magical}
George~A Miller. 1956.
\newblock The magical number seven, plus or minus two: Some limits on our capacity for processing information.
\newblock \emph{Psychological review}, 63(2):81.

\bibitem[{Mills(1958)}]{mills1958maa}
Allen~William Mills. 1958.
\newblock On the minimum audible angle.
\newblock \emph{The Journal of the Acoustical Society of America}, 30(4):237--246.

\bibitem[{{OpenAI}(2024)}]{gpt4o_blog}
{OpenAI}. 2024.
\newblock \href {https://openai.com/index/hello-gpt-4o/} {Hello gpt-4o}.

\bibitem[{Peng et~al.(2025)Peng, Wang, Li, Guo, Wang, Fang, Xi, Li, Li, Zhang, Wang, and Yu}]{peng2025surveyspeechllm}
Jing Peng, Yucheng Wang, Bohan Li, Yiwei Guo, Hankun Wang, Yangui Fang, Yu~Xi, Haoyu Li, Xu~Li, Ke~Zhang, Shuai Wang, and Kai Yu. 2025.
\newblock \href {https://arxiv.org/abs/2410.18908} {A survey on speech large language models for understanding}.
\newblock \emph{Preprint}, arXiv:2410.18908.

\bibitem[{Piczak(2015)}]{esc50}
Karol~J. Piczak. 2015.
\newblock \href {https://doi.org/10.1145/2733373.2806390} {{ESC}: {Dataset} for {Environmental Sound Classification}}.
\newblock In \emph{Proceedings of the 23rd {Annual ACM Conference} on {Multimedia}}, pages 1015--1018. {ACM Press}.

\bibitem[{Radford et~al.(2022)Radford, Kim, Xu, Brockman, McLeavey, and Sutskever}]{radford2022robustspeechrecognitionlargescale}
Alec Radford, Jong~Wook Kim, Tao Xu, Greg Brockman, Christine McLeavey, and Ilya Sutskever. 2022.
\newblock \href {https://arxiv.org/abs/2212.04356} {Robust speech recognition via large-scale weak supervision}.
\newblock \emph{Preprint}, arXiv:2212.04356.

\bibitem[{Rammsayer and Ulrich(2012)}]{rammsayer2012greater}
Thomas Rammsayer and Rolf Ulrich. 2012.
\newblock The greater temporal acuity in the reminder task than in the 2afc task is independent of standard duration and sensory modality.
\newblock \emph{Canadian Journal of Experimental Psychology/Revue canadienne de psychologie exp{\'e}rimentale}, 66(1):26.

\bibitem[{Rammsayer(2010)}]{rammsayer2010differences}
Thomas~H Rammsayer. 2010.
\newblock Differences in duration discrimination of filled and empty auditory intervals as a function of base duration.
\newblock \emph{Attention, Perception, \& Psychophysics}, 72(6):1591--1600.

\bibitem[{Saitis and Siedenburg(2020)}]{saitis2020brightness}
Charalampos Saitis and Kai Siedenburg. 2020.
\newblock Brightness perception for musical instrument sounds: Relation to timbre dissimilarity and source-cause categories.
\newblock \emph{The Journal of the Acoustical Society of America}, 148(4):2256--2266.

\bibitem[{Sakshi et~al.(2024)Sakshi, Tyagi, Kumar, Seth, Selvakumar, Nieto, Duraiswami, Ghosh, and Manocha}]{Sakshi2024mmau}
S~Sakshi, Utkarsh Tyagi, Sonal Kumar, Ashish Seth, Ramaneswaran Selvakumar, Oriol Nieto, Ramani Duraiswami, Sreyan Ghosh, and Dinesh Manocha. 2024.
\newblock \href {https://arxiv.org/abs/2410.19168} {Mmau: A massive multi-task audio understanding and reasoning benchmark}.
\newblock \emph{Preprint}, arXiv:2410.19168.

\bibitem[{Schubert and Wolfe(2006)}]{schubert2006does}
Emery Schubert and Joe Wolfe. 2006.
\newblock Does timbral brightness scale with frequency and spectral centroid?
\newblock \emph{Acta acustica united with acustica}, 92(5):820--825.

\bibitem[{Shimada et~al.(2023)Shimada, Politis, Sudarsanam, Krause, Uchida, Adavanne, Hakala, Koyama, Takahashi, Takahashi, Virtanen, and Mitsufuji}]{shimada2023starss23}
Kazuki Shimada, Archontis Politis, Parthasaarathy Sudarsanam, Daniel Krause, Kengo Uchida, Sharath Adavanne, Aapo Hakala, Yuichiro Koyama, Naoya Takahashi, Shusuke Takahashi, Tuomas Virtanen, and Yuki Mitsufuji. 2023.
\newblock \href {https://arxiv.org/abs/2306.09126} {Starss23: An audio-visual dataset of spatial recordings of real scenes with spatiotemporal annotations of sound events}.
\newblock \emph{Preprint}, arXiv:2306.09126.

\bibitem[{Stewart et~al.(2005)Stewart, Brown, and Chater}]{stewart2005absolute}
Neil Stewart, Gordon~DA Brown, and Nick Chater. 2005.
\newblock Absolute identification by relative judgment.
\newblock \emph{Psychological review}, 112(4):881.

\bibitem[{Sun et~al.(2017)Sun, Lu, Ho, and Thompson}]{sun2017pitch}
Yanan Sun, Xuejing Lu, Hao~Tam Ho, and William~Forde Thompson. 2017.
\newblock Pitch discrimination associated with phonological awareness: Evidence from congenital amusia.
\newblock \emph{Scientific Reports}, 7(1):44285.

\bibitem[{Sun et~al.(2025)Sun, Geng, Wei, Chen, Yang, Chen, Zhang, and Shen}]{sun2025llaso}
Yirong Sun, Yizhong Geng, Peidong Wei, Yanjun Chen, Jinghan Yang, Rongfei Chen, Wei Zhang, and Xiaoyu Shen. 2025.
\newblock \href {https://arxiv.org/abs/2508.15418} {Llaso: A foundational framework for reproducible research in large language and speech model}.
\newblock \emph{Preprint}, arXiv:2508.15418.

\bibitem[{Tang et~al.(2024)Tang, Yu, Sun, Chen, Tan, Li, Lu, Ma, and Zhang}]{tang2024salmonn}
Changli Tang, Wenyi Yu, Guangzhi Sun, Xianzhao Chen, Tian Tan, Wei Li, Lu~Lu, Zejun Ma, and Chao Zhang. 2024.
\newblock \href {https://arxiv.org/abs/2310.13289} {Salmonn: Towards generic hearing abilities for large language models}.
\newblock \emph{Preprint}, arXiv:2310.13289.

\bibitem[{Team(2024)}]{qwen2.5}
Qwen Team. 2024.
\newblock \href {https://qwenlm.github.io/blog/qwen2.5/} {Qwen2.5: A party of foundation models}.

\bibitem[{ten Hoopen et~al.(2004)ten Hoopen, van~den Berg, Memelink, Bocanegra, and Boon}]{ten2004multiple}
Gert ten Hoopen, Stephanie van~den Berg, Jiska Memelink, B~Bocanegra, and R~Boon. 2004.
\newblock Multiple looks on temporal discrimination in sound sequences.
\newblock \emph{Transactions of Technical Committee of Psychological and Physiological Acoustics—The Acoustical Society of Japan}, 34:693--700.

\bibitem[{Tian et~al.(2025)Tian, Zhang, Zhang, Zhang, Li, Liu, Deng, Wu, Chen, Zhao et~al.}]{tian2025step}
Fei Tian, Xiangyu~Tony Zhang, Yuxin Zhang, Haoyang Zhang, Yuxin Li, Daijiao Liu, Yayue Deng, Donghang Wu, Jun Chen, Liang Zhao, and 1 others. 2025.
\newblock Step-audio-r1 technical report.
\newblock \emph{arXiv preprint arXiv:2511.15848}.

\bibitem[{Town and Bizley(2013)}]{town2013neural}
Stephen~M Town and Jennifer~K Bizley. 2013.
\newblock Neural and behavioral investigations into timbre perception.
\newblock \emph{Frontiers in systems neuroscience}, 7:88.

\bibitem[{V{\"a}lim{\"a}ki and Reiss(2016)}]{valimaki2016all}
Vesa V{\"a}lim{\"a}ki and Joshua~D Reiss. 2016.
\newblock All about audio equalization: Solutions and frontiers.
\newblock \emph{Applied Sciences}, 6(5):129.

\bibitem[{Wang et~al.(2025{\natexlab{a}})Wang, Zou, Lin, Sun, Liu, Zhang, Liu, Aw, and Chen}]{wang2024audiobench}
Bin Wang, Xunlong Zou, Geyu Lin, Shuo Sun, Zhuohan Liu, Wenyu Zhang, Zhengyuan Liu, AiTi Aw, and Nancy~F Chen. 2025{\natexlab{a}}.
\newblock Audiobench: A universal benchmark for audio large language models.
\newblock \emph{NAACL}.

\bibitem[{Wang et~al.(2025{\natexlab{b}})Wang, Wu, Li, Yang, Chen, Zhang, and Meng}]{wang2025mmsu}
Dingdong Wang, Jincenzi Wu, Junan Li, Dongchao Yang, Xueyuan Chen, Tianhua Zhang, and Helen Meng. 2025{\natexlab{b}}.
\newblock Mmsu: A massive multi-task spoken language understanding and reasoning benchmark.
\newblock \emph{arXiv preprint arXiv:2506.04779}.

\bibitem[{Warren et~al.(2005)Warren, Jennings, and Griffiths}]{warren2005analysis}
Joe~D Warren, AR~Jennings, and Timothy~D Griffiths. 2005.
\newblock Analysis of the spectral envelope of sounds by the human brain.
\newblock \emph{Neuroimage}, 24(4):1052--1057.

\bibitem[{Weck et~al.(2024)Weck, Manco, Benetos, Quinton, Fazekas, and Bogdanov}]{weck2024muchomusic}
Benno Weck, Ilaria Manco, Emmanouil Benetos, Elio Quinton, George Fazekas, and Dmitry Bogdanov. 2024.
\newblock \href {https://arxiv.org/abs/2408.01337} {Muchomusic: Evaluating music understanding in multimodal audio-language models}.
\newblock \emph{Preprint}, arXiv:2408.01337.

\bibitem[{wen Yang et~al.(2021)wen Yang, Chi, Chuang, Lai, Lakhotia, Lin, Liu, Shi, Chang, Lin, Huang, Tseng, tik Lee, Liu, Huang, Dong, Li, Watanabe, Mohamed, and yi~Lee}]{superb}
Shu wen Yang, Po-Han Chi, Yung-Sung Chuang, Cheng-I~Jeff Lai, Kushal Lakhotia, Yist~Y. Lin, Andy~T. Liu, Jiatong Shi, Xuankai Chang, Guan-Ting Lin, Tzu-Hsien Huang, Wei-Cheng Tseng, Ko~tik Lee, Da-Rong Liu, Zili Huang, Shuyan Dong, Shang-Wen Li, Shinji Watanabe, Abdelrahman Mohamed, and Hung yi~Lee. 2021.
\newblock \href {https://arxiv.org/abs/2105.01051} {Superb: Speech processing universal performance benchmark}.
\newblock \emph{Preprint}, arXiv:2105.01051.

\bibitem[{Wu et~al.(2025)Wu, Yan, Hu, Yi, Feng, Tian, Shen, Yu, Zhang, Li, Chen, Liu, You, Zhang, Li, Yang, Deng, Huang, Li, Zhang, You, Li, Wan, Hu, Zhen, Chen, Yuan, Zhang, Jiang, Zhou, Yang, Li, Ma, Song, Pang, Hu, Sun, An, Wang, Gao, Ji, Li, Sun, Wen, Ren, Ma, Lu, Wang, Li, Miao, Liu, Xu, Shi, Hu, Wu, Liu, Huang, Yan, Zhang, Nie, Jia, Zhou, Sun, Wu, Wu, Yang, Yang, Lin, Li, Yang, Shi, Zhou, Gu, Li, Li, Li, Wu, Han, Tan, Pang, Fan, Liu, Cao, Lu, He, Xie, Zhao, Li, Yu, Yang, Liu, Lu, Wang, Ding, Liang, Lu, Luo, Yin, Zhan, Zhang, Yang, Zhang, Jiao, Jiang, Shum, Chen, Li, Zhang, and Zhu}]{stepaudio2}
Boyong Wu, Chao Yan, Chen Hu, Cheng Yi, Chengli Feng, Fei Tian, Feiyu Shen, Gang Yu, Haoyang Zhang, Jingbei Li, Mingrui Chen, Peng Liu, Wang You, Xiangyu~Tony Zhang, Xingyuan Li, Xuerui Yang, Yayue Deng, Yechang Huang, Yuxin Li, and 90 others. 2025.
\newblock \href {https://arxiv.org/abs/2507.16632} {Step-audio 2 technical report}.
\newblock \emph{Preprint}, arXiv:2507.16632.

\bibitem[{Wüst et~al.(2025)Wüst, Woydt, Helff, Ibs, Stammer, Dhami, Rothkopf, and Kersting}]{wüst2025bongard}
Antonia Wüst, Tim Woydt, Lukas Helff, Inga Ibs, Wolfgang Stammer, Devendra~S. Dhami, Constantin~A. Rothkopf, and Kristian Kersting. 2025.
\newblock \href {https://arxiv.org/abs/2410.19546} {Bongard in wonderland: Visual puzzles that still make ai go mad?}
\newblock \emph{Preprint}, arXiv:2410.19546.

\bibitem[{Xiaomi(2025)}]{mimoaudio}
LLM-Core-Team Xiaomi. 2025.
\newblock \href {https://github.com/XiaomiMiMo/MiMo-Audio} {Mimo-audio: Audio language models are few-shot learners}.

\bibitem[{Xie et~al.(2024)Xie, Xu, Wu, and Wu}]{picoaudio}
Zeyu Xie, Xuenan Xu, Zhizheng Wu, and Mengyue Wu. 2024.
\newblock \href {https://arxiv.org/abs/2407.02869} {Picoaudio: Enabling precise timestamp and frequency controllability of audio events in text-to-audio generation}.
\newblock \emph{Preprint}, arXiv:2407.02869.

\bibitem[{Xie et~al.(2025)Xie, Lin, Liu, Wu, Yan, and Miao}]{audioreasoner}
Zhifei Xie, Mingbao Lin, Zihang Liu, Pengcheng Wu, Shuicheng Yan, and Chunyan Miao. 2025.
\newblock \href {https://arxiv.org/abs/2503.02318} {Audio-reasoner: Improving reasoning capability in large audio language models}.
\newblock \emph{Preprint}, arXiv:2503.02318.

\bibitem[{Xu et~al.(2025{\natexlab{a}})Xu, Guo, He, Hu, He, Bai, Chen, Wang, Fan, Dang, Zhang, Wang, Chu, and Lin}]{qwen25omni}
Jin Xu, Zhifang Guo, Jinzheng He, Hangrui Hu, Ting He, Shuai Bai, Keqin Chen, Jialin Wang, Yang Fan, Kai Dang, Bin Zhang, Xiong Wang, Yunfei Chu, and Junyang Lin. 2025{\natexlab{a}}.
\newblock \href {https://arxiv.org/abs/2503.20215} {Qwen2.5-omni technical report}.
\newblock \emph{Preprint}, arXiv:2503.20215.

\bibitem[{Xu et~al.(2025{\natexlab{b}})Xu, Guo, Hu, Chu, Wang, He, Wang, Shi, He, Zhu, Lv, Wang, Guo, Wang, Ma, Zhang, Zhang, Hao, Guo, Yang, Zhang, Ma, Wei, Bai, Chen, Liu, Wang, Yang, Liu, Ren, Zheng, Men, Zhou, Yu, Yang, Yu, Zhou, and Lin}]{xu2025qwen3omni}
Jin Xu, Zhifang Guo, Hangrui Hu, Yunfei Chu, Xiong Wang, Jinzheng He, Yuxuan Wang, Xian Shi, Ting He, Xinfa Zhu, Yuanjun Lv, Yongqi Wang, Dake Guo, He~Wang, Linhan Ma, Pei Zhang, Xinyu Zhang, Hongkun Hao, Zishan Guo, and 19 others. 2025{\natexlab{b}}.
\newblock \href {https://arxiv.org/abs/2509.17765} {Qwen3-omni technical report}.
\newblock \emph{Preprint}, arXiv:2509.17765.

\bibitem[{Xu et~al.(2025{\natexlab{c}})Xu, Guo, Hu, Chu, Wang, He, Wang, Shi, He, Zhu, Lv, Wang, Guo, Wang, Ma, Zhang, Zhang, Hao, Guo, Yang, Zhang, Ma, Wei, Bai, Chen, Liu, Wang, Yang, Liu, Ren, Zheng, Men, Zhou, Yu, Yang, Yu, Zhou, and Lin}]{Qwen3-Omni}
Jin Xu, Zhifang Guo, Hangrui Hu, Yunfei Chu, Xiong Wang, Jinzheng He, Yuxuan Wang, Xian Shi, Ting He, Xinfa Zhu, Yuanjun Lv, Yongqi Wang, Dake Guo, He~Wang, Linhan Ma, Pei Zhang, Xinyu Zhang, Hongkun Hao, Zishan Guo, and 19 others. 2025{\natexlab{c}}.
\newblock Qwen3-omni technical report.
\newblock \emph{arXiv preprint arXiv:2509.17765}.

\bibitem[{Xu et~al.(2025{\natexlab{d}})Xu, Xie, Tang, and Hu}]{xu2025fireredasr}
Kai-Tuo Xu, Feng-Long Xie, Xu~Tang, and Yao Hu. 2025{\natexlab{d}}.
\newblock Fireredasr: Open-source industrial-grade mandarin speech recognition models from encoder-decoder to llm integration.
\newblock \emph{arXiv preprint arXiv:2501.14350}.

\bibitem[{Yang et~al.(2024{\natexlab{a}})Yang, Quan, Wang, Wang, Yang, Fang, Shao, Bu, Xu, and Li}]{realman}
Bing Yang, Changsheng Quan, Yabo Wang, Pengyu Wang, Yujie Yang, Ying Fang, Nian Shao, Hui Bu, Xin Xu, and Xiaofei Li. 2024{\natexlab{a}}.
\newblock \href {https://arxiv.org/abs/2406.19959} {Realman: A real-recorded and annotated microphone array dataset for dynamic speech enhancement and localization}.
\newblock \emph{Preprint}, arXiv:2406.19959.

\bibitem[{Yang et~al.(2024{\natexlab{b}})Yang, Xu, Liu, Chu, Jiang, Zhou, Leng, Lv, Zhao, Zhou, and Zhou}]{yang2024airbench}
Qian Yang, Jin Xu, Wenrui Liu, Yunfei Chu, Ziyue Jiang, Xiaohuan Zhou, Yichong Leng, Yuanjun Lv, Zhou Zhao, Chang Zhou, and Jingren Zhou. 2024{\natexlab{b}}.
\newblock \href {https://arxiv.org/abs/2402.07729} {Air-bench: Benchmarking large audio-language models via generative comprehension}.
\newblock \emph{Preprint}, arXiv:2402.07729.

\bibitem[{yu~Huang et~al.(2025)yu~Huang, Chen, wen Yang, Liu, Li, Lin, Tseng, Diwan, Shih, Shi, Chen, Yang, Ren, Chen, Hsiao, Peng, Wang, Kuan, Lu, Chang, Ritter-Gutierrez, Huang, Arora, Lin, Chuang, Yeo, Chang, Chien, Choi, Wang, Hsieh, Lin, Yu, Chiu, Guimarães, Han, Lin, Lin, Chang, Chang, Chen, Chen, Chen, Cheng, Dhawan, Fang, Fang, Chiang, Fu, Hsiao, Hsu, Huang, Wei, Lin, Lin, Lin, Lin, Liu, Lu, Pai, Pasad, Kuan, Shon, Tang, Tsai, Wei, Wei, Wu, Wu, Yang, Yang, Yip, Yuan, Noroozi, Chen, Wu, Livescu, Harwath, Watanabe, and yi~Lee}]{huang2025dynamicsuperbphase2}
Chien yu~Huang, Wei-Chih Chen, Shu wen Yang, Andy~T. Liu, Chen-An Li, Yu-Xiang Lin, Wei-Cheng Tseng, Anuj Diwan, Yi-Jen Shih, Jiatong Shi, William Chen, Chih-Kai Yang, Wenze Ren, Xuanjun Chen, Chi-Yuan Hsiao, Puyuan Peng, Shih-Heng Wang, Chun-Yi Kuan, Ke-Han Lu, and 61 others. 2025.
\newblock \href {https://arxiv.org/abs/2411.05361} {Dynamic-superb phase-2: A collaboratively expanding benchmark for measuring the capabilities of spoken language models with 180 tasks}.
\newblock \emph{Preprint}, arXiv:2411.05361.

\bibitem[{Zahorik(2002)}]{zahorik2002direct}
Pavel Zahorik. 2002.
\newblock Direct-to-reverberant energy ratio sensitivity.
\newblock \emph{The Journal of the Acoustical Society of America}, 112(5):2110--2117.

\bibitem[{Zahorik et~al.(2005)Zahorik, Brungart, and Bronkhorst}]{zahorik2005auditory}
Pavel Zahorik, Douglas~S Brungart, and Adelbert~W Bronkhorst. 2005.
\newblock Auditory distance perception in humans: A summary of past and present research.
\newblock \emph{ACTA Acustica united with Acustica}, 91(3):409--420.

\bibitem[{Zarate et~al.(2012)Zarate, Ritson, and Poeppel}]{zarate2012pitch}
Jean~Mary Zarate, Caroline~R Ritson, and David Poeppel. 2012.
\newblock Pitch-interval discrimination and musical expertise: Is the semitone a perceptual boundary?
\newblock \emph{The Journal of the Acoustical Society of America}, 132(2):984--993.

\bibitem[{Zeng et~al.(2024)Zeng, Du, Liu, Wang, Jiang, Zhao, Dong, and Tang}]{glm4voice}
Aohan Zeng, Zhengxiao Du, Mingdao Liu, Kedong Wang, Shengmin Jiang, Lei Zhao, Yuxiao Dong, and Jie Tang. 2024.
\newblock \href {https://arxiv.org/abs/2412.02612} {Glm-4-voice: Towards intelligent and human-like end-to-end spoken chatbot}.
\newblock \emph{Preprint}, arXiv:2412.02612.

\bibitem[{Zhang et~al.(2022)Zhang, Lv, Guo, Shao, Yang, Xie, Xu, Bu, Chen, Zeng, Wu, and Peng}]{zhang2022wenetspeech10000hours}
Binbin Zhang, Hang Lv, Pengcheng Guo, Qijie Shao, Chao Yang, Lei Xie, Xin Xu, Hui Bu, Xiaoyu Chen, Chenchen Zeng, Di~Wu, and Zhendong Peng. 2022.
\newblock \href {https://arxiv.org/abs/2110.03370} {Wenetspeech: A 10000+ hours multi-domain mandarin corpus for speech recognition}.
\newblock \emph{Preprint}, arXiv:2110.03370.

\bibitem[{Zhao et~al.(2025)Zhao, Yang, Peng, Bai, Yao, Sun, Chen, Fu, chen, Wei, and Bo}]{humanomni}
Jiaxing Zhao, Qize Yang, Yixing Peng, Detao Bai, Shimin Yao, Boyuan Sun, Xiang Chen, Shenghao Fu, Weixuan chen, Xihan Wei, and Liefeng Bo. 2025.
\newblock \href {https://arxiv.org/abs/2501.15111} {Humanomni: A large vision-speech language model for human-centric video understanding}.
\newblock \emph{Preprint}, arXiv:2501.15111.

\bibitem[{Zheng et~al.(2023)Zheng, Chiang, Sheng, Zhuang, Wu, Zhuang, Lin, Li, Li, Xing, Zhang, Gonzalez, and Stoica}]{zheng2023vicuna}
Lianmin Zheng, Wei-Lin Chiang, Ying Sheng, Siyuan Zhuang, Zhanghao Wu, Yonghao Zhuang, Zi~Lin, Zhuohan Li, Dacheng Li, Eric~P. Xing, Hao Zhang, Joseph~E. Gonzalez, and Ion Stoica. 2023.
\newblock \href {https://arxiv.org/abs/2306.05685} {Judging llm-as-a-judge with mt-bench and chatbot arena}.
\newblock \emph{Preprint}, arXiv:2306.05685.

\bibitem[{Zheng et~al.(2025{\natexlab{a}})Zheng, Peng, Ma, Chen, Choi, and Harwath}]{zheng2025bat}
Zhisheng Zheng, Puyuan Peng, Ziyang Ma, Xie Chen, Eunsol Choi, and David Harwath. 2025{\natexlab{a}}.
\newblock \href {https://arxiv.org/abs/2402.01591} {Bat: Learning to reason about spatial sounds with large language models}.
\newblock \emph{Preprint}, arXiv:2402.01591.

\bibitem[{Zheng et~al.(2025{\natexlab{b}})Zheng, Peng, Ma, Chen, Choi, and Harwath}]{bat}
Zhisheng Zheng, Puyuan Peng, Ziyang Ma, Xie Chen, Eunsol Choi, and David Harwath. 2025{\natexlab{b}}.
\newblock \href {https://arxiv.org/abs/2402.01591} {Bat: Learning to reason about spatial sounds with large language models}.
\newblock \emph{Preprint}, arXiv:2402.01591.

\end{thebibliography}

\appendix

\section{Attribute Details}
\label{appendix:attri_details}
In this section, we provide a comprehensive breakdown of the 12 physical attributes that constitute the SonicBench taxonomy. To ensure a systematic evaluation of auditory perception, we organize these attributes into five core dimensions. Each attribute is selected not merely for its acoustic measurability, but for its foundational role in how humans parse and interpret the auditory world, ranging from identifying source proximity to tracking rhythmic patterns. Below, we detail the physical definition, psychophysical relevance, and specific task implementation for each dimension, clarifying how we isolate these signal-level properties from high-level semantic context.

\subsection{Spectral \& Amplitude}
This dimension targets the low-level physical structure of sound in the frequency and energy domain cues that are present in virtually every everyday acoustic event and form the perceptual bedrock of human hearing. Humans continuously rely on these spectral and level patterns to recognise sources, separate overlapping streams, and make rapid safety-critical judgements e.g., detecting sharp, high-pitched alarms or unusually loud impacts. For audio-enabled systems and embodied agents, robust representations of these properties are therefore not a niche skill such as music transcription but a foundational substrate on which a wide range of downstream behaviours must build. In SonicBench, we instantiate this dimension with following four attributes:

\paragraph{Pitch.} 
Pitch denotes the perceived height of a sound and varies approximately logarithmically with frequency. It underpins a wide range of behaviours. Beyond music-related tasks such as melody following, humans rely on pitch reused across many contexts to interpret speech prosody, detect high-pitched alarms or sharp impacts, and make safety-critical decisions in interactive settings. Psychophysical studies show that pitch sensitivity is remarkably fine-grained. In pure-tone discrimination, typical adults exhibit JNDs of around 10-20 cents, and trained listeners can reach 3-4 cents near 1~kHz roughly 0.2\%~\cite{bianchi2016pitch,mathias2010individual,zarate2012pitch,sun2017pitch}. In SonicBench, we deliberately work in a much easier regime, using 1-semitone (100-cent) contrasts that are trivial for normal-hearing listeners. This coarse, forced-choice design avoids borderline detectability and turns pitch into a foundational probe of whether models encode stable, human-like pitch representations that can in principle support more complex downstream reasoning.

\paragraph{Brightness.} 
Brightness characterizes how sharp or dull a sound is, reflecting how much spectral energy is concentrated at higher versus lower frequencies. This cue is central to timbre and voice perception, equalization and mixing decisions, and everyday source or scene identification~\cite{warren2005analysis,valimaki2016all,town2013neural}. Psychoacoustic work shows that perceived brightness is strongly linked to spectral centroid and related high-frequency energy measures, and that listeners can reliably order and discriminate changes along this dimension across a wide range of instruments and environmental sounds~\cite{schubert2006does,almeida2017brightness,saitis2020brightness}. In SonicBench, we do not analytically manipulate the centroid; instead, we draw on community-endorsed corpora with brightness annotations and apply manual screening to ensure that paired items are clearly separable to human listeners~\cite{nsynth}. This yields a coarse, forced-choice probe of whether models encode a stable notion of spectral balance, rather than merely reacting to extreme or degenerate spectral conditions.

\paragraph{Loudness.} 
Loudness characterizes the perceived intensity of a sound rather than its raw signal level. It is central to speech intelligibility, mix balance, and attentional salience in everyday listening, for example when a listener must follow one talker in noise or react to a sudden, loud alert. Psychophysical studies show that normal-hearing listeners are highly sensitive to level differences: for broadband or complex stimuli, just-noticeable changes are typically on the order of 0.5--1~dB, and for pure tones the difference limen can drop to roughly 0.2--0.6~dB at higher sensation levels~\cite{florentine1987level,hall2008auditory}. In SonicBench, we operationalize loudness using integrated LUFS with K-weighting following ITU-R BS.1770 and EBU R128~\cite{iturbs1770,ebur128}, and construct contrasts that sit well above these thresholds (e.g., $\geq$2--3~LU). This coarse, forced-choice design ensures that the task is trivial for human listeners and probes whether models encode a robust representation of perceived loudness, rather than merely skirting the boundary of detectability.

\paragraph{Velocity.} 
Velocity captures how forcefully a sound-producing action is executed, shaping both the sharpness of its onset and its overall energy. In everyday listening, many impact- and contact-driven sounds e.g., knocking, footsteps, percussion, keyboard instruments change systematically with the speed and force of the underlying motion, and listeners rely on these cues to infer effort, material, and even affect~\cite{giordano2005impactsounds,giordano2006materials}. In practice, this notion is often operationalized as a control parameter that co-regulates loudness and spectral brightness, for example MIDI velocity (1-127) in digital or computer-controlled instruments, which correlates closely with hammer speed and radiated level on systems such as the Yamaha Disklavier~\cite{goebl2001disklavier}. Behavioural studies on impact and performance sounds show that humans are highly sensitive to such variations and can reliably distinguish differences in playing effort and dynamic strength well within this control range~\cite{giordano2005impactsounds,lemaitre2012action}. In SonicBench, we instantiate this attribute using recordings with explicit velocity-related control and construct pairs that differ by a substantial margin in excitation strength while keeping pitch and other factors as stable as possible, yielding a coarse, forced-choice probe of whether models encode a robust representation of dynamic impact strength rather than merely reacting to incidental loudness fluctuations.

\subsection{Temporal}
This dimension captures how acoustic events unfold over time, from the duration of individual segments to the pacing of repeated patterns. Temporal cues are a core low-level property of the signal, yet humans routinely recruit them for higher-level judgments, using duration and pacing to segment actions, perceive urgency, and distinguish slow, heavy motion from fast, agile motion. For audio understanding models, such temporal understanding is increasingly a foundational requirement, beyond answering explicit timing questions, models in complex settings such as interactive assistants should be able to infer changes in speed, intent, or urgency from patterns like footsteps gradually accelerating or alarms changing rate. We therefore instantiate this dimension with two complementary attributes, Duration and Tempo, which targets sensitivity to absolute interval length and probes how well models track relative pacing across repeated or patterned sounds.

\paragraph{Duration.}
Duration denotes the physical time span between the onset and offset of an isolated sound event and serves as a basic probe of temporal awareness. Psychophysical studies of auditory interval discrimination in the hundreds-of-milliseconds to seconds range suggest that sensitivity is well approximated by a roughly constant Weber fraction on the order of 10--15\%, i.e., a 600~ms reference typically requires a change of about 60--90~ms to be reliably discriminated~\cite{grondin1993duration,rammsayer2010differences,rammsayer2012greater}. This stable, low-level acuity makes duration a natural baseline for assessing whether models have acquired a human-like temporal scaffold on which more complex timing judgements can build. In SonicBench, we therefore construct duration contrasts that sit well above this regime e.g., differences of 30--50\% relative to the base interval, turning the task into a coarse, forced-choice probe of temporal grounding: we test whether models can represent absolute time, segment events, and compare intervals in a robust way, rather than merely responding to fine-grained threshold differences.

\paragraph{Tempo.} 
Tempo captures the rate at which a sequence of events unfolds, i.e., how quickly successive onsets occur over time. It is a property of the temporal pattern rather than an absolute interval: changing tempo alters the spacing between events while preserving their order. Humans routinely recruit tempo cues for higher-level inference-for example, using accelerating footsteps to infer a transition from walking to running, or changes in machine cycles to detect shifts in operating state-so robust tempo perception forms a foundational layer of temporal understanding on which more complex reasoning about actions and scenes can build. Psychophysical studies indicate that tempo discrimination approximately follows Weber’s law, with just-noticeable differences on the order of a few percent of the base rate roughly 4-8\% at moderate tempi~\cite{michon1964studies,drake1993tempo,boltz1998tempo,ten2004multiple}. In SonicBench, we operationalize tempo using rhythmic and musical excerpts with explicit beat annotations and construct pairs whose tempo differences are substantially larger than these human JNDs. This yields a coarse, forced-choice probe of whether models encode stable representations of relative pacing, rather than merely reacting to marginally detectable speed changes.

\subsection{Spatial \& Environment}
This dimension captures how sounds are embedded in space and in their surrounding acoustic environment whether a source is in front or behind, near or far, and in a dry booth or a reverberant hall. For humans, such cues are part of the low-level acoustics of every signal, yet they are constantly recruited for higher-level judgements, orienting toward a talker, judging whether an approaching sound is dangerously close, or inferring that a voice is echoing in a large indoor space. For LALMs and embodied agents, spatial and environmental grounding is therefore a foundational capacity rather than a niche skill, without a basic sense of direction, depth, and room context, models cannot reliably follow situated instructions, coordinate movement, or assess risk in real-world audio scenes. SonicBench instantiates this dimension with three attributes, Direction, Distance, and Reverberation, each framed as a simple binary judgement as Figure~\ref{fig:benchmark_overview} shown that isolates one spatial or environmental property while leaving other cues as controlled as possible.

\paragraph{Direction.}
Direction denotes the perceived azimuth of a sound source where it is located around the listener in the horizontal plane. Humans are highly sensitive to directional cues, especially for frontal sources, achieving minimum audible angles of only a few degrees under favourable conditions~\cite{mills1958maa,makous1990twodim}, supported by interaural time and level differences together with direction-dependent spectral filtering by the head and pinnae~\cite{blauert1997psychophysics}. This acuity underlies everyday behaviours such as orienting toward a talker, avoiding approaching vehicles, and monitoring events outside the visual field, and is equally crucial for LALMs deployed in embodied or interactive settings. In SonicBench, we treat direction as a coarse, foundational spatial attribute rather than a fine-grained localisation task. We restrict the horizontal plane to two clearly separated regions: a front sector spanning $\pm 60^\circ$ around $0^\circ$ and a back sector spanning $\pm 60^\circ$ around $180^\circ$. For each stimulus, we sample an azimuth uniformly within one sector and map it to the corresponding front/back label. These large angular separations far exceed human just-noticeable differences, yielding a coarse, forced-choice probe of whether models exhibit stable front-back awareness at all, rather than near-threshold localisation acuity.
    
\paragraph{Distance.} 
Distance denotes the perceived proximity of a sound source along the source-listener axis-how near or far it seems, rather than its physical distance in metres. Human distance perception draws on a combination of cues, including overall level falloff, spectral changes, and, in typical rooms, the direct-to-reverberant energy ratio (DRR). Psychophysical studies show that listeners can discriminate relatively small changes in source distance at near ranges, and that DRR becomes a dominant cue as distance increases in reverberant environments~\cite{zahorik2002direct,kolarik2016auditory}. Such proximity judgments are crucial for everyday behaviour and for LALMs in embodied settings, for example when deciding whether a sound source is close enough to warrant attention or action. In SonicBench, we therefore treat distance as a coarse, foundational attribute rather than a fine-grained ranging task. We partition stimuli into two broad bands, near versus far, separated by substantial changes in nominal source distance and DRR, and label them accordingly. These large separations lie well above typical human discrimination thresholds and yield a simple forced-choice probe of whether models encode any stable sense of auditory proximity.

\paragraph{Reverberation.} 
Reverberation captures whether a sound is perceived as occurring in a reflective environment or in an acoustically dry, near-anechoic setting. Psychophysically, reverberation arises from dense patterns of early reflections and late decay, and listeners are highly sensitive to these cues: even modest changes in reverberation time, direct-to-reverberant ratio (DRR), or early reflection structure can alter the perceived room size, source–room distance, and overall sense of envelopment~\cite{zahorik2005auditory,kuttruff2016room}. Such environmental impressions are critical for everyday reasoning about whether a sound is indoors or outdoors, in a small room or a large hall, or mediated by a strongly reflective space. In SonicBench, we therefore treat reverberation as a coarse, foundational environmental attribute rather than a fine-grained RT60 or DRR estimation task. Each item is constructed either as a dry signal with negligible room contribution or as a reverberant version obtained by convolving the same source with a realistic room impulse response, while keeping source content, level, and direction as stable as possible. This binary, far-above-threshold contrast turns the task into a robust forced-choice probe of whether models can reliably distinguish “in-room” versus “dry” conditions, a prerequisite for downstream behaviours such as inferring indoor/outdoor context, room scale, and whether a sound is likely mediated by an enclosed space.

\subsection{Timbre}
In contrast to the largely low-level physical cues above, the timbre dimension targets how spectral and temporal structure are integrated into mid-level representations of “what kind of sound this is.” Timbre allows listeners to tell a flute from a violin at the same pitch and loudness, or to distinguish crackling fire from rustling leaves, by jointly using spectral shape, temporal envelope, and fluctuations. This ability underpins everyday source and material recognition and acts as a bridge between raw acoustics and semantic understanding where an agent that cannot reliably infer “what is making this sound” or “what kind of background this is” will struggle to interpret scenes or follow natural, audio-grounded instructions. In SonicBench, we therefore instantiate this dimension with two complementary attributes, Timbre, which probes source-centred identity under controlled pitch and loudness, and Texture, which targets the statistical structure of dense, background-like sound fields such as rain, fire, or crowd noise.

\paragraph{Timbre.} 
Timbre is the attribute of auditory sensation that allows listeners to judge two sounds as different even when their pitch and loudness are matched. Perceptually, it reflects a mid-level integration of spectral shape, temporal envelope, and fine-structure cues, and underlies our ability to recognise “what is sounding” (e.g., instrument type, material, or mode of excitation) from very short excerpts. Human listeners are highly proficient at this kind of judgement: even brief, isolated notes often suffice to distinguish familiar instruments or everyday sound sources. In SonicBench, we operationalise timbre as source-centred discrimination under controlled conditions. We draw on existing corpora with reliable instrument labels, normalise pitch and loudness within each pair, and construct contrasts where the primary difference lies in timbral character rather than in fundamental frequency or overall level. Our goal is not to introduce yet another general-purpose instrument-classification benchmark, but to provide a pragmatic, well-controlled proxy for timbre perception within our broader taxonomy of physical attributes. This coarse, forced-choice setup probes whether LALMs encode robust representations of source identity from timbral cues alone, while remaining flexible enough to be extended beyond the specific instrument categories used here.
    
\paragraph{Texture.} 
Texture refers to aggregate, noise-like sound patterns produced by the superposition of many similar acoustic events, such as rainfall, crackling fire, or dense insect choruses~\cite{MCDERMOTT2011926}. Unlike timbre, which is tied to the identity of a single, salient source, auditory texture perception is inherently statistical: listeners do not track individual droplets or clicks, but summarise ensemble properties over time e.g., density, spectral spread, modulation patterns and use these summaries to recognise and categorise backgrounds and environments in everyday listening. This mid-level representation provides a bridge from raw acoustics to semantic scene understanding, supporting judgements such as whether a setting sounds “rainy”, “crowded”, or “mechanical” even when no single event is isolated. In SonicBench, we operationalise texture as texture-centred environment discrimination. We draw on existing corpora with reliable labels for prototypical textures e.g., rain, fire, crowd noise and construct pairs in which the dominant difference lies in their texture statistics, while overall level and other basic factors are normalised as far as practicable. Our goal is not to build a full-scale environmental sound classification benchmark, but to provide a coarse, forced-choice probe of whether LALMs encode stable statistical representations of background texture that can support higher-level scene reasoning.

\subsection{Scene-Level}
This dimension targets properties of the entire auditory scene rather than individual sources or isolated events. Whereas the previous dimensions probe local cues such as spectrum, timing, spatial layout, and timbre, scene-level perception asks how many distinct things are happening and how they are organised over time. In human hearing, this corresponds to classic auditory scene analysis~\cite{bregman1994auditory}, grouping sounds into events and streams, then summarising them at an abstract level e.g., “three knocks before the door opens” or “several dogs barking at once”. For LALMs and embodied agents, such holistic structure is a foundational layer above low-level detection, supporting decisions about complexity, crowding, and how many entities may require attention or action. In SonicBench, we instantiate this dimension with one key attribute, Counting which probes whether models can move beyond simple “is it there?” detection to track how many target events occur in a soundscape.

\paragraph{Counting.} 
Counting assesses the ability to enumerate repeated sound events of a given type within a short clip e.g., door knocks, dog barks. This goes beyond simple detection: a system must segment the waveform into discrete events, maintain them in a working representation, and compare their total number. Human studies suggest that auditory numerosity follows a pattern similar to visual “subitizing”: small numerosities around 1--3 or 4 events can be judged rapidly and accurately, while performance and response times degrade as the count increases and attention and working memory become limiting resources~\cite{cowan2001magical,anobile2021groupitizing}. Guided by these findings, SonicBench uses stimuli containing between 1 and 6 target events with controlled event type and spacing, and formulates the task as a coarse multiple-choice judgement over candidate counts. This design keeps the contrasts well within the range that is straightforward for human listeners, while providing a scene-level probe of whether models can move beyond binary detection to robustly track \emph{how many} distinct events occur in a soundscape.

\section{Toolbox Details}
\label{appendix:toolbox}
\begin{figure*}[t]
    \centering
    \begin{minipage}[t]{0.49\linewidth}
        \vspace{0pt} %
        \centering
        \includegraphics[width=\linewidth]{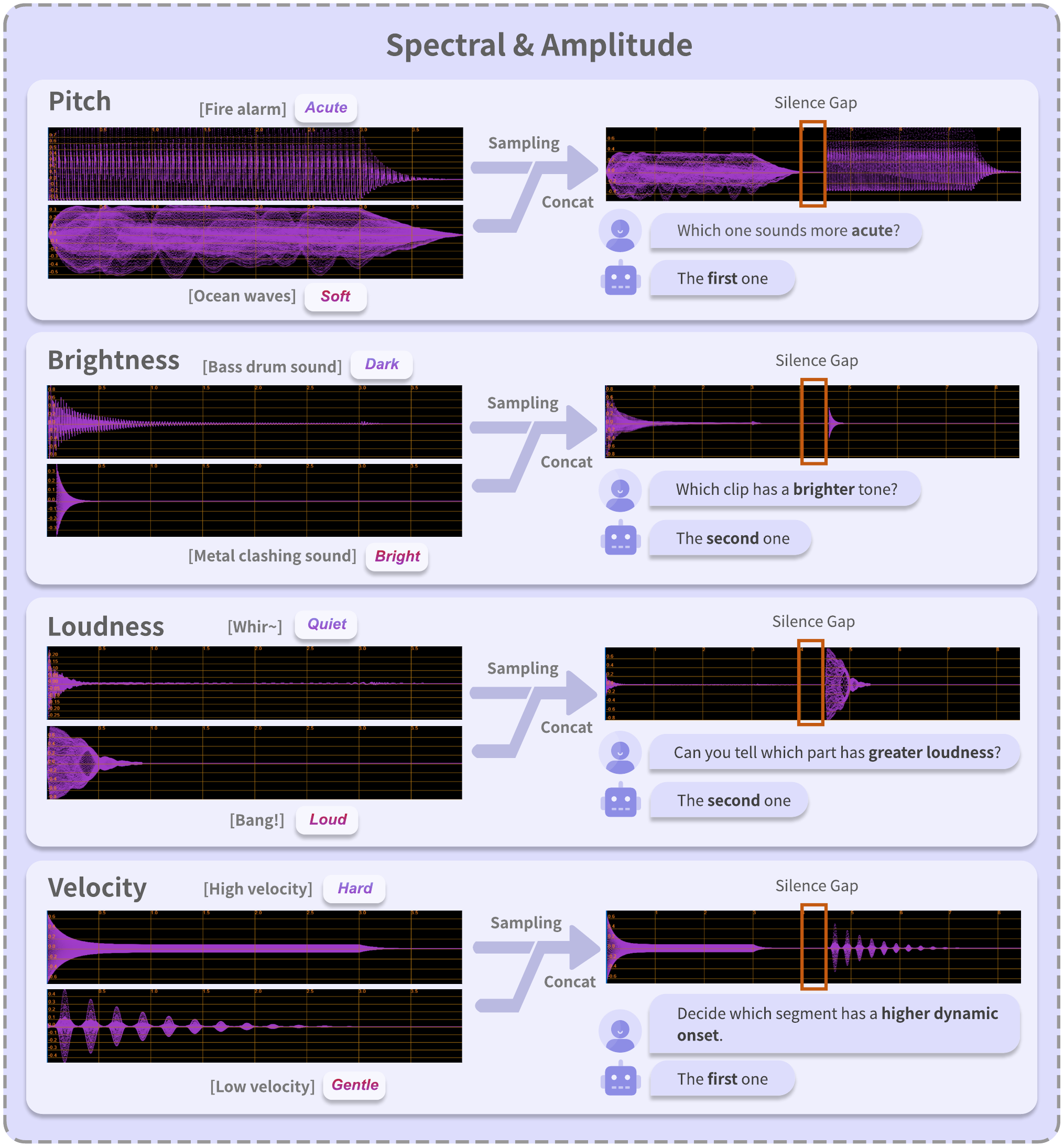}
        \vspace{2mm} 
        \includegraphics[width=\linewidth]{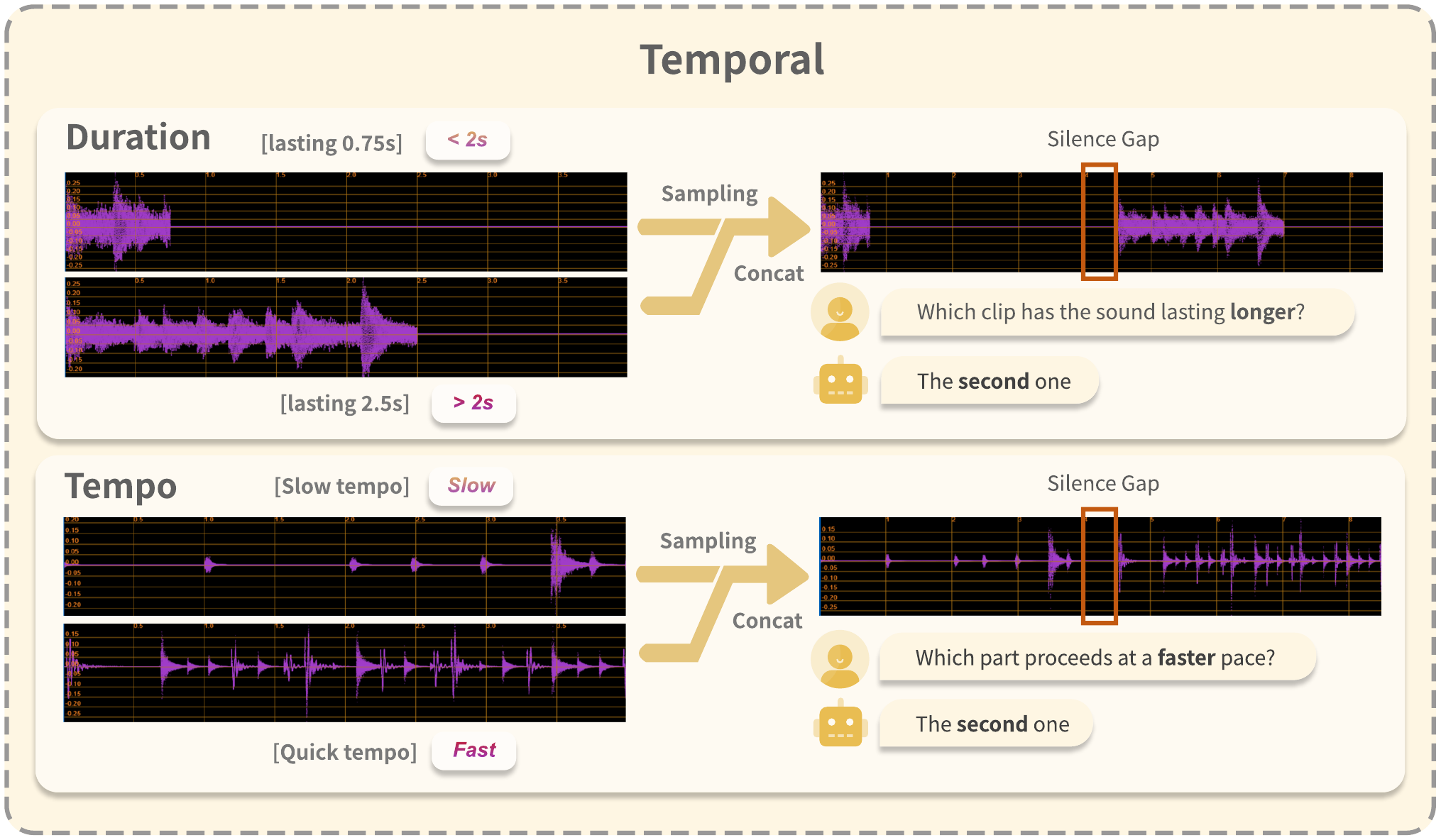}
    \end{minipage}
    \hfill %
    \begin{minipage}[t]{0.49\linewidth}
        \vspace{0pt} %
        \centering
        \includegraphics[width=\linewidth]{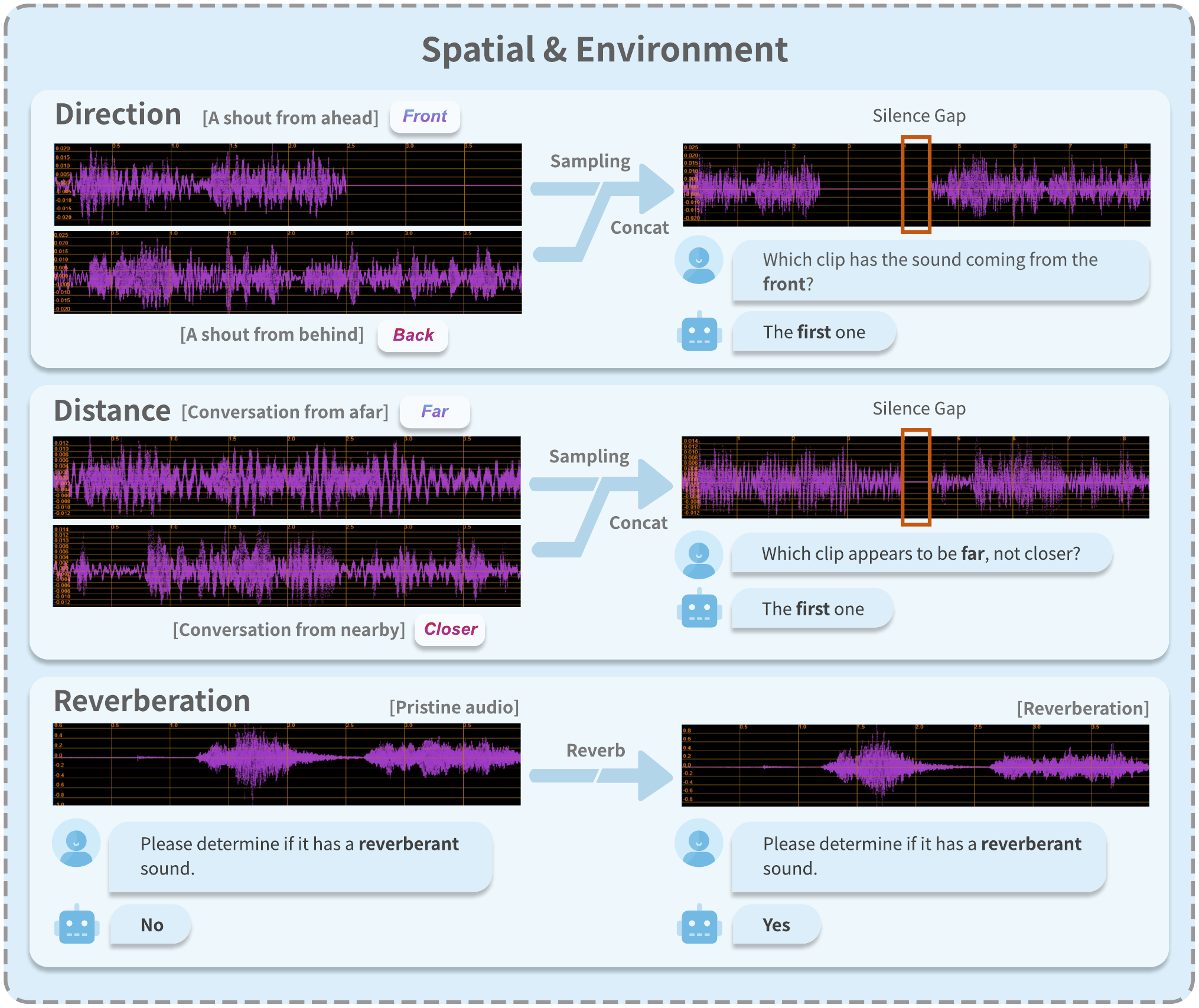}
        \vspace{0mm}
        \includegraphics[width=\linewidth]{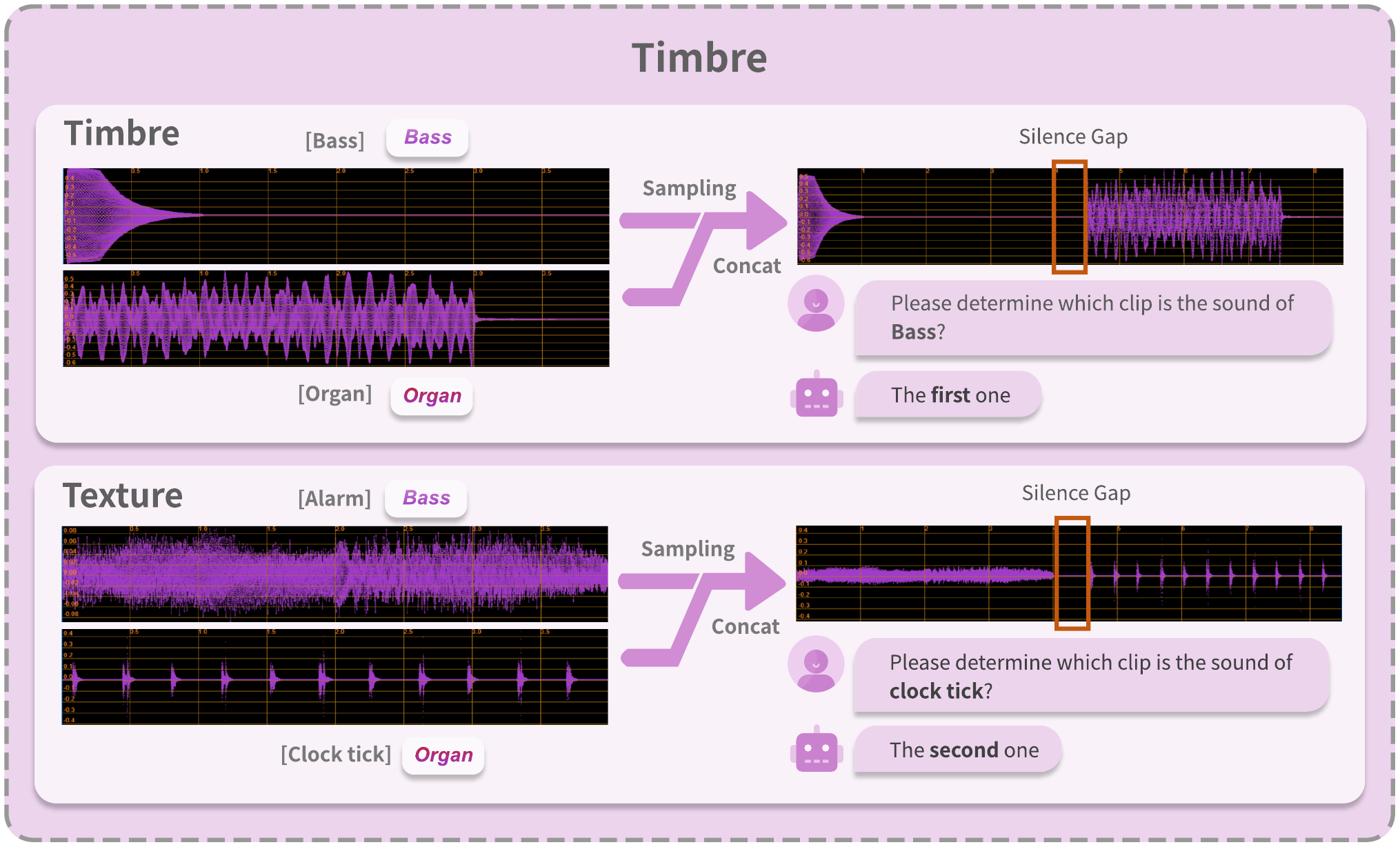}
        \includegraphics[width=\linewidth]{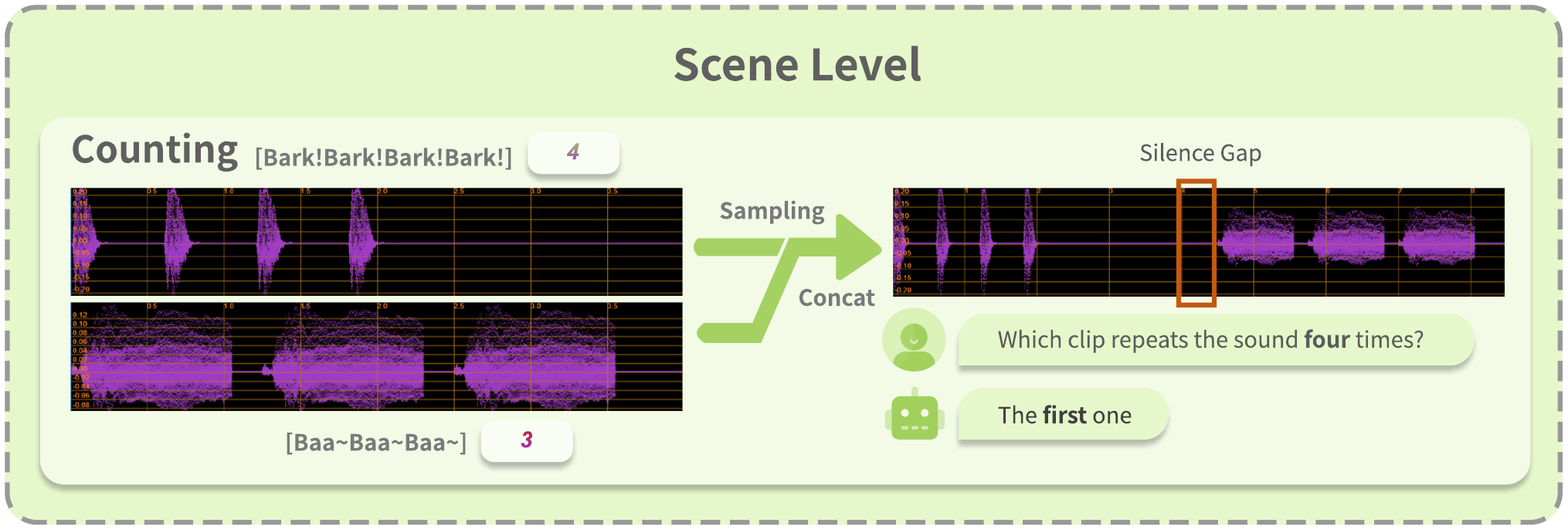}
    \end{minipage}
    \caption{Overview of large scale sampling modes from toolbox used in the dataset construction.}
    \label{fig:combined_attributes} 
\end{figure*}
\subsection{Mode Selection}
The SonicBench Toolbox supports both modes of use: it accepts short raw audio clips as the primary acoustic input for user customized generation, and it also enables large scale sampling from extensive raw audio collections under user specified constraints for automatic batch stimulus creation. Both modes share a unified input specification and automatic preprocessing pipeline that ensure consistency, experimental control, and reproducibility. Unless otherwise noted, input clips must be mono, 16-bit PCM, and between 0.5 and 5.0 seconds in duration. The clips should contain a single, stable sound event such as an instrument note or isolated vocal token to serve as a reliable reference. Upon ingestion, the toolbox automatically resamples, gain normalizes, and trims or pads the signal to a 4.0 second, 48 kHz reference waveform. This removes the need for manual editing in a DAW and guarantees uniform downstream processing.

\paragraph{User Customized Mode.} In this mode, we provides reference clips together with a target configuration, such as the desired pitch shift, duration, or loudness level. The toolbox then applies deterministic rule based signal processing recipes that modify only the specified attribute while holding all other acoustic dimensions constant. Users are encouraged to provide clean and monophonic reference recordings to ensure that intended single attribute contrasts remain unconfounded.

\paragraph{Large Scale Sampling Mode. (Used in This Work)} In this mode, the user supplies a task specification instead of reference audio. The minimal specification includes the task type, such as comparison or recognition, the target attribute or attributes selected from the twelve item taxonomy (pitch, brightness, loudness, velocity, duration, tempo, direction, distance, reverberation, timbre, texture, counting), value ranges or sampling constraints, and any language or format requirements, as shown in Figure~\ref{fig:combined_attributes}. The toolbox samples attribute values within the defined space and synthesizes the corresponding audio automatically, producing example pairs, evaluation prompts, and gold annotations suitable for large experimental sets.

\subsection{Rule-Based Attribute Control}
\paragraph{Fine-grained Attribute Control.} 
In user customized mode, the toolbox supports fine-grained adjustment of all twelve target attributes while keeping other acoustic dimensions constant. Pitch is shifted using PSOLA or phase vocoder, brightness is adjusted via spectral envelope transformations, and loudness is controlled using EBU R128 calibrated gain. Velocity is modulated through amplitude envelope control, duration and tempo are modified via WSOLA or time stretching, and directional cues are changed with HRTF or panning. Distance is adjusted by controlling the direct-to-early-reflection ratio with loudness compensation, reverberation is altered via convolution and artificial reverb, and timbre is shaped through filtering. Texture is modified using noise enhancement, modulation, or granular processing, and counting is controlled by duplicating, trimming, or rhythmically arranging discrete events. These rule-based operations ensure that each generated stimulus isolates the intended attribute for precise perceptual comparisons.

\paragraph{Attribute-controlled Sampling.} 
For six continuous and categorical attributes including pitch, loudness, velocity, tempo, duration, and distance, we first compute feature statistics for all candidate samples and discretize them into bins using perceptually motivated thresholds. Specifically, pitch is divided using a boundary at 65, loudness is thresholded at -15 LUFS, velocity at 75, duration at 2.4s, and tempo at 100 BPM. Stratified sampling is then performed based on these bins. Candidate clips that meet the specified criteria are selected, and samples are drawn across multiple bins to ensure perceptible differences while preserving the pre-sampling bin distribution. This procedure eliminates potential sampling biases and ensures that non-target attributes remain strictly controlled, avoiding confounding effects in downstream evaluation.

For brightness, candidate clips are filtered from a large pool to include only clearly perceptible bright or dark samples, covering a wide variety of sound types such as musical instruments, environmental sounds, human speech, and animal vocalizations. Duration-controlled sampling extracts valid audio segments from diverse sources including music, animal calls, daily conversations, and speeches, which are then zero-padded to create uniform 4-second clips.

For reverberation, clean audio is convolved with different room impulse responses to produce varying reverb levels. During evaluation, all other attributes and content are held constant, with only reverberation varying. For counting, the number of sound events is controlled by replicating audio segments to achieve between 1 and 7 occurrences, such as repeated animal calls or impact sounds, while keeping all other attributes fixed.

Texture control involves automated sampling from audio with distinct spectral or temporal roughness patterns, producing a rich set of clips that span smooth, granular, and noisy textures, suitable for perceptual discrimination tasks. For directional attributes, we define a broad front sector as within ±60° from the center and a broad rear sector similarly, ensuring that source positions fall within these regions while allowing controlled variation in angular difficulty.

\subsection{Task Construction} 
For recognition tasks, the toolbox generates a single transformed audio clip according to the specified target attribute and automatically produces accompanying textual instructions, categorical labels, and ground truth answers. This fully automated process allows models to accurately identify the direction, magnitude, or category of the attribute change without manual intervention. For comparison tasks, the system automatically constructs paired datasets consisting of the original and transformed audio clips and generates explicit comparative prompts, such as “Which clip is brighter,” “Which clip is longer,” or “Which clip has a higher pitch.” Corresponding ground truth answers are also provided, enabling straightforward, standardized, and reproducible evaluation of model performance. The design supports flexible experimentation across all twelve defined attributes and can be scaled to generate large volumes of controlled stimuli for both recognition and comparison scenarios.

\subsection{Output} 
Each run of the toolbox produces the following outputs. First, audio files including both reference and transformed samples, with filenames explicitly indicating the attribute changes. Second, a task annotation file in JSON format that records the task type, target attribute, audio file paths, textual instructions, and ground truth answers. Third, a batch-level index in CSV or JSONL format to facilitate downstream evaluation, large-scale processing, and dataset release. These outputs ensure full traceability, reproducibility, and ease of integration into automated evaluation pipelines.

\subsection{Reproducibility} 
All signal processing operations in the toolbox are executed deterministically, ensuring that given the same input audio and configuration, the generated outputs are bitwise reproducible. This guarantees that SonicBench produces consistent stimuli across repeated runs, enabling fair and reliable evaluation. Such reproducibility provides a solid foundation for comparing different models, languages, or dataset scales, and ensures that experimental conclusions remain consistent regardless of the computational environment or deployment scenario.

\section{Annotator Qualifications}
\label{appendix:annotator_qualifications}
Although SonicBench builds upon curated open-source corpora, the critical phases of data filtration, concatenation, and final perceptual verification required expert human judgment to ensure physical correctness. To maintain the highest standards of quality control, our team was composed entirely of researchers with deep academic backgrounds. Every team member held at least a bachelor’s degree, with over a half currently pursuing doctoral studies in relevant fields. Crucially, all personnel were screened for normal hearing function, ensuring they possessed the auditory acuity necessary to validate JNDs and confirm the perceptual salience of attribute contrasts across all samples.

\section{Details of Data Sources}
\label{appendix:details_data_sources}
In this section, we provide detailed specifications of the data sources utilized to construct SonicBench. The selection logic prioritizes datasets that offer verifiable ground-truth labels and distinct signal properties relevant to the target dimension. Table~\ref{tab:task_design_data_source} provides a systematic mapping between the evaluation dimensions and their corresponding source datasets. The specific characteristics of each dataset are detailed below.

\begin{table}[h!]
\centering
\begin{adjustbox}{max width=\columnwidth}
\begin{tabular}{lll}
    \toprule
        Dimensions & Attributes & Data Source \\
    \midrule
        \multirow{4}{*}{\makecell[l]{Spectral \&\\Amplitude}} 
        & Pitch & \multirow{4}{*}{NSynth~\cite{nsynth}} \\
        & Brightness &   \\ 
        & Loundness &   \\
        & Velocity &   \\
    \midrule
        \multirow{2}*{Temporal} 
        & Duration & TAU~\cite{dcase2019}\\
        & Tempo & Groove MIDI~\cite{groovemidi} \\
    \midrule
        \multirow{3}*{\makecell[l]{Spatial \&\\Environment}}
        & Direction & LOCATA~\cite{locata} \\
        & Distance & RealMAN~\cite{realman} \\
        & Reverberation & SRIRS~\cite{srirs} \\
    \midrule
        \multirow{2}*{Timbre} & Timbre & NSynth~\cite{nsynth} \\
        & Texture & ESC-50~\cite{esc50} \\
    \midrule
        \multirow{1}*{Scene Level} & Counting & PicoAudio~\cite{picoaudio} \\
    \bottomrule
\end{tabular}
\end{adjustbox}
\caption{Overview of Data Source.}
\label{tab:task_design_data_source}
\end{table}

\paragraph{NSynth.~\cite{nsynth}} It's a large-scale, high-quality dataset for neural audio synthesis of musical notes, consisting of approximately 306043 four-second monophonic audio snippets (sampled at 16kHz) from around 1000 musical instruments. Each note is annotated with pitch, quality, velocity, brightness, etc.

\paragraph{TAU.~\cite{dcase2019}} This dataset is a multi-room reverberant dataset designed for Sound Event Localization and Detection (SELD) in the DCASE 2019 challenge. It includes two subsets, each with a 400-recording development set and a 100-recording evaluation set. It is synthesized by convolving isolated sound events from DCASE 2016 Task 2 with real impulse responses from 5 indoor environments, adding ambient noise from these environments. This dataset is labeled with azimuth, elevation angle, start and end time.

\paragraph{Groove MIDI.~\cite{groovemidi}} It's a large-scale dataset for expressive drum performance research, containing 13.5 hours of recordings of drummers playing electronic drum kits. It includes detailed metadata (drummer IDs, musical style, tempo), and is an order of magnitude larger than the largest previously publicly available comparable dataset for drum performance modeling.

\paragraph{LOCATA.~\cite{locata}} This dataset is an open-access corpus tailored for benchmarking acoustic source localization and tracking algorithms, containing multichannel audio recordings from four distinct microphone arrays. It is fully annotated with critical information including ground-truth positions/orientations of sources and sensors, and hand-labelled voice activity periods.

\paragraph{RealMAN.~\cite{realman}} RealMAN is a real-recorded and annotated microphone array dataset designed for dynamic speech enhancement and source localization, which uses a 32-channel high-fidelity microphone array to record speech across 32 scenes and background noise across. It provides key annotations such as direct-path target clean speech and speaker location.

\paragraph{SRIRS.~\cite{srirs}} It's created by Detmold University of Music, comprises around 600 multichannel Spatial Room Impulse Responses (SRIRs) captured in three spaces. It includes unique setups such as artificial reverberation in Detmold Konzerthaus, Wave Field Synthesis (WFS) focused sources, and stage measurements with music stands.

\paragraph{ESC-50.~\cite{esc50}} The ESC Dataset, created for environmental sound classification research. It derives from Freesound and standardized to 44.1 kHz, single-channel Ogg Vorbis format. It also provides human classification accuracy benchmarks and baseline machine learning results, along with replication code via a Jupyter notebook, supporting open research in auditory recognition.

\paragraph{PicoAudio.~\cite{picoaudio}} The dataset used in the PicoAudio project is constructed by crawling audio clips from Freesound using sound event keywords, then segmenting them with a text-to-audio grounding model and filtering via the LAION-CLAP model. Each sample is paired with timestamp captions and frequency captions to support training and evaluating temporally controllable audio generation models.

\section{Data Selection and Filtering}
\label{appendix:filtering}
\subsection{Data Selection}
In constructing SonicBench, we first perform a systematic data selection process over existing public audio datasets to identify samples that are suitable for evaluating fundamental physical auditory perception. The primary selection criterion is the definability and controllability of target physical attributes. Although the audio samples are not directly generated by us, we only retain those for which key physical properties, can be reliably estimated or are annotated by the original datasets. Moreover, the values of these attributes must be well-defined and stable, enabling consistent formulation of both absolute (recognition) and relative (comparison) judgment tasks. For continuous attributes, we prioritize samples whose values exhibit clear distributions and smooth variations that do not rely on semantic interpretation. For discrete or graded attributes, category boundaries must be grounded in consistent physical criteria rather than subjective or semantic definitions, ensuring that all labels retain clear physical meaning.

Building upon this, the data selection process further follows the principles of single-attribute dominance and task compatibility. Each selected sample is used to evaluate only one target physical attribute, minimizing confounding effects from simultaneous variations in multiple cues. For comparison tasks, samples are organized into pairs such that only the target attribute differs systematically, while all other physical attributes are held constant or remain statistically symmetric. In addition, we retain only samples that naturally support both absolute (recognition) and relative (comparison) paradigms, allowing the same physical attribute to be evaluated consistently under single-stimulus identification and pairwise discrimination settings. 

\subsection{Data Filtering}
After completing data selection, we apply a systematic data filtering process to the retained samples as a form of secondary validation, removing instances that may still compromise the effectiveness of the evaluation. We first filter the data from the perspective of perceptual validity, with particular emphasis on excluding audio samples in which the target physical attribute is insufficiently salient or exhibits unstable values. For continuous physical attributes, we remove samples whose attribute differences are too small, lie close to perceptual boundary conditions, or show high uncertainty in the time or frequency domains. For spatial, environmental, and timbral attributes, we exclude cases where the available acoustic cues are insufficient to induce stable and reliable perceptual differences. This step ensures that, for all retained samples, the target physical attribute is clearly and discriminably expressed at the signal level, preventing model errors from being erroneously attributed to ambiguities in the input audio itself.

The data filtering process further enforces strict control over confounding factors, label consistency, and task structure. For comparison tasks, we remove paired samples in which systematic differences remain in non-target attributes, ensuring that model decisions cannot rely on spurious cues. For absolute judgment tasks, we discard samples whose labels do not align cleanly with the corresponding attribute value ranges or exhibit category overlap, thereby preserving a one-to-one correspondence between physical parameters and labels. In addition, we filter out audio samples containing salient semantic cues, abnormal energy distributions, clipping, or other signal artifacts that could enable shortcut solutions. 

\section{Review \& Double Check}
\label{appendix:review_double_check}

As illustrated in the ``Review \& Double Check'' stage of Figure~\ref{fig:benchmark_constr_pipeline}, we implemented a rigorous, multi-stage verification pipeline to ensure that every sample in SonicBench adheres to both physical correctness and perceptual validity. This process involves three distinct checkpoints, executed by independent domain experts to enforce role separation.

\paragraph{1. Correction (Metadata \& Logic Verification).}
The first line of defense involves a manual review of the generated metadata and Question-Answer (QA) pairs. In this phase, an independent annotator (distinct from the author) verifies the consistency between the audio filename parameters and the JSON labels. Common errors targeted in this stage include label mismatches (e.g., a file generated with high pitch labeled as low) or grammatical inconsistencies in the instruction templates. Samples containing correctable errors are routed to revision, while those with fundamental logical flaws are flagged for rejection.

\paragraph{2. Significant Difference Check (Perceptual Validation).}
Passing samples proceed to the most critical phase: the Significant Difference Check. Here, expert annotators listen to the audio to confirm that the target attribute manipulation is perceptually salient. 
For comparison tasks, this step rigorously enforces the JND constraint. Annotators must verify that the contrast between the two clips is immediately distinguishable to a healthy human ear without ambiguity. 
For recognition tasks, annotators verify that the attribute value (e.g., ``High Brightness'') clearly aligns with the auditory sensation. 
As shown in the pipeline diagram, any sample failing this perceptual threshold, where the difference is too subtle or masked by other acoustic features, is immediately routed to the ``Rejected Samples'' pool to prevent ``guessing games'' in evaluation.

\paragraph{3. Quality Check (Signal Integrity).}
The final gate is a technical Quality Check. This step focuses on the acoustic fidelity of the file. Annotators inspect the waveform for generation artifacts, such as:
\begin{itemize}
    \item \textbf{Boundary Artifacts:} Unnatural clicks or pops at the splicing points (0.5s silence intervals) in comparison pairs.
    \item \textbf{Digital Distortion:} Clipping or excessive synthetic noise introduced during the rendering process.
    \item \textbf{Format Compliance:} Ensuring duration (4s/8.5s) and sampling rate consistency.
\end{itemize}
Only samples that pass this final inspection move to the final benchmark

\paragraph{Rejection Policy.}
As depicted by the red dashed paths in Figure~\ref{fig:benchmark_constr_pipeline}, a ``Fail'' at any of the three stages triggers a rejection mechanism. To ensure high quality, we adopted a strict policy: while minor metadata issues allowed for one round of correction, any failure regarding perceptual salience (Stage 2) or irreversible signal degradation (Stage 3) resulted in permanent discarding. After this process, we retained 2,400 high-quality questions in the final version. We provide our benchmark statistics in Table~\ref{tab:benchmark_statistics}.

\onecolumn
\newpage
\section{Details of Task Instruction}
\label{appendix:task_instruction}
\begin{figure*}[h]
\centering
\includegraphics[width=0.4\linewidth]{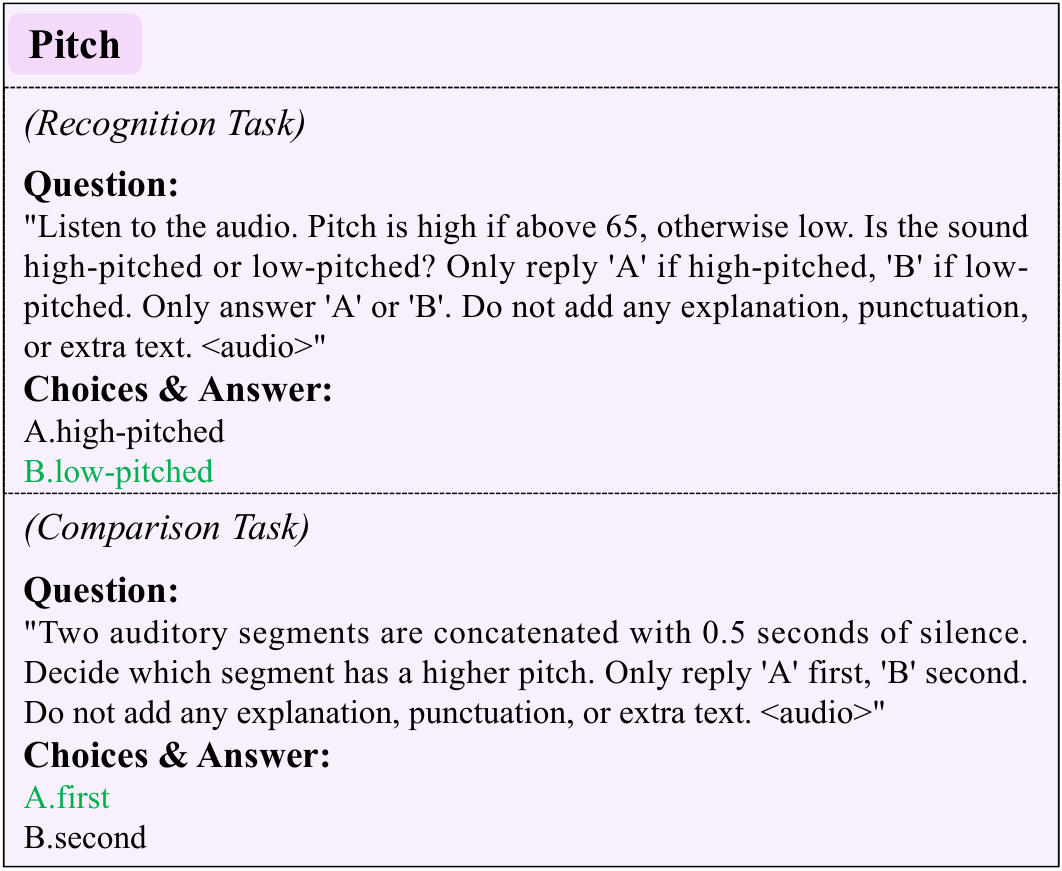}
\includegraphics[width=0.4\linewidth]{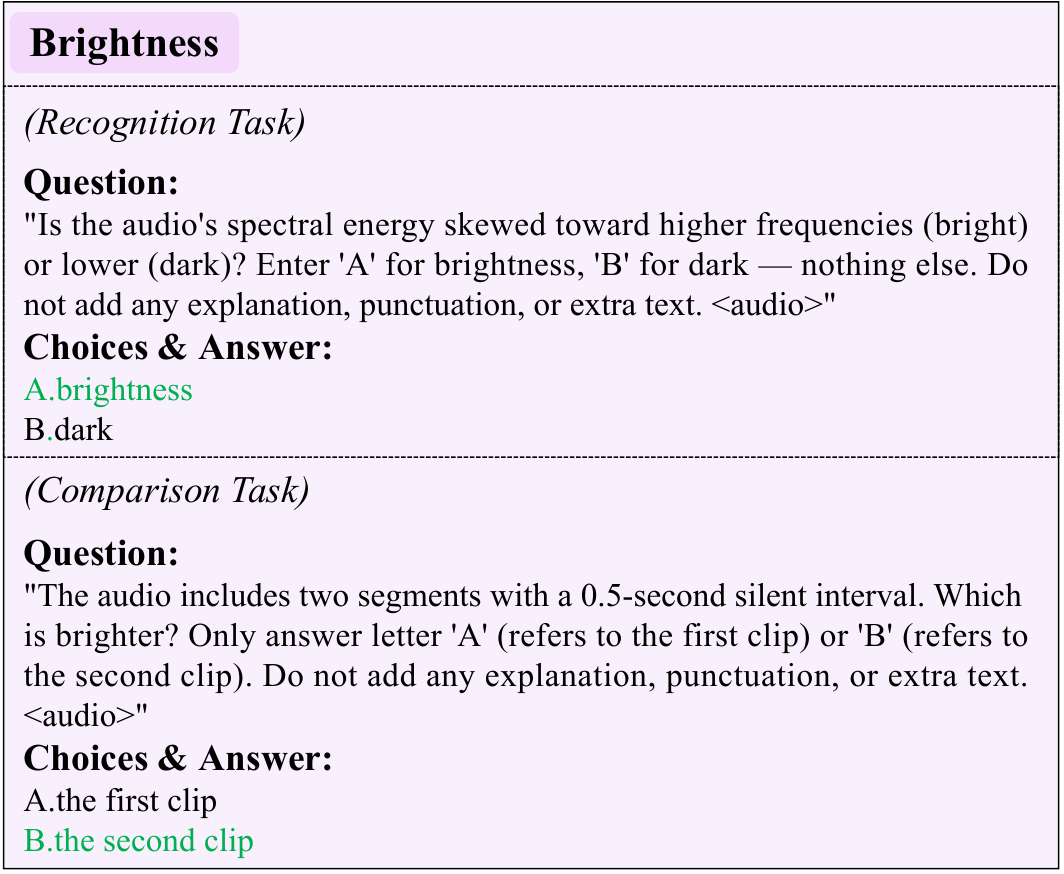}
\includegraphics[width=0.4\linewidth]{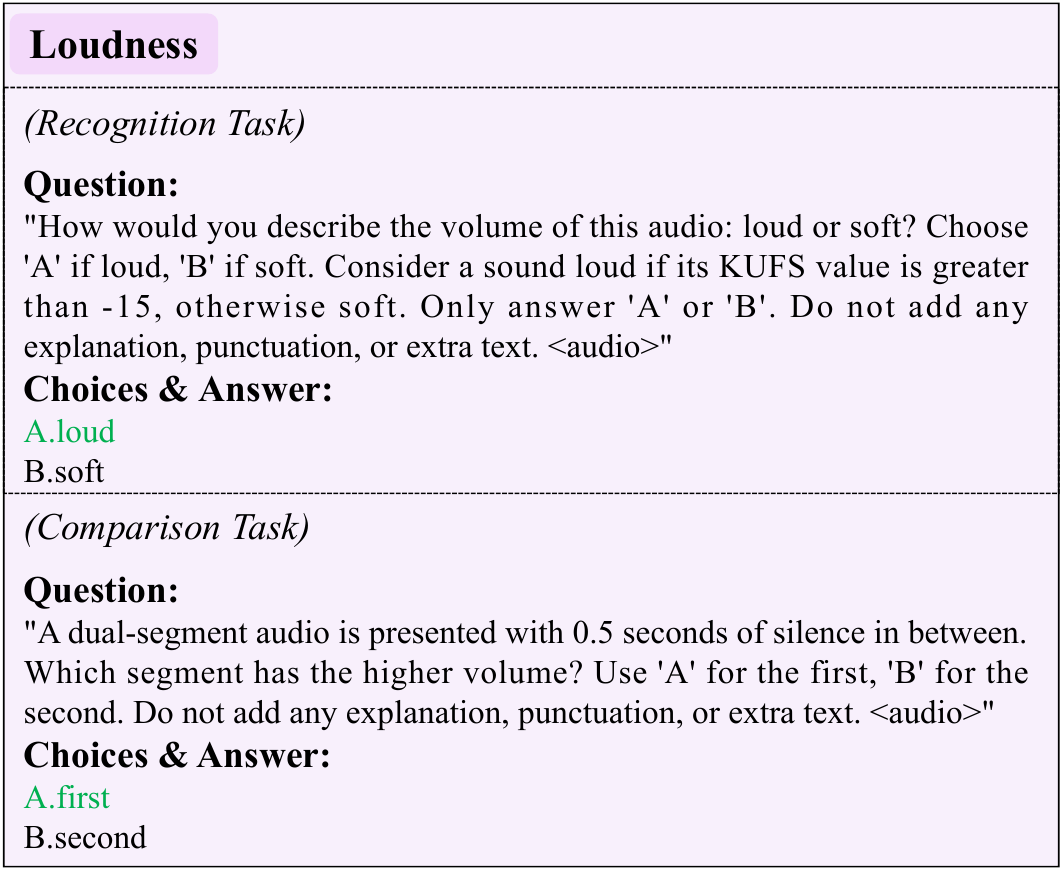}
\includegraphics[width=0.4\linewidth]{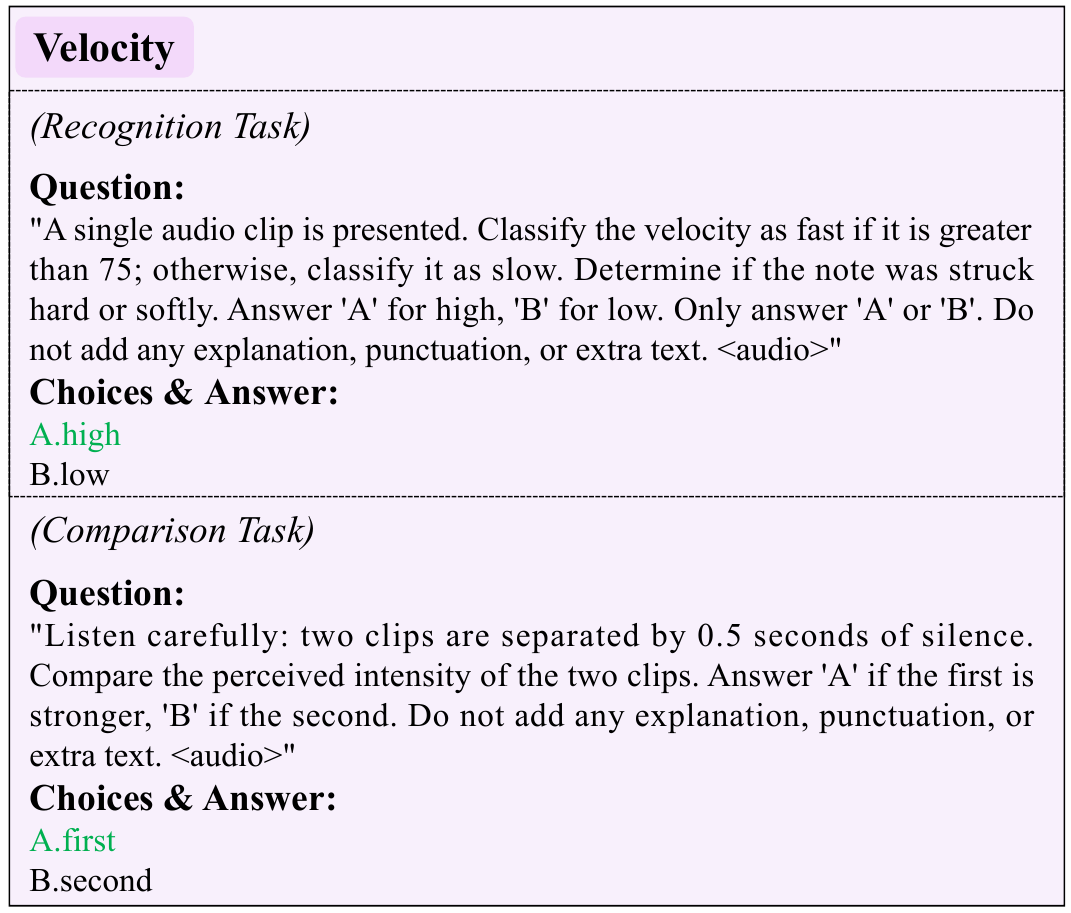}
\caption{Examples of Spectral \& Amplitude Dimension. Shown are representative samples for Pitch, Brightness, Loudness, and Velocity attributes across both Recognition and Comparison tasks.}
\end{figure*}

\begin{figure*}[h]
\centering
\includegraphics[width=0.4\linewidth]{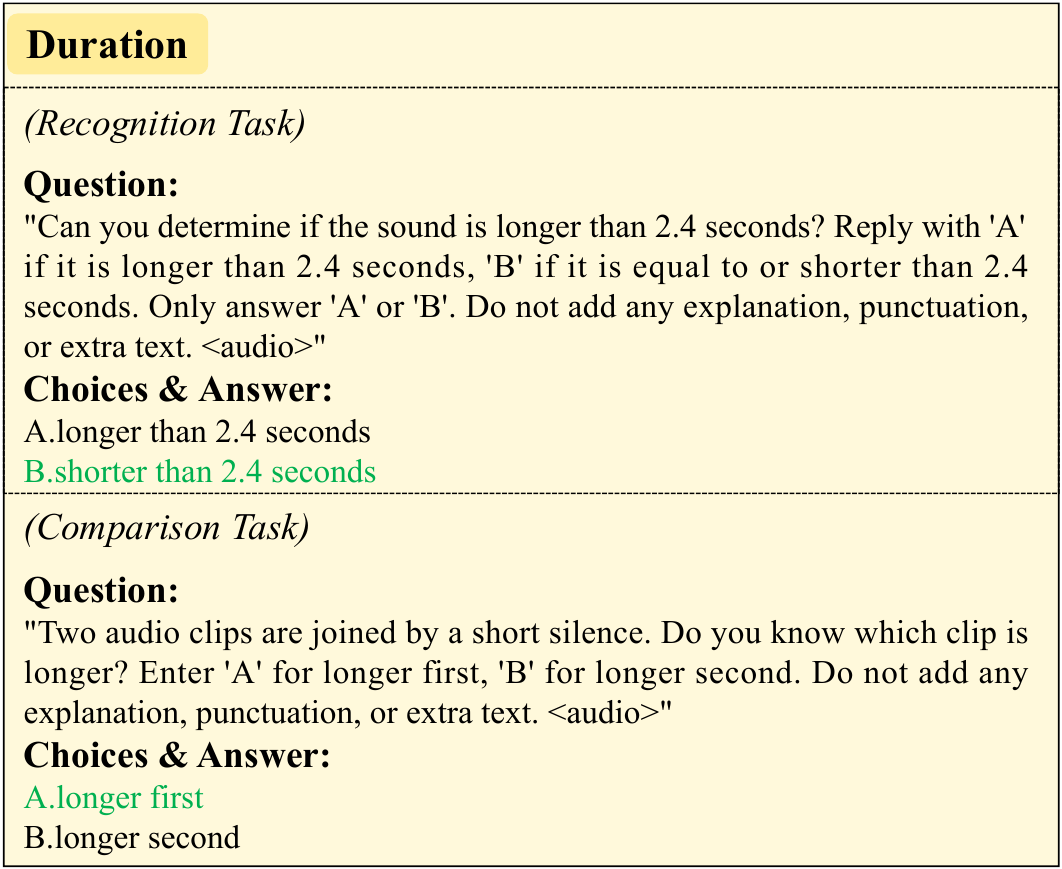}
\includegraphics[width=0.4\linewidth]{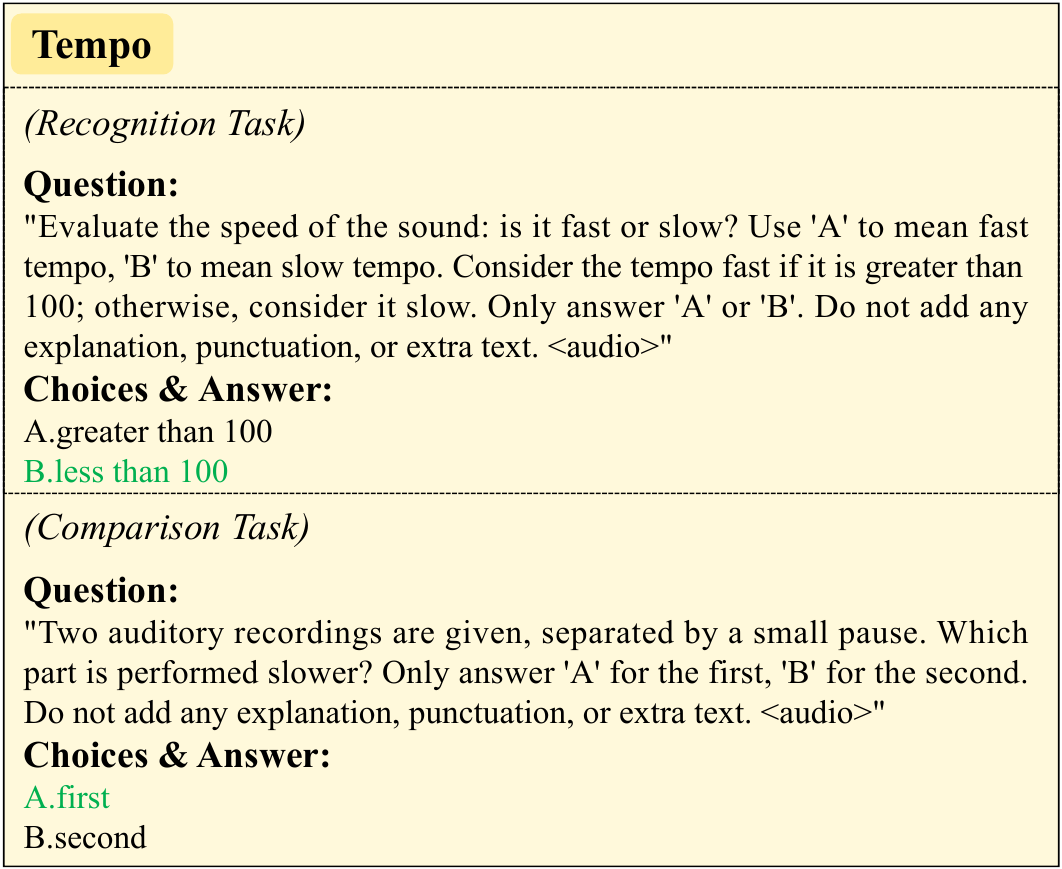}
\caption{ Examples of Temporal Dimension. Shown are representative samples for Duration and Tempo attributes across both Recognition and Comparison tasks.}
\end{figure*}

\begin{figure*}[h]
\centering
\includegraphics[width=0.4\linewidth]{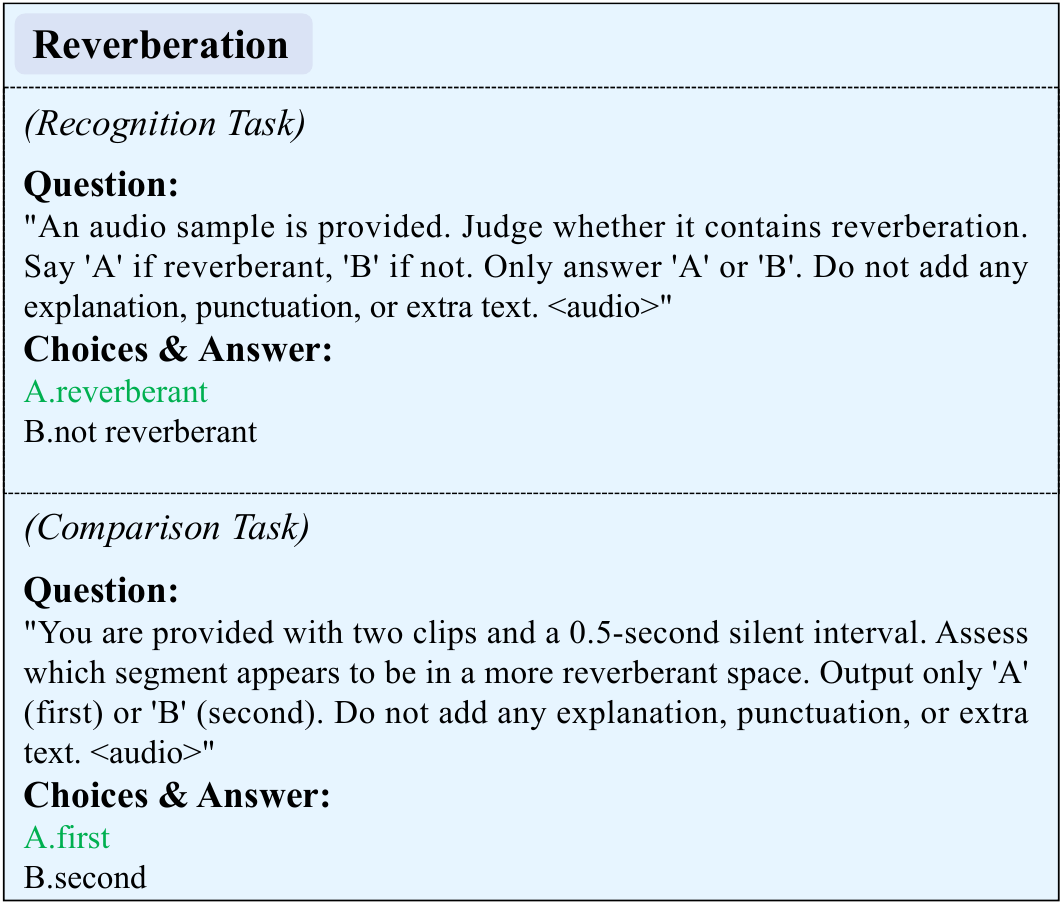}
\includegraphics[width=0.4\linewidth]{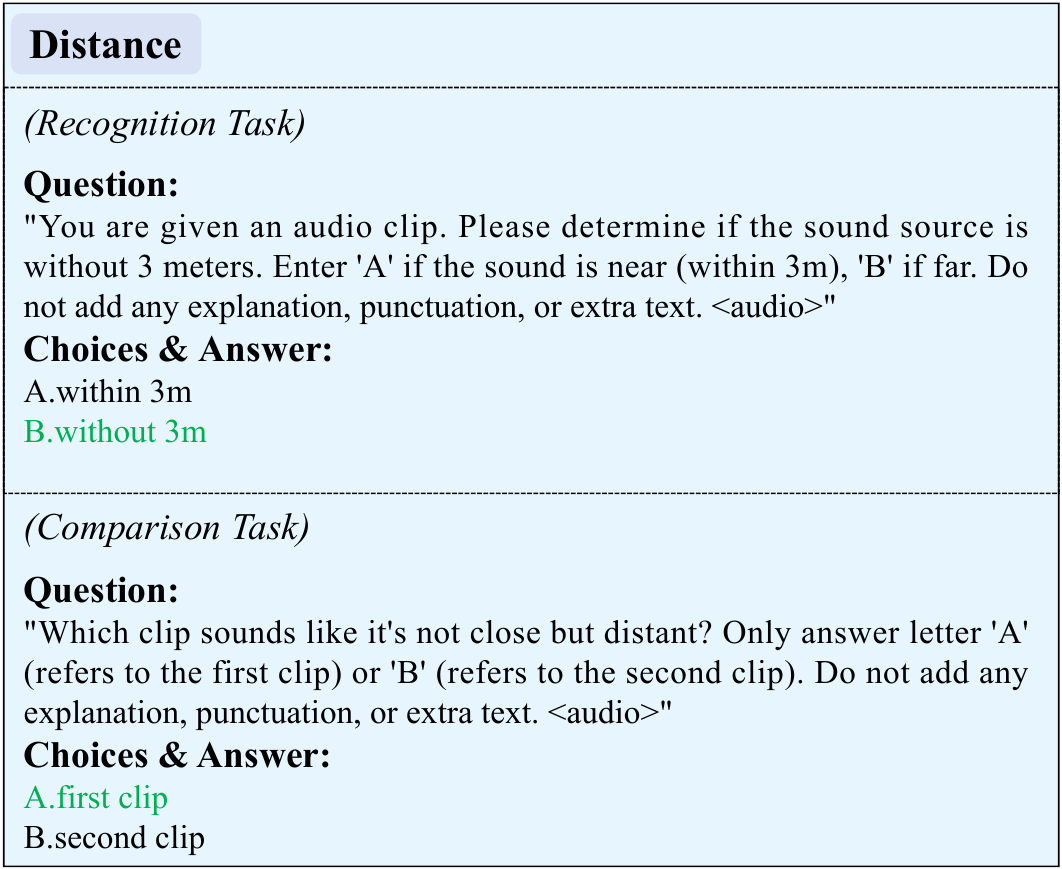}
\includegraphics[width=0.4\linewidth]{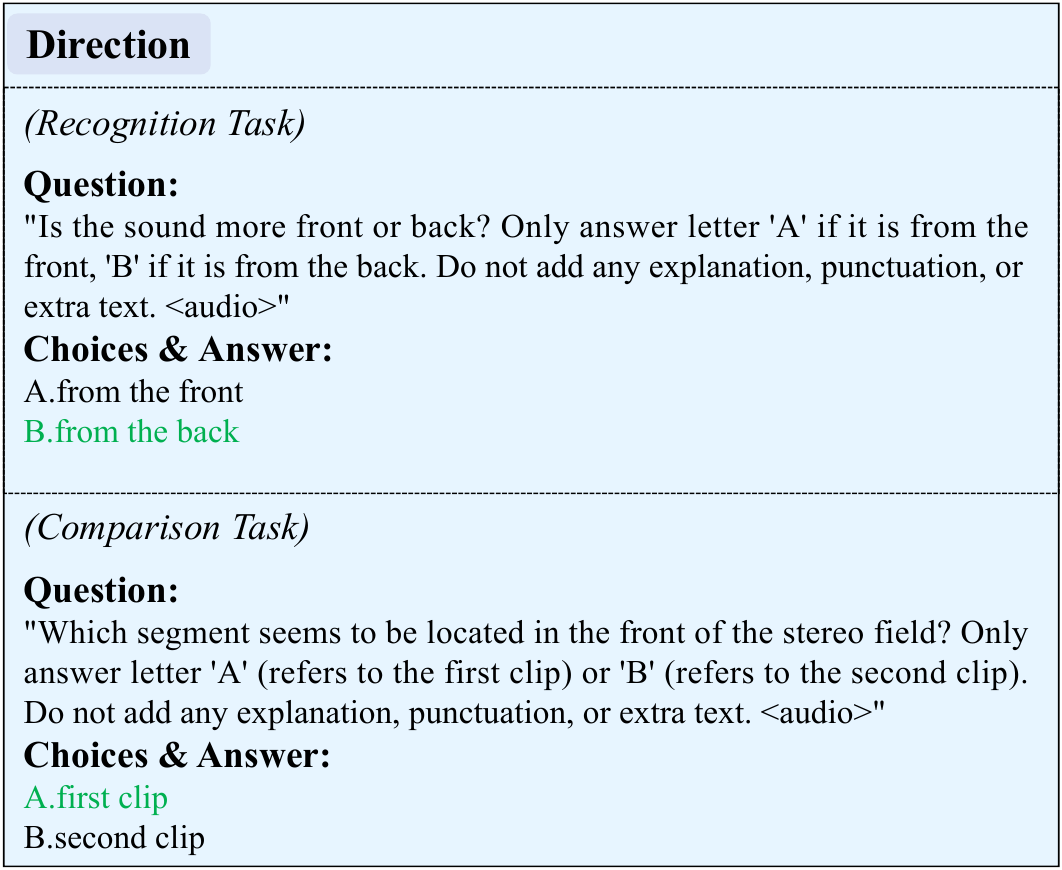}
\includegraphics[width=0.4\linewidth]{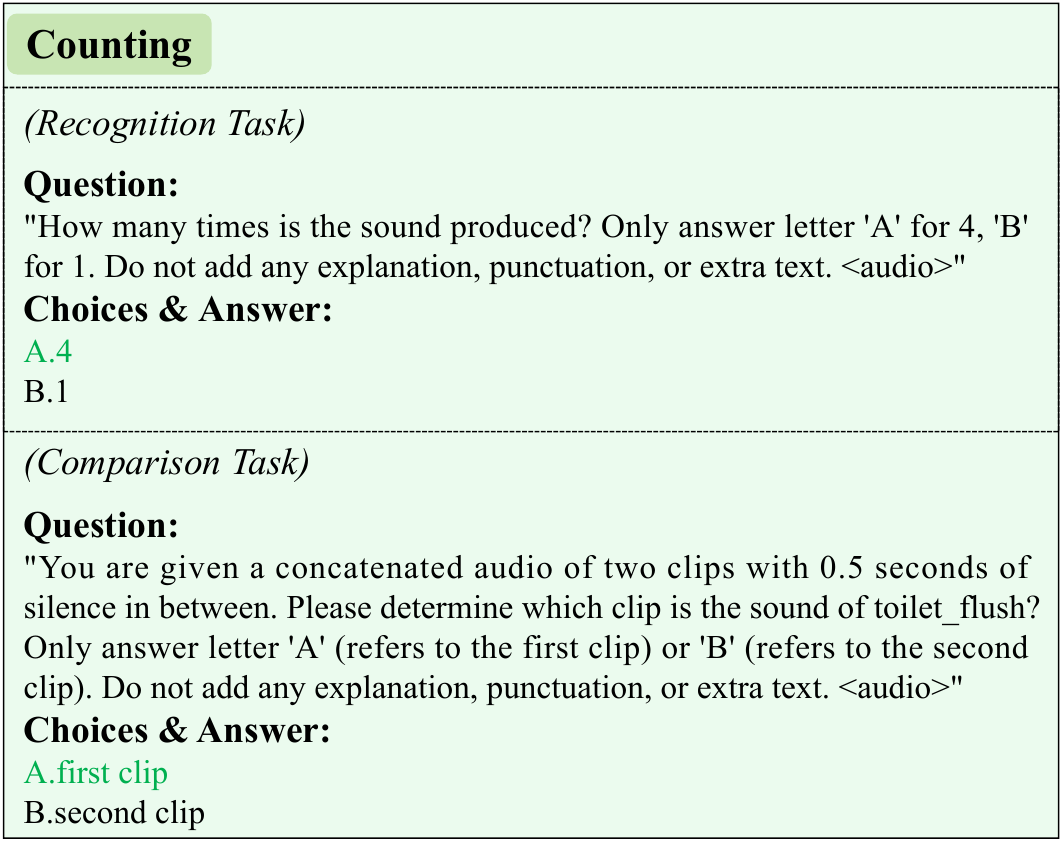}
\caption{Examples of Spatial \& Environment and Scene Level Dimensions. Shown are representative samples for Reverberation, Distance, Direction, and Counting attributes across both Recognition and Comparison tasks.}
\end{figure*}

\begin{figure*}[h]
\centering
\includegraphics[width=0.4\linewidth]{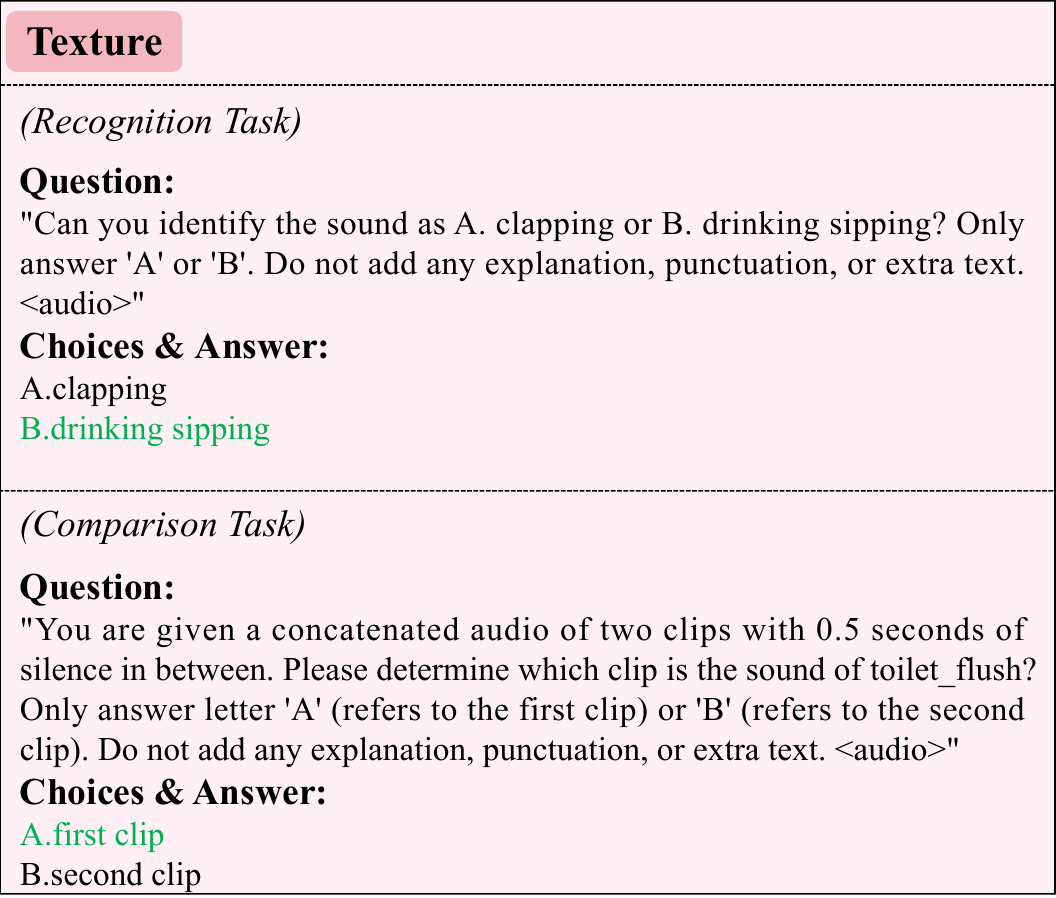}
\includegraphics[width=0.4\linewidth]{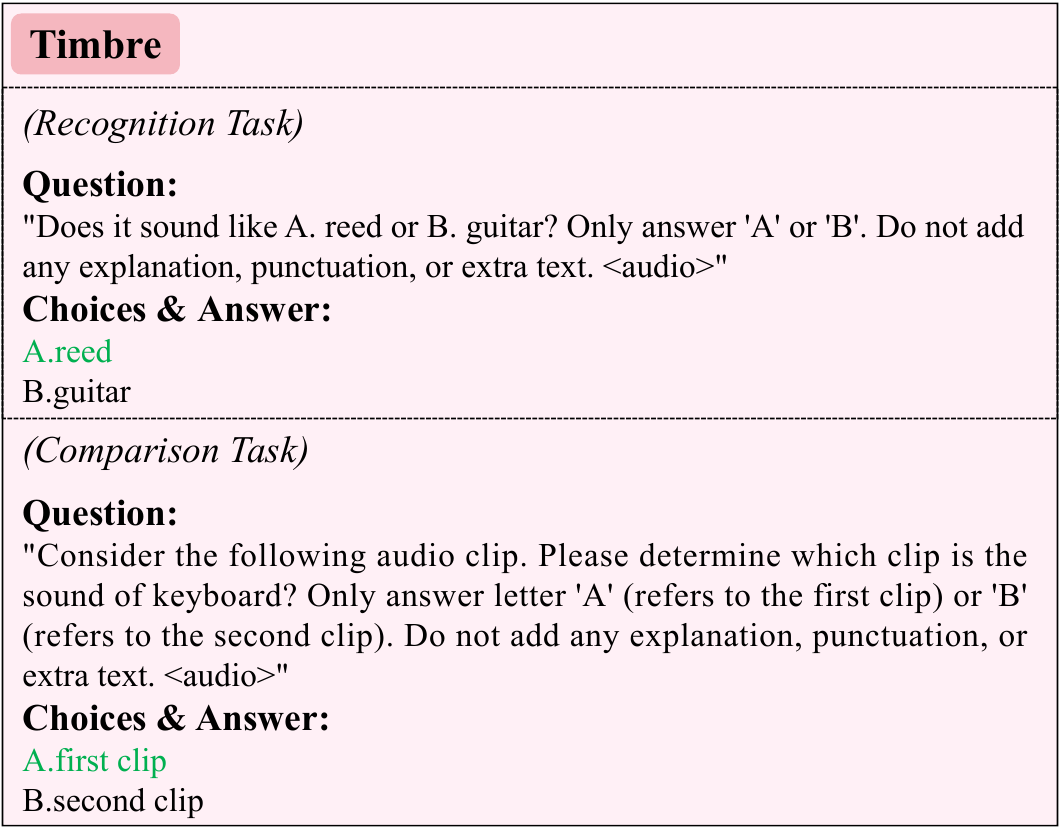}
\caption{Examples of Timbre Dimension. Shown are representative samples for Timbre and Texture attributes across both Recognition and Comparison tasks.}
\end{figure*}
\twocolumn

\newpage
\section{Benchmarking Candidates}
\label{appendix:benchmarking_candidates}
In this section, we provide a comprehensive overview of the models evaluated in SonicBench. Our assessment covers a diverse spectrum of architectures, categorized into LALMs, LARMs, and OLMs, encompassing both open-source community checkpoints and state-of-the-art proprietary services. Detailed specifications for each candidate, including specific model names, weight versions, access URLs, and parameter counts, are systematically tabulated in Table~\ref{tab:benchmarking_candidate}.

\begin{table*}[h!]
\begin{center}
\small
\resizebox{1.0\linewidth}{!}{
    \begin{tabularx}{\textwidth}{lcccc}
        \toprule \toprule 
        \textbf{Model Names} & \textbf{Weight Versions} & \textbf{URLs} & \textbf{\#Params}  \\
        
        \midrule \midrule
        \multicolumn{5}{c}{\textbf{Large Audio Language Models (LALMs)}} \\ 
        \midrule \midrule
        
        Kimi-Audio-Instruct~\cite{kimiaudio} & moonshotai/Kimi-Audio-7B-Instruct & \href{https://huggingface.co/moonshotai/Kimi-Audio-7B-Instruct}{model card} & 7B  \\
        
        Qwen2-Audio-Instruct~\cite{qwen2audio} & Qwen/Qwen2-Audio-7B-Instruct & \href{https://huggingface.co/Qwen/Qwen2-Audio-7B-Instruct}{model card} & 7B  \\
        
        VITA-Audio-Plus-Vanilla~\cite{vitaaudio} & VITA-MLLM/VITA-Audio-Plus-Vanilla & \href{https://huggingface.co/VITA-MLLM/VITA-Audio-Plus-Vanilla}{model card} & 8B  \\

        Audio Flamingo 2~\cite{audioflamingo2} & nvidia/audio-flamingo-2 & \href{https://huggingface.co/nvidia/audio-flamingo-2}{model card} & 3B   \\

        Audio Flamingo 3~\cite{audioflamingo3} & nvidia/audio-flamingo-3 & \href{https://huggingface.co/nvidia/audio-flamingo-3}{model card} & 7B   \\
        
        Voxtral-Mini~\cite{liu2025voxtral} & mistralai/Voxtral-Mini-3B-2507 & \href{https://huggingface.co/mistralai/Voxtral-Mini-3B-2507}{model card} & 3B  \\
        
        Baichuan-Audio-Instruct~\cite{li2025baichuanaudio} & baichuan-inc/Baichuan-Audio-Instruct & \href{https://huggingface.co/baichuan-inc/Baichuan-Audio-Instruct}{model card} & 7B \\
        
        MiDashengLM~\cite{midashenglm} & mispeech/midashenglm-7b-0804-fp32 &\href{https://huggingface.co/mispeech/midashenglm-7b-0804-fp32}{model card} & 7B   \\
        
        Llama-Omni~\cite{llamaomni} & ICTNLP/Llama-3.1-8B-Omni & \href{https://huggingface.co/ICTNLP/Llama-3.1-8B-Omni}{model card} & 8B   \\
        
        Llama-Omni2~\cite{llamaomni2} & ICTNLP/LLaMA-Omni2-7B & \href{https://huggingface.co/ICTNLP/LLaMA-Omni2-7B}{model card} & 7B   \\
        
        GLM-4-Voice~\cite{glm4voice} & zai-org/glm-4-voice-9b & \href{https://huggingface.co/zai-org/glm-4-voice-9b}{model card} & 9B  \\
        
        SALMONN~\cite{tang2024salmonn} & tsinghua-ee/SALMONN-7B & \href{https://huggingface.co/tsinghua-ee/SALMONN-7B}{model card} & 7B  \\
        
        MU-LLaMA~\cite{mullama} & mu-llama/MU-LLaMA & \href{https://huggingface.co/mu-llama/MU-LLaMA}{model card} & 7B  \\
        
        BAT~\cite{bat} & - & \href{https://github.com/X-LANCE/SLAM-LLM/tree/main/examples/seld_spatialsoundqa}{model card} & 7B  \\
        
        R1-AQA~\cite{li2025r1_aqa} & mispeech/r1-aqa & \href{https://huggingface.co/mispeech/r1-aqa}{model card} & 7B   \\
        
        Step-Audio 2 mini~\cite{stepaudio2} & stepfun-ai/Step-Audio-2-mini & \href{https://huggingface.co/stepfun-ai/Step-Audio-2-mini}{model card} & 7B   \\
        
        MiMo-Audio-Instruct~\cite{mimoaudio} & XiaomiMiMo/MiMo-Audio-7B-Instruct & \href{https://huggingface.co/XiaomiMiMo/MiMo-Audio-7B-Instruct}{model card} & 7B   \\
        
        \cdashline{1-4}
        \noalign{\vskip 0.5mm}
        
        GPT-4o-Audio~\cite{gpt4o_blog} & gpt-4o-audio-preview-2025-06-03 & \href{https://platform.openai.com/docs/models/gpt-4o-audio-preview}{model card} & -  \\
        
        \midrule \midrule
        \multicolumn{5}{c}{\textbf{Large Audio Reasoning Models (LARMs)}} \\ 
        \midrule \midrule
        
        Mellow~\cite{mellow} & soham97/mellow & \href{https://huggingface.co/soham97/mellow}{model card} & 167M   \\

        Audio-Reasoner~\cite{audioreasoner} & zhifeixie/Audio-Reasoner & \href{https://huggingface.co/zhifeixie/Audio-Reasoner}{model card} & 7B   \\
        
        GAMA~\cite{ghosh2024gama} & sonalkum/GAMA & \href{https://huggingface.co/spaces/sonalkum/GAMA/tree/main}{model card} & 7B   \\
        
        R1-AQA (think mode)~\cite{li2025r1_aqa} & mispeech/r1-aqa & \href{https://huggingface.co/mispeech/r1-aqa}{model card} & 7B   \\

        Audio Flamingo 2 Sound-CoT~\cite{audioflamingosoundcot} & nvidia/audio-flamingo-2-SoundCoT & \href{https://huggingface.co/nvidia/audio-flamingo-2-SoundCoT}{model card} & 3B   \\
                
        Audio Flamingo 3 (think mode) & nvidia/audio-flamingo-3-hf & \href{https://huggingface.co/nvidia/audio-flamingo-3-hf}{model card} & 7B   \\
        
        Step-Audio 2 mini Think & stepfun-ai/Step-Audio-2-mini-Think & \href{https://huggingface.co/stepfun-ai/Step-Audio-2-mini-Think}{model card} & 7B   \\
        
        MiMo-Audio (think mode) & XiaomiMiMo/MiMo-Audio-7B-Instruct & \href{https://huggingface.co/XiaomiMiMo/MiMo-Audio-7B-Instruct}{model card} & 7B   \\
        Step-Audio-R1 & stepfun-ai/Step-Audio-R1 &\href{https://huggingface.co/stepfun-ai/Step-Audio-R1}{model card} & 33B \\
        \midrule \midrule
        \multicolumn{5}{c}{\textbf{Omni Language Models (OLMs)}} \\ 
        \midrule \midrule
        
        Qwen2.5-Omni~\cite{qwen25omni} & Qwen/Qwen2.5-Omni-7B & \href{https://huggingface.co/Qwen/Qwen2.5-Omni-7B}{model card} & 7B   \\
        
        Baichuan-Omni-1.5~\cite{li2025baichuanomni15} & baichuan-inc/Baichuan-Omni-1d5 & \href{https://huggingface.co/baichuan-inc/Baichuan-Omni-1d5}{model card} & 7B   \\
        
        VITA-1.5~\cite{vita} & VITA-MLLM/VITA-1.5 & \href{https://huggingface.co/VITA-MLLM/VITA-1.5}{model card} & 7B   \\
        
        Ola~\cite{ola} & THUdyh/Ola-7b & \href{https://huggingface.co/THUdyh/Ola-7b}{model card} & 7B   \\
        
        HumanOmni~\cite{humanomni} & StarJiaxing/HumanOmni-7B & \href{https://huggingface.co/StarJiaxing/HumanOmni-7B}{model card} & 7B   \\

        OpenOmni~\cite{openomni} & Tongyi-ConvAI/OpenOmni & \href{https://huggingface.co/Tongyi-ConvAI/OpenOmni}{model card} & 7B   \\

        Qwen3-Omni-Instruct~\cite{xu2025qwen3omni} & Qwen/Qwen3-Omni-30B-A3B-Instruct & \href{https://huggingface.co/Qwen/Qwen3-Omni-30B-A3B-Instruct}{model card} & 30B (A3B)   \\

        Ming-Lite-Omni-1.5~\cite{ai2025mingflashomni} & inclusionAI/Ming-Lite-Omni-1.5 & \href{https://huggingface.co/inclusionAI/Ming-Lite-Omni-1.5}{model card} & 20B (A3B)   \\
        \cdashline{1-4}
        \noalign{\vskip 0.5mm}
        Gemini-2.5-Flash~\cite{comanici2025gemini25} & gemini-2.5-flash (June 17, 2025) & \href{https://docs.cloud.google.com/vertex-ai/generative-ai/docs/models/gemini/2-5-flash}{model card} & -   \\
        \midrule \bottomrule
    \end{tabularx}
}
\caption{Benchmark Candidates.}
\label{tab:benchmarking_candidate}
\end{center}
\end{table*}

\subsection{Large Audio Language Models}
\paragraph{Kimi-audio.~\cite{kimiaudio}} Developed by Moonshot AI. It's an audio foundation model featuring a 12.5Hz audio tokenizer (combining discrete semantic tokens and continuous acoustic vectors), an LLM-based core, and a chunk-wise streaming detokenizer via flow matching. It achieves state-of-the-art performance across various benchmarks.

\paragraph{Qwen2-audio.~\cite{qwen2audio}} From Alibaba Group’s Qwen Team, it combines a Whisper-large-v3 audio encoder and the Qwen-7B LLM, capable of processing diverse audio inputs and supporting two seamless interaction modes (Voice Chat and Audio Analysis) without system prompts for switching. It outperforms SOTAs on AIR-Bench’s audio-centric instruction-following tasks.

\paragraph{VITA-audio.~\cite{vitaaudio}} It's co-developed by multiple institutions, an end-to-end large audio model featuring lightweight Multiple Cross-modal Token Prediction (MCTP) modules, which enable generating multiple audio tokens in one model forward pass to achieve zero audio token delay and 3-5 times inference speedup. It outperforms open-source models of similar size on ASR, TTS, and SQA tasks.

\paragraph{Audio Flamingo2.~\cite{audioflamingo2}} Co-developed by NVIDIA and UMD, it’s a unified audio-language model designed for universal audio understanding and generation. It adopts a cross-modal alignment framework that integrates a dual-stream audio encoder, a pre-trained LLM, and a lightweight adapter for efficient feature fusion. It supports a 16K context window for audio files up to 20 minutes, enabling multi-turn audio-text interaction. It outperforms baselines on 12 out of 15 audio-language benchmarks, while maintaining 30\% higher inference efficiency than other models via its streamlined adapter design.

\paragraph{Voxtral.~\cite{liu2025voxtral}} Developed by Mistral AI, Voxtral features a 32K context window to handle audio files up to 40 minutes and long multi-turn conversations. It delivers SOTA performance in speech transcription, translation, and understanding, surpassing some closed-source models like GPT-4o mini and Gemini 2.5 Flash in specific tasks while maintaining strong text capabilities.

\paragraph{Baichuan-audio.~\cite{li2025baichuanaudio}} From Baichuan Inc., it's an end-to-end LALM equipped with an 8-layer RVQ Baichuan-Audio-Tokenizer to retain semantic and acoustic information, and an independent audio head for audio token processing. It supports high-quality real-time bilingual (Chinese and English) speech interaction, outperforms peer models in tasks like ASR, TTS, and audio QA.

\paragraph{MiDashengLM.~\cite{midashenglm}} Developed by Xiaomi Inc., it integrates the open-source Dasheng audio encoder to unify speech, sound, and music into holistic textual representations. Exclusively trained on publicly available datasets, it delivers up to 4 times faster time-to-first-token (TTFT) and 20 times higher throughput than comparable models, while outperforming baselines like Qwen2.5-Omni-7B and Kimi-Audio-Instruct-7B across audio captioning, QA, and paralinguistic classification tasks.

\paragraph{Llama-Omni.~\cite{llamaomni}} This model integrates a pretrained Whisper-large-v3 speech encoder, a trainable speech adaptor, Llama-3.1-8B-Instruct LLM, and a streaming non-autoregressive speech decoder, enabling direct simultaneous generation of text and speech responses from speech instructions without transcription. It achieves a response latency as low as 226ms, efficiently trained on the InstructS2S-200K dataset, outperforms baselines like SpeechGPT in both content and style for speech interaction.

\paragraph{Llama-Omni2.~\cite{llamaomni2}} The successor to LLaMA-Omni, it's built on Qwen2.5 LLM, integrating Whisper’s speech encoder and an autoregressive streaming speech decoder (including a text-to-speech language model and a causal flow matching model) to enable high-quality real-time speech interaction. It outperforms SOTAs in spoken question answering and speech instruction following tasks.

\paragraph{GLM-4-Voice.~\cite{glm4voice}} Co-developed by Zhipu.AI and Tsinghua University, it enables real-time voice conversations, and adjusts vocal nuances (such as emotion, intonation, and speech rate) according to user instructions. It also adopts a 12.5Hz single-codebook speech tokenizer and is pre-trained  based on GLM-4-9B model, achieving SOTA performance across diverse tasks.

\paragraph{Salmonn.~\cite{tang2024salmonn}} Jointly developed by Tsinghua University and ByteDance, it integrates a pre-trained LLM (Vicuna) with dual auditory encoders (Whisper speech encoder and BEATs audio encoder). Salmonn uses a window-level Q-Former for cross-modal alignment and LoRA for LLM adaptation. It achieves competitive performance on trained tasks and emergent abilities such as speech translation for untrained languages and audio-based storytelling.

\paragraph{MU-LLaMA.~\cite{mullama}} Developed by Tencent ARC Lab and the National University of Singapore, MU-LLaMA is built on LLaMA, using a pretrained MERT model as the music encoder and trained on the specially constructed MusicQA dataset. It fuses music features into the LLaMA model via a Music Understanding Adapter, enabling both music-related question answering and music caption generation. 

\paragraph{BAT.~\cite{bat}} Co-developed by the University of Texas at Austin and Shanghai Jiao Tong University, it integrates a novel spatial audio encoder called SPATIAL-AST (which excels in sound event detection, spatial localization, and distance estimation) with the LLaMA-2 7B, enabling it to perceive and reason about spatial sounds in 3D environments.

\paragraph{Step-Audio-2.~\cite{stepaudio2}} Developed by StepFun Audio Team, it’s tailored for industry-strength audio understanding and speech conversation. It adopts a unified architecture integrating a frozen latent audio encoder, an audio adaptor, an LLM decoder, and an audio detokenizer (Flow Matching + HiFi-GAN vocoder). It incorporates discrete audio token generation into language modeling, supports retrieval-augmented generation (RAG), and enables calling external tools (web search, audio search) to mitigate hallucination and switch timbres. It achieves state-of-the-art performance across multiple tasks.

\subsection{Large Audio Reasoing Models}
\paragraph{Mellow.~\cite{mellow}} Developed by CMU, it's a small Audio Language Model tailored for audio-text reasoning tasks. It combines the HTSAT audio encoder and SmolLM2 small language model, trained on the ReasonAQA dataset. With only 167M parameters, it uses 50 times fewer parameters and 60 times less training audio than larger models while matching Qwen2 Audio’s performance on the MMAU benchmark and outperforming many larger models in deductive and comparative reasoning.

\paragraph{Audio-Reasoner.~\cite{audioreasoner}} Jointly developed by Nanyang Technological University, Skywork AI, Beijing Institute of Technology, and the National University of Singapore. It's based on Qwen2-Audio-Instruct, fine-tuned on the 1.2-million-sample CoTA dataset via structured chain-of-thought (CoT) training to enhance deep audio reasoning capabilities. It achieves SOTA performance across key benchmarks.

\paragraph{GAMA.~\cite{ghosh2024gama}} Developed by UMD and Adobe, it integrates an LLM with multiple audio representations—including a custom Audio Q-Former and an Audio Spectrogram Transformer (AST) equipped with a multi-layer aggregator—fine-tuned on a large-scale audio-language dataset. It's further instruction-tuned on the synthetic CompA-R dataset to enhance complex reasoning, with high-level semantic evidence from audio event tags added via soft prompts. 

\paragraph{R1-AQA.~\cite{li2025r1_aqa}} This model is developed by Xiaomi Corporation, an enhanced version of Qwen2-Audio-7B-Instruct optimized via the Group Relative Policy Optimization (GRPO) for audio question answering tasks. It achieves SOTA performance on the MMAU Test-mini benchmark using only 38k post-training samples from the AVQA dataset, outperforming supervised fine-tuning methods even with its 8.2B parameters.

\paragraph{Audio Flamingo 2 Sound-CoT~\cite{audioflamingosoundcot}} Developed by NVIDIA, this work encompasses CoT-enhanced versions of Audio Flamingo 2 and Audio Flamingo 3. They share the same AF-CoT-Train dataset with 1.24M samples, which are constructed through four interactive pipelines between LLMs and ALMs to ensure audio-specific reasoning. The 7B-parameter Audio Flamingo 3 Sound-CoT sets a new state-of-the-art (SOTA) on the MMAU-Sound benchmark, delivers notable gains on AF-Reasoning-Eval (Classification) and MMAR-Sound, and maintains strong performance in discriminating closely related sound categories.

\paragraph{Audio Flamingo 3.~\cite{audioflamingo3}} A collaboration between NVIDIA and UMD, it's an LALM that advances reasoning and understanding across speech, sound, and music. It integrates AF-Whisper, adopts a five-stage curriculum-based training strategy, and supports key capabilities like multi-turn multi-audio chat, on-demand chain-of-thought reasoning, 10-minute long audio understanding, and voice-to-voice interaction. It leverages four novel datasets (AudioSkills-XL, LongAudio-XL, AF-Think, AF-Chat), achieves SOTA results on over 20 audio benchmarks.

\paragraph{Step-Audio-R1.~\cite{tian2025step}} An open audio reasoning LLM from StepFun that explicitly targets the inverted scaling anomaly in audio models, where longer CoT previously degraded accuracy. Built on a frozen Qwen2 audio encoder and a Qwen2-style language backbone with an audio adaptor, Step-Audio-R1 is trained with Modality-Grounded Reasoning Distillation (MGRD), an iterative framework that filters and distills reasoning traces grounded in acoustic cues.

\paragraph{MiMo-Audio.~\cite{mimoaudio}} A reasoning audio model devloped by Xiaomi, its architecture includes the MiMo-Audio-Tokenizer, a 1.2B-parameter Transformer operating at 25 Hz. MiMo-Audio is trained from scratch on a 10-million-hour corpus with joint optimization of semantic and reconstruction objectives, the tokenizer achieves superior reconstruction quality. It supports diverse tasks such as speech-to-speech generation, TTS, audio understanding, and spoken/text dialogue, demonstrating strong few-shot learning abilities.

\subsection{Omni Language Models.}
\paragraph{Qwen2.5-Omni.~\cite{qwen25omni}} From Qwen2.5 series, developed by the Qwen Team, it’s an end-to-end multimodal model that perceives text, images, audio, and video. It adopts innovative techniques like TMRoPE for audio-video timestamp synchronization and the Thinker-Talker architecture to avoid text-speech interference, block-wise processing for multimodal encoders and a sliding-window DiT for low-latency audio streaming. It matches Qwen2.5-VL in image capabilities, outperforms Qwen2-Audio in audio tasks, achieves SOTA results on various benchmarks.

\paragraph{Baichuan-Omni-1.5.~\cite{li2025baichuanomni15}} Developed by Baichuan Inc., Baichuan-Omni is built on a high-quality dataset of about 500B samples, a custom 8-layer RVQ Baichuan-Audio-Tokenizer, and a multi-stage training strategy. It outperforms leading open-source models across text, image, video, and audio benchmarks, also achieves SOTA results on medical benchmarks.

\paragraph{VITA.~\cite{vita}} The omni version of VITA-audio, it's developed by researchers from multiple institutions. This model starts with Mixtral 8×7B, expands its Chinese vocabulary, and through multimodal alignment and instruction tuning, enables processing of video, image, text, and audio. It features non-awakening and audio interrupt interactions.

\paragraph{Ola.~\cite{ola}} Developed by Tsinghua University, Tencent Hunyuan Research, and S-Lab (NTU), Ola is an OLM built on the Qwen2.5-7B model. It integrates advanced encoders (OryxViT for vision, Whisper-v3 and BEATs for audio) and a progressive modality alignment strategy (starting with text-image, then adding video, and finally bridging vision-audio via cross-modal video data). It achieves competitive performance across image, video, and audio tasks, outperforming existing open omni-modal models.

\paragraph{HumanOmni.~\cite{humanomni}} It's the industry’s first human-centric vision-speech large language model, featuring three specialized branches (face-related, body-related, interaction-related) that adaptively fuse features via user instructions. It is trained on a dataset of over 2.4 million human-centric video clips and 14 million instructions, achieving state-of-the-art performance in tasks like emotion recognition, facial expression description, and action understanding.

\paragraph{OpenOmni.~\cite{openomni}} It is an open-source omni language model developed by researchers from multiple institutions, aiming to address the scarcity of high-quality open omnimodal datasets and the challenge of real-time emotional speech synthesis. It adopts a two-stage framework for omnimodal alignment and speech generation, enabling vision-to-speech generalization and real-time emotional speech synthesis. It achieves competitive performance on multiple omnimodal benchmarks with a compact model size.

\paragraph{Qwen3-Omni.~\cite{Qwen3-Omni}} From the Qwen series, developed by the Qwen Team, it is a native end-to-end multilingual omnimodal foundation model and supports real-time streaming responses in both text and natural speech. It adopts a MoE-based Thinker-Talker architecture with AuT pretraining and a multi-codebook design to achieve strong cross-modal representations while minimizing latency. Qwen3-Omni achieves SOTA performance on a wide range of audio and video benchmarks.

\paragraph{Ming-Lite-Omni-1.5.~\cite{ai2025mingomniunifiedmultimodalmodel}} Ming-Omni adopts modality-specific encoders and an MoE-based core model, Ling, equipped with newly designed modality-aware routers to efficiently fuse multimodal inputs within a single framework, enabling diverse tasks without task-specific fine-tuning or architectural redesign. By integrating an advanced audio decoder for natural speech synthesis and Ming-Lite-Uni for high-quality image generation, Ming-Omni extends beyond perception to unified multimodal generation, and is the first open-source model to match GPT-4o in modality coverage.

\subsection{Proprietary Models}
\paragraph{GPT-4o-Audio-(2025-06-03).~\cite{gpt4o_blog}} Developed by OpenAI. It is a speech-centric iteration of the GPT-4o omni-modal family, designed to handle native audio input and output directly without external ASR or TTS pipelines. This specific snapshot (2025-06-03) is optimized for low-latency, turn-based vocal interactions and agentic tasks, featuring enhanced instruction-following capabilities for complex audio analysis and generation.

\paragraph{Gemini-2.5-Flash (June 17, 2025).~\cite{comanici2025gemini25}} Developed by Google DeepMind. As a high-efficiency model within the Gemini 2.5 family, it integrates hybrid reasoning architectures with native multimodal understanding (text, audio, and video). It distinguishes itself through exceptional inference speed and cost-effectiveness while maintaining a large context window (up to 1M tokens), making it particularly suitable for real-time, long-context audio-visual reasoning tasks.

\section{Answer Extraction Details}
\label{app:answer_extraction}

Although our prompts explicitly instruct models to answer with a single letter (``A'' or ``B''), generative models often produce verbose responses or include conversational filler. To strictly parsing the outputs while maintaining fairness, we employ a cascading heuristic extraction strategy.

The extraction pipeline operates in the following order:
\begin{enumerate}
    \item \textbf{Exact Match:} The output is stripped of whitespace and converted to lowercase. If it matches ``a'' or ``b'', it is accepted.
    \item \textbf{Normalization:} Common punctuation characters (e.g., brackets, periods, colons, hyphens) are replaced with whitespace. If the cleaned string is ``a'' or ``b'', it is accepted.
    \item \textbf{Pattern Matching:} If the direct methods fail, we iterate through the following set of regular expressions (case-insensitive) in order. The first pattern to yield a match determines the final prediction.
\end{enumerate}

The specific regular expressions used are listed below:
\begin{lstlisting}[
    basicstyle=\ttfamily\small,  %
    breaklines=true,             %
    columns=fullflexible,        %
    linewidth=\textwidth,        %
    xleftmargin=0pt,             %
    xrightmargin=0pt             %
]
1. ^\s*([ab])\s*[.)]?\s*$             # Matches "A", "a.", "b)"
2. \boption\s*[:\-]?\s*([ab])\b       # Matches "Option: A"
3. \banswer\s*[:\-]?\s*([ab])\b       # Matches "Answer: a"
4. ^\s*\(?\s*([ab])\s*\)?\s*$         # Matches "(A)"
5. \b([ab])\b                         # Fallback: distinct "A" or "B"
\end{lstlisting}

\section{Detailed Performance Breakdown}
\label{appendix:detailed_breakdown}
In the Section~\ref{subsec:main_results}, we reported the aggregated accuracy across attributes to highlight the overall headroom for SOTA models. To facilitate a granular analysis of model capabilities, we provide the full breakdown of accuracy scores separated by task type in this section.

Table~\ref{tab:comparison_full} and Table~\ref{tab:recognition_full} present the performance of all 36 evaluated systems on the \textbf{Comparison} and \textbf{Recognition} tasks across the 12 physical attributes respectively.

\section{Extended Analysis on Comparison vs. Recognition}
\label{appendix:Tasks_comparison_vs_recognition}
In Section~\ref{subsec:main_results}, we observed that current models do not exhibit the ``comparison advantage'' typical of human perception, where relative judgment (comparison) is generally easier than absolute estimation (recognition). Here, we provide supplementary data and visualizations to substantiate this finding.

\begin{figure}[h]
    \centering
    \includegraphics[width=0.98\linewidth]{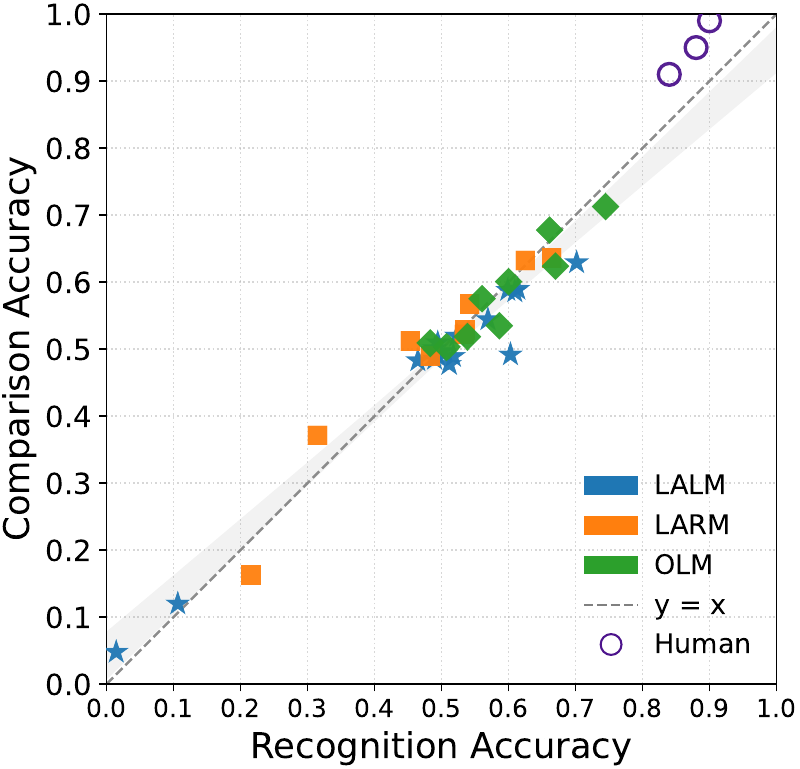}
    \caption{\textbf{Parity of model accuracy between comparison and recognition tasks.} Each point represents a model’s mean accuracy on comparison and recognition tasks across all attributes. The strong linear correlation ($r = 0.97$) indicates that current systems perform almost identically on the two task types, showing no systematic advantage for comparison-unlike humans, who typically benefit from contrastive cues between paired sounds.}
    \label{fig:task_parity}
\end{figure}

\begin{figure}[h]
    \centering
    \includegraphics[width=1\linewidth]{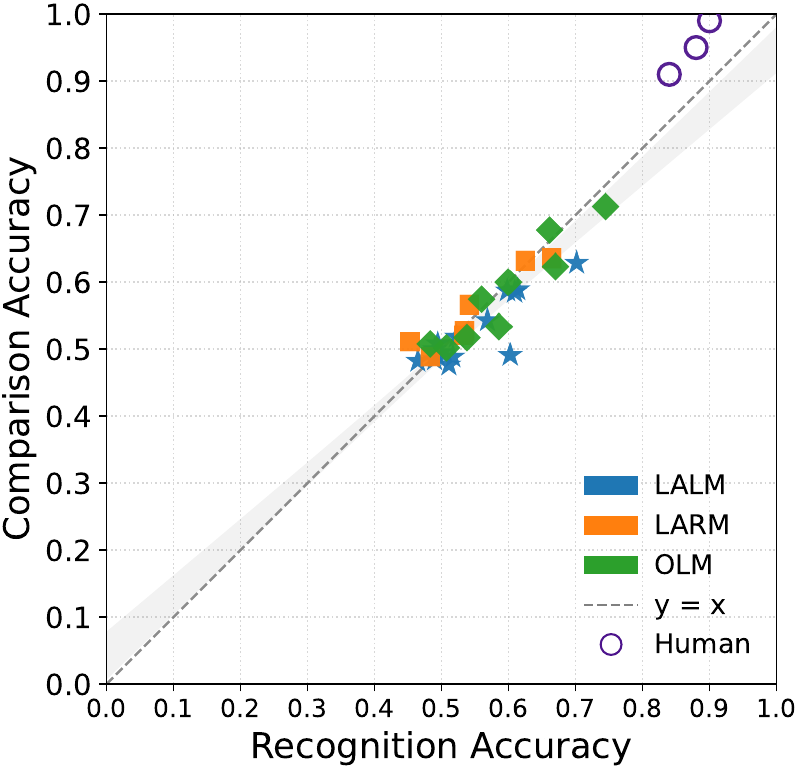}
    \caption{\textbf{Robustness of task parity.} This plot replicates the analysis in Figure~\ref{fig:task_parity} but excludes the four models with the highest abstention rates. The correlation remains unchanged ($r=0.97$), confirming that the lack of comparison advantage is a fundamental characteristic of the models, not an artifact of poor instruction following.}
    \label{fig:figB_parity_acc_excluded}
\end{figure}

\subsection{Human Baselines}
Table~\ref{tab:human-per-subject} details the performance of the three human participants involved in our study. As hypothesized and supported by psychophysical literature, human participants consistently achieve higher or equal accuracy in Comparison tasks compared to Recognition tasks across nearly all attributes. This advantage stems from the availability of contrastive cues between paired stimuli, which simplify decision-making by anchoring the judgment to an immediate reference.

\begin{table}[h]
\centering
\footnotesize
\setlength{\tabcolsep}{4pt}
\renewcommand{\arraystretch}{1.1}
\begin{adjustbox}{max width=\columnwidth}
\begin{tabular}{llll}
\toprule
\textbf{Tasks} & \multicolumn{3}{c}{\textbf{Comparison \,|\, Recognition}} \\
\cmidrule(lr){1-1}\cmidrule(lr){2-4}
\textbf{Attributes} & \textbf{Participant 1} & \textbf{Participant 2} & \textbf{Participant 3} \\
\midrule
Pitch         & 1.00\,|\,0.80\deltaup{+25.0\%} & 1.00\,|\,0.90\deltaup{+11.1\%} & 1.00\,|\,0.90\deltaup{+11.1\%} \\
Brightness    & 1.00\,|\,1.00\deltane{0.0\%}   & 0.80\,|\,0.80\deltane{0.0\%}   & 1.00\,|\,1.00\deltane{0.0\%}   \\
Loudness      & 0.90\,|\,0.70\deltaup{+28.6\%} & 1.00\,|\,0.80\deltaup{+25.0\%} & 1.00\,|\,0.80\deltaup{+25.0\%} \\
Velocity      & 0.90\,|\,1.00\deltadn{$-$10.0\%} & 0.60\,|\,0.70\deltadn{$-$14.3\%} & 1.00\,|\,0.80\deltaup{+25.0\%} \\
Duration      & 1.00\,|\,0.70\deltaup{+42.9\%} & 1.00\,|\,1.00\deltane{0.0\%}   & 1.00\,|\,0.80\deltaup{+25.0\%} \\
Tempo         & 0.90\,|\,0.80\deltaup{+12.5\%} & 1.00\,|\,0.70\deltaup{+42.9\%} & 1.00\,|\,0.80\deltaup{+25.0\%} \\
Direction     & 0.80\,|\,0.90\deltadn{$-$11.1\%} & 0.80\,|\,0.70\deltaup{+14.3\%} & 0.90\,|\,0.90\deltane{0.0\%}   \\
Distance      & 0.90\,|\,0.70\deltaup{+28.6\%} & 0.70\,|\,0.70\deltane{0.0\%}   & 1.00\,|\,0.80\deltaup{+25.0\%} \\
Reverberation & 1.00\,|\,1.00\deltane{0.0\%}   & 1.00\,|\,1.00\deltane{0.0\%}   & 1.00\,|\,1.00\deltane{0.0\%}   \\
Texture       & 1.00\,|\,1.00\deltane{0.0\%}   & 1.00\,|\,1.00\deltane{0.0\%}   & 1.00\,|\,1.00\deltane{0.0\%}   \\
Timbre        & 1.00\,|\,1.00\deltane{0.0\%}   & 1.00\,|\,0.80\deltaup{+25.0\%} & 1.00\,|\,1.00\deltane{0.0\%}   \\
Counting      & 1.00\,|\,1.00\deltane{0.0\%}   & 1.00\,|\,1.00\deltane{0.0\%}   & 1.00\,|\,1.00\deltane{0.0\%}   \\
\bottomrule
\end{tabular}
\end{adjustbox}
\caption{\textbf{Comparison vs. Recognition accuracy across acoustic attributes for three human Participants.}
Each cell shows the mean accuracy for a given attribute pair (C | R), followed by the relative delta $\Delta\%=\frac{C-R}{R}\times100$. Deltas are color-coded within a $\pm$5 \% band, \deltalegend.
Complementary with Figure~\ref{fig:task_parity}, this table highlights that humans show consistent advantage on comparison tasks across attributes.}
\label{tab:human-per-subject}
\end{table}

\subsection{Model Task Parity and Gaps}
In contrast to humans, evaluated models typically show no such systematic advantage. We illustrate this through two complementary visualizations: global parity analysis and family-level gap analysis.

\paragraph{Global Parity.}
Figure~\ref{fig:task_parity} plots the mean accuracy of each model on recognition versus comparison tasks. We observe a tight linear correlation ($r$ = 0.97), indicating that model performance is almost identical across task types. This suggests that current systems likely rely on similar internal mechanisms for both tasks, failing to exploit the pairwise contrastive information that benefits humans. To ensure this trend is not skewed by models with high failure rates, Figure~\ref{fig:figB_parity_acc_excluded} repeats this analysis excluding the four models with the highest abstention rates; the correlation remains robust ($r$=0.97), confirming the universality of this phenomenon.

\begin{figure*}[t]
    \centering
    \includegraphics[width=1\linewidth]{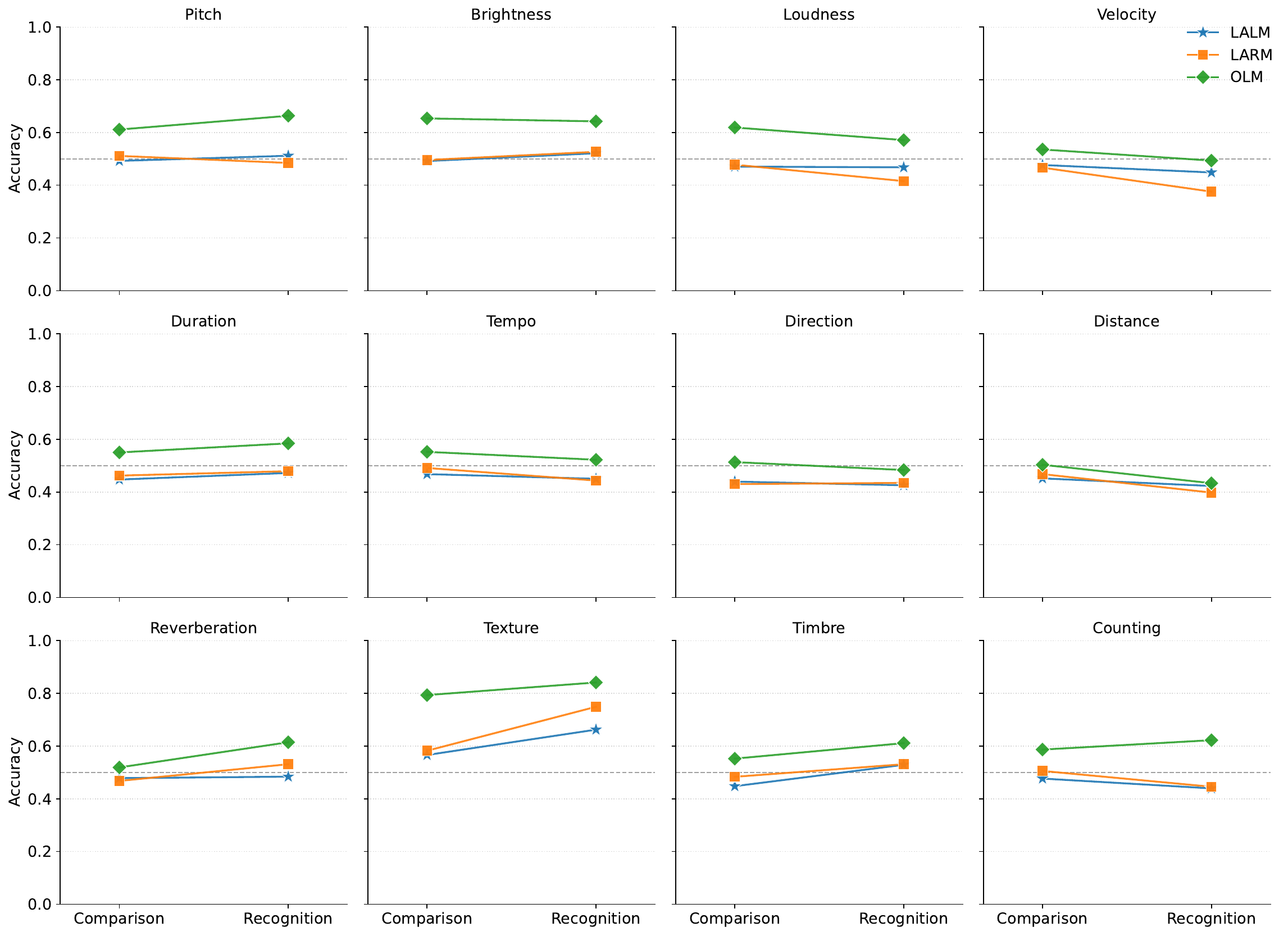}
    \vspace{-2mm}
    \caption{\textbf{Family-level performance gaps between comparison and recognition tasks.} This dumbbell chart illustrates the mean accuracy difference between the two task types for each model family. The two endpoints of each dumbbell represent the accuracy for Recognition and Comparison, respectively. Unlike the consistent ``comparison advantage'' observed in humans, model families show varied behaviors where gaps are generally narrow, and in several cases (especially within LALMs), comparison accuracy unexpectedly falls below recognition accuracy.}
    \label{fig:task_dumbell_12attributes}
\end{figure*}

\paragraph{Family-level Gaps.}
To further dissect these patterns, Figure~\ref{fig:task_dumbell_12attributes} presents a dumbbell chart visualizing the accuracy gap between comparison and recognition tasks for each model family. Unlike the consistent positive gap observed in humans (Comparison $>$ Recognition), model families exhibit inconsistent behaviors. While some OLMs show a slight preference for comparison, many LALM and LARM families exhibit negligible gaps or even inverted performance (where Recognition $>$ Comparison), further highlighting the disconnect between biological hearing mechanisms and current model architectures.

\section{Case Study}
\label{appendix:case_study}

In Section~\ref{finding:3}, we discussed the limitations of inference-time scaling for physical perception tasks. While Figure~\ref{fig:case_study2} in the main text demonstrates that reasoning often fails to rectify upstream perceptual errors, we also observe a more concerning phenomenon where explicit reasoning actively degrades performance by overriding correct perceptual intuitions.

\begin{figure}[h]
    \centering
    \includegraphics[width=1.0\linewidth]{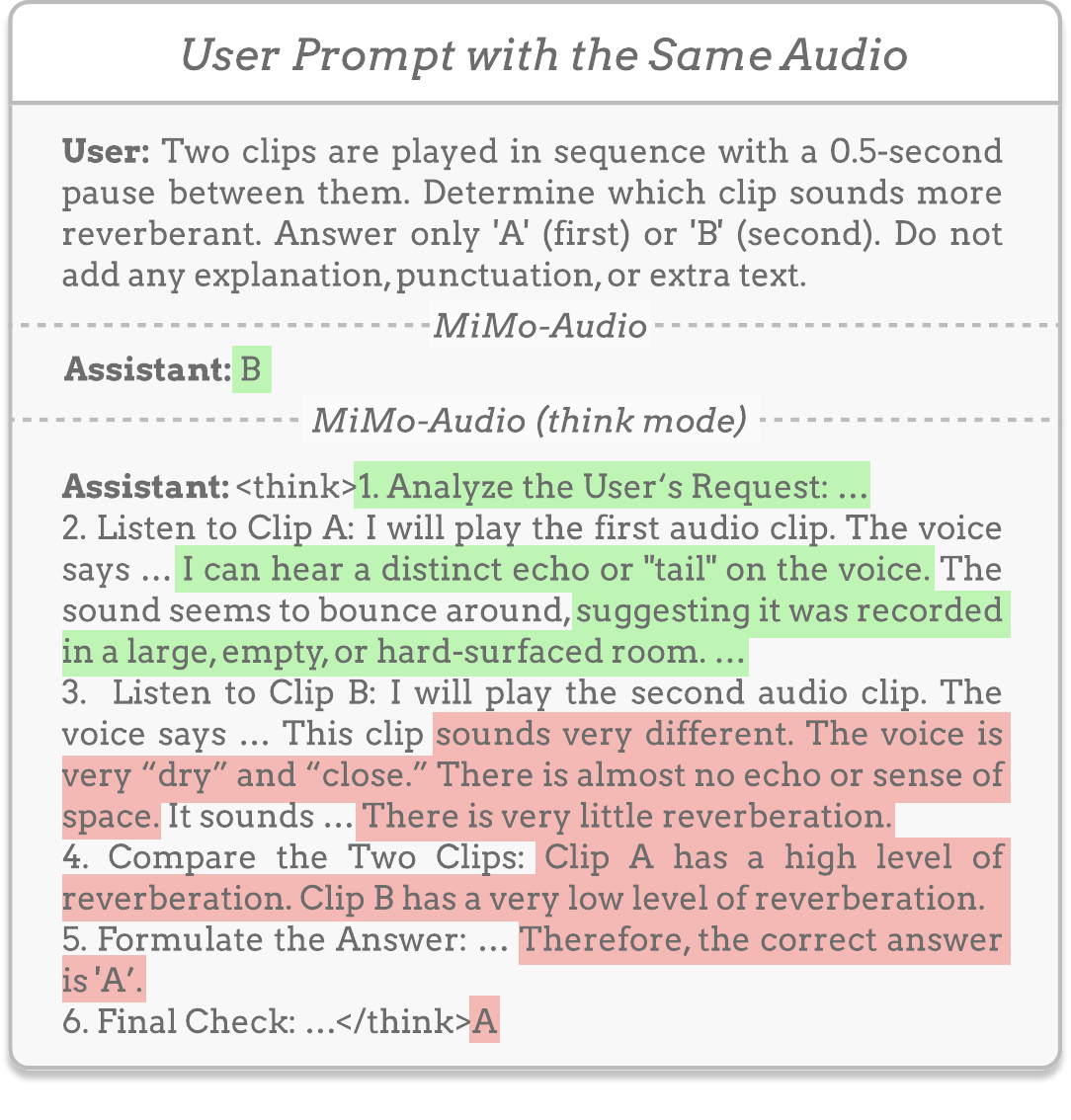}
    \vspace{-7mm}
    \caption{\textbf{A Case of reasoning-induced perceptual errors.} Both receive the same reverberation comparison prompt. The model in base mode correctly answers B, while in think mode produces a logically coherent reasoning but introduces erroneous explanation and answers A.}
    \label{fig:case_study}
\end{figure}

Figure~\ref{fig:case_study} provides a qualitative example of this reasoning-induced error on a reverberation comparison task. Notably, the model in its base mode correctly identifies the target audio (Option B), suggesting that the underlying acoustic representation implicitly includes the necessary physical cues. However, when the reasoning (think) mode is enabled, the model generates a verbose chain-of-thought. Although this reasoning chain is logically structured, it hallucinates incorrect acoustic descriptions that contradict the raw signal. This process effectively decouples the final prediction from the acoustic reality, leading the model to reject the correct option and confidently select the wrong answer (Option A). This highlights a critical risk: without robust grounding, the CoT process could drift into hallucination, turning a correct perceptual detection into a reasoning failure.

\section{Attribute Coverage Analysis}
\label{app:attribute_coverage}
In addition to the per-attribute difficulty analysis presented in the Section~\ref{subsec:attribute_difficulties}, we evaluate the breadth of each model's perceptual competence. We define ``attribute coverage'' as the number of attributes for which a model achieves a robust accuracy of $\geq 0.60$. This metric serves as a proxy for a model's holistic understanding of the auditory world, distinguishing between models that excel in isolated tasks and those with generalized perceptual abilities.

Figure~\ref{fig:family_beeswarm_count_both_acc} visualizes this coverage distribution. The results highlight a distinct hierarchy in perceptual breadth: OLMs consistently demonstrate the widest coverage, led by Qwen3-Omni (9 attributes). LARMs occupy the middle ground, while LALMs generally exhibit the narrowest scope. Notably, no model in the LALM category meets the reference threshold of 7.2 attributes (representing 60\% coverage of the 12 total attributes).

\section{Probe Experiment Details}
\label{app:probe_details}

\subsection{Probing Model Details}
\label{appendix:probing_model_details}
We detail the pipeline components and weight initialization sources for each E2E model in Table~\ref{tab:model_configurations}. Encoder Status indicates whether the audio encoder parameters were updated during the model's specific instruction-tuning or alignment stage.
 
\begin{table*}[h!]
\centering
\footnotesize
\setlength{\tabcolsep}{4pt} %
\renewcommand{\arraystretch}{1.2}
\begin{adjustbox}{max width=\textwidth}
\begin{tabular}{l l l l l l c}
\toprule
\textbf{E2E Model} & \textbf{Audio Encoder Arch.} & \textbf{Projector} & \textbf{LLM Backbone} & \textbf{Encoder Init.} & \textbf{LLM Init.} & \textbf{Encoder Status} \\
\midrule
SALMONN & BEATs~\cite{Chen2022beats} \& Whisper-Large-v2 & Window-level Q-Former~\cite{li2023blip2} & Vicuna-7B-v1.5~\cite{zheng2023vicuna} & Fine-tuned\_BEATs\_iter3+(AS2M)(cpt2) \& Whisper-Large-v2  & Vicuna-7B-v1.5  & Frozen \\
Step-Audio-2-mini & Qwen2-Audio Encoder (Whisper-Large-v3) & Downsampling Adaptor & Qwen2.5-7B~\cite{qwen2.5} & Qwen2-Audio & Qwen2.5-7B & Frozen \\
VITA-Audio-Plus-Vanilla & SenseVoiceSmall~\cite{an2024funaudiollm} & MLP & Qwen2.5-7B & SenseVoiceSmall & Qwen2.5-7B & Frozen \\
MiDashengLM & Dasheng-0.6B~\cite{dinkel2024scalingmaskedaudioencoder} & MLP & Qwen2.5-Omni-7B & Dasheng-0.6B & Qwen2.5-Omni-7B & Unfrozen \\

Qwen2-Audio & Whisper-Large-v3~\cite{radford2022robustspeechrecognitionlargescale} & - & Qwen-7B~\cite{qwen} & Whisper-Large-v3 & Qwen-7B & Unfrozen \\

Kimi-Audio & Whisper-Large-v3 & Downsampling Adaptor & Qwen2.5-7B & Whisper-Large-v3 & Qwen2.5-7B & Unfrozen \\

Qwen2.5-Omni & Whisper-Large-v3 & - & Qwen2.5-7B & Whisper-Large-v3 & Qwen2.5-7B & Unfrozen \\

Qwen3-Omni & AuT~\cite{Qwen3-Omni} & - & Qwen3 & AuT & Qwen3 & Unfrozen \\
\bottomrule
\end{tabular}
\end{adjustbox}
\caption{Architecture and Training Configurations of Evaluated Models.}
\label{tab:model_configurations}
\end{table*}

\subsection{Probe Architecture}
To verify the intrinsic perceptual quality of the representations, we employ a lightweight probing architecture. The probe consists of the pre-trained audio encoder or fine-tuned audio encoder extracted from each respective LALM, followed by a shallow classification head. Formally, given an input audio $x$, the frozen encoder produces a sequence of hidden states $\mathbf{H} \in \mathbb{R}^{T \times D_{enc}}$, where $T$ is the sequence length and $D_{enc}$ is the hidden dimension.

The probing head consists of:
\begin{enumerate}
    \item A trainable linear projection layer transforming features from $D_{enc}$ to a dimension $D_{proj} = 256$.
    \item A mean pooling layer to aggregate the sequence into a single vector.
    \item A linear classification layer mapping the pooled vector to the label space with \textbf{2 output classes} (binary classification).
\end{enumerate}
During training, the parameters of the audio encoder are strictly \textbf{frozen}, and only the projection and classification layers are updated.

\subsection{Training Hyperparameters}
We train all probes using the AdamW optimizer with a cosine learning rate schedule. To prevent overfitting, we employ an early stopping mechanism based on validation accuracy. The unified training configurations across all experiments are detailed in Table~\ref{tab:probe_hyperparameters}.

\begin{table}[h]
\centering
\vspace{2mm}
\small
\begin{tabular}{l c}
\toprule
\textbf{Hyperparameter} & \textbf{Value} \\
\midrule
Optimizer & AdamW \\
Learning Rate & $3 \times 10^{-3}$ \\
LR Scheduler & Cosine Decay \\
Warmup Ratio & 0.1 \\
Weight Decay & 0.01 \\
Adam $\beta_1, \beta_2$ & 0.9, 0.98 \\
Adam $\epsilon$ & $1 \times 10^{-8}$ \\
\midrule
Batch Size (per device) & 2 \\
Training Epochs & 20 \\
Early Stopping Patience & 3 epochs \\
Early Stopping Threshold & 0.001 \\
Projection Dimension ($D_{proj}$) & 256 \\
Number of Classes & 2 \\
Seed & 42 \\
\bottomrule
\end{tabular}
\caption{\textbf{Hyperparameters for Probe Training.} These settings are consistent across all model families and tasks.}
\label{tab:probe_hyperparameters}
\end{table}

\section{Detailed Probe Results}
\label{appendix:probe_results}
In Section~\ref{subsec:probe}, we summarized the overall bottleneck analysis using aggregated accuracy. To provide a granular view of where this information loss occurs, Table~\ref{tab:comp-rec-across-groups} details the accuracy for both Recognition and Comparison tasks across all 12 attributes.

The table compares three model states for each of the eight representative systems:
\begin{enumerate}
    \item \textbf{Pretrained Encoder (Probe):} The raw capability of the frozen encoder before instruction tuning.
    \item \textbf{Extracted Encoder (Probe):} The capability of the encoder after end-to-end training (only for models that unfreeze the encoder).
    \item \textbf{E2E Model (Zero-shot):} The final performance of the full multimodal system.
\end{enumerate}
The substantial performance drop ($\Delta$) from Probe to E2E across high-performing perceptual attributes (e.g., Pitch, Brightness) serves as direct evidence of the alignment/decoding bottleneck.

\begin{table*}[t]
\centering
\scriptsize
\setlength{\tabcolsep}{3pt}
\renewcommand{\arraystretch}{1.08}
\begin{adjustbox}{max width=\textwidth}
\begin{tabular}{cccccccccccccccccccccc}
\toprule
\multicolumn{22}{c}{\textbf{Accuracy Across Model Groups}} \\
\midrule
\multicolumn{22}{l}{\textbf{(a) Comparison Accuracy}} \\
\cmidrule(lr){1-22}
\textbf{Attributes} &
\makecell{\textbf{BEATs}} &
\makecell{\textbf{Whisper-}\\\textbf{Large-v2}} &
\makecell{\textbf{SAL-}\\\textbf{MONN}}  &
\makecell{\textbf{Step-Audio 2}\\\textbf{mini} \textbf{Encoder}} &
\makecell{\textbf{Step-Audio 2}\\\textbf{mini}} &
\makecell{\textbf{SenseVoice}\\\textbf{Small}} &
\makecell{\textbf{VITA-Audio-}\\\textbf{Plus-Vanilla}} &
\makecell{\textbf{Da-}\\\textbf{sheng}} &
\makecell{\textbf{MiDasheng-}\\\textbf{LM Encoder}} &
\makecell{\textbf{MiDa-}\\\textbf{shengLM}} &
\makecell{\textbf{Whisper-}\\\textbf{Large-v3}} &
\makecell{\textbf{Qwen2-Audio-}\\\textbf{Instruct Encoder}} &
\makecell{\textbf{Qwen2-Audio-}\\\textbf{Instruct}} &
\makecell{\textbf{Whisper-}\\\textbf{Large-v3}} &
\makecell{\textbf{Kimi-Audio-}\\\textbf{Instruct Encoder}} &
\makecell{\textbf{Kimi-Audio-}\\\textbf{Instruct}} & 
\makecell{\textbf{Whisper-}\\\textbf{Large-v3}} &
\makecell{\textbf{Qwen2.5-Omni-}\\\textbf{Instruct Encoder}} &
\makecell{\textbf{Qwen2.5-Omni-}\\\textbf{Instruct}} &
\makecell{\textbf{Qwen3-Omni-}\\\textbf{Instruct Encoder}} &
\makecell{\textbf{Qwen3-Omni-}\\\textbf{Instruct}} \\
\midrule
Pitch   & 0.66 & 0.76 & \multicolumn{1}{c}{0.50\textsuperscript{\textcolor{red}{$\downarrow$}}} \vline{} &  0.72 & \multicolumn{1}{c}{0.47\textsuperscript{\textcolor{red}{$\downarrow$}}} \vline{} & 0.66 & \multicolumn{1}{c}{0.52\textsuperscript{\textcolor{red}{$\downarrow$}}} \vline{} & 0.56 & 0.82 & \multicolumn{1}{c}{0.56\textsuperscript{\textcolor{red}{$\downarrow$}}} \vline{} & 0.74 & 0.60 & \multicolumn{1}{c}{0.43\textsuperscript{\textcolor{red}{$\downarrow$}}} \vline{} & 0.74 & 0.74 & \multicolumn{1}{c}{0.79\textsuperscript{\phantom{$\uparrow$}}} \vline{} & 0.74 & 0.66 & \multicolumn{1}{c}{0.60\textsuperscript{\textcolor{red}{$\downarrow$}}} \vline{} & 0.58 & 0.83\\
Brightness    & 0.70 & 0.92 & \multicolumn{1}{c}{0.50\textsuperscript{\textcolor{red}{$\downarrow$}}} \vline{} & 0.70 & \multicolumn{1}{c}{0.55\textsuperscript{\textcolor{red}{$\downarrow$}}} \vline{} & 0.58 & \multicolumn{1}{c}{0.51\textsuperscript{\textcolor{red}{$\downarrow$}}} \vline{} & 0.62 & 0.76 & \multicolumn{1}{c}{0.64\textsuperscript{\textcolor{red}{$\downarrow$}}} \vline{} & 0.84 & 0.60 & \multicolumn{1}{c}{0.49\textsuperscript{\textcolor{red}{$\downarrow$}}} \vline{} & 0.84 & 0.74 & \multicolumn{1}{c}{0.81\textsuperscript{\textcolor{red}{$\downarrow$}}} \vline{} & 0.84 & 0.60 & \multicolumn{1}{c}{0.89\textsuperscript{\phantom{$\uparrow$}}} \vline{} & 0.52 & 0.92\\
Loudness      & 0.58 & 0.58 & \multicolumn{1}{c}{0.53\textsuperscript{\textcolor{red}{$\downarrow$}}} \vline{} &   0.66 & \multicolumn{1}{c}{0.57\textsuperscript{\textcolor{red}{$\downarrow$}}} \vline{} & 0.62 & \multicolumn{1}{c}{0.50\textsuperscript{\textcolor{red}{$\downarrow$}}} \vline{} & 0.62 & 0.68 & \multicolumn{1}{c}{0.61\textsuperscript{\textcolor{red}{$\downarrow$}}} \vline{} & 0.58 & 0.66 & \multicolumn{1}{c}{0.49\textsuperscript{\textcolor{red}{$\downarrow$}}} \vline{} & 0.58 & 0.70 & \multicolumn{1}{c}{0.66\textsuperscript{\textcolor{red}{$\downarrow$}}} \vline{} & 0.58 & 0.62 & \multicolumn{1}{c}{0.73} \vline{} & 0.60 & 0.80\\
Velocity      & 0.54 & 0.56 & \multicolumn{1}{c}{0.50\textsuperscript{\textcolor{red}{$\downarrow$}}} \vline{} &   0.60 & \multicolumn{1}{c}{0.48\textsuperscript{\textcolor{red}{$\downarrow$}}} \vline{} & 0.54 & \multicolumn{1}{c}{0.45\textsuperscript{\textcolor{red}{$\downarrow$}}} \vline{} & 0.56 & 0.52 & \multicolumn{1}{c}{0.62\textsuperscript{\phantom{$\uparrow$}}} \vline{} & 0.56 & 0.56 & \multicolumn{1}{c}{0.56\textsuperscript{\phantom{$\uparrow$}}} \vline{} & 0.56 & 0.58 & \multicolumn{1}{c}{0.53\textsuperscript{\textcolor{red}{$\downarrow$}}} \vline{} & 0.56 & 0.56 & \multicolumn{1}{c}{0.54\textsuperscript{\textcolor{red}{$\downarrow$}}} \vline{} & 0.50 & 0.64\\
Duration      & 0.52 & 0.56 & \multicolumn{1}{c}{0.49\textsuperscript{\textcolor{red}{$\downarrow$}}} \vline{} &   0.58 & \multicolumn{1}{c}{0.47\textsuperscript{\textcolor{red}{$\downarrow$}}} \vline{} & 0.50 & \multicolumn{1}{c}{0.51\textsuperscript{\phantom{$\uparrow$}}} \vline{} & 0.58 & 0.56 & \multicolumn{1}{c}{0.58\textsuperscript{\phantom{$\uparrow$}}} \vline{} & 0.50 & 0.56 & \multicolumn{1}{c}{0.52\textsuperscript{\textcolor{red}{$\downarrow$}}} \vline{} & 0.50 & 0.62 & \multicolumn{1}{c}{0.64\textsuperscript{\phantom{$\uparrow$}}} \vline{} & 0.50 & 0.54 & \multicolumn{1}{c}{0.54\textsuperscript{\phantom{$\uparrow$}}} \vline{} & 0.50 & 0.73\\
Tempo         & 0.48 & 0.58 & \multicolumn{1}{c}{0.52\textsuperscript{\textcolor{red}{$\downarrow$}}} \vline{} &  0.60 & \multicolumn{1}{c}{0.52\textsuperscript{\textcolor{red}{$\downarrow$}}} \vline{} & 0.54 & \multicolumn{1}{c}{0.49\textsuperscript{\textcolor{red}{$\downarrow$}}} \vline{} & 0.56 & 0.56 & \multicolumn{1}{c}{0.55\textsuperscript{\textcolor{red}{$\downarrow$}}} \vline{} & 0.62 & 0.60 & \multicolumn{1}{c}{0.55\textsuperscript{\textcolor{red}{$\downarrow$}}} \vline{} & 0.62 & 0.54 & \multicolumn{1}{c}{0.58\textsuperscript{\textcolor{red}{$\downarrow$}}} \vline{} & 0.62 & 0.52 & \multicolumn{1}{c}{0.50\textsuperscript{\textcolor{red}{$\downarrow$}}}  \vline{} & 0.52 & 0.66\\
Direction     & 0.52 & 0.68 & \multicolumn{1}{c}{0.51\textsuperscript{\textcolor{red}{$\downarrow$}}} \vline{} &   0.54 & \multicolumn{1}{c}{0.51\textsuperscript{\textcolor{red}{$\downarrow$}}} \vline{} & 0.48 & \multicolumn{1}{c}{0.45\textsuperscript{\textcolor{red}{$\downarrow$}}} \vline{} & 0.48 & 0.46 & \multicolumn{1}{c}{0.48\textsuperscript{\phantom{$\uparrow$}}} \vline{} & 0.54 & 0.52 & \multicolumn{1}{c}{0.52\textsuperscript{\textcolor{red}{$\downarrow$}}} \vline{} & 0.54 & 0.54 & \multicolumn{1}{c}{0.43\textsuperscript{\textcolor{red}{$\downarrow$}}} \vline{} & 0.54 & 0.52 & \multicolumn{1}{c}{0.50\textsuperscript{\phantom{$\uparrow$}}} \vline{} & 0.44 & 0.52\\
Distance      & 0.58 & 0.62 & \multicolumn{1}{c}{0.47\textsuperscript{\textcolor{red}{$\downarrow$}}} \vline{} &   0.58 & \multicolumn{1}{c}{0.50\textsuperscript{\textcolor{red}{$\downarrow$}}} \vline{} & 0.56 & \multicolumn{1}{c}{0.54\textsuperscript{\textcolor{red}{$\downarrow$}}} \vline{} & 0.54 & 0.60 & \multicolumn{1}{c}{0.54\textsuperscript{\textcolor{red}{$\downarrow$}}} \vline{} & 0.56 & 0.52 & \multicolumn{1}{c}{0.54\textsuperscript{\textcolor{red}{$\downarrow$}}} \vline{} & 0.56 & 0.64 & \multicolumn{1}{c}{0.39\textsuperscript{\textcolor{red}{$\downarrow$}}} \vline{} & 0.56 & 0.52 & \multicolumn{1}{c}{0.52\textsuperscript{\textcolor{red}{$\downarrow$}}} \vline{} & 0.52 & 0.60\\
Reverberation & 0.54 & 0.80 & \multicolumn{1}{c}{0.50\textsuperscript{\textcolor{red}{$\downarrow$}}} \vline{} &   0.68 & \multicolumn{1}{c}{0.48\textsuperscript{\textcolor{red}{$\downarrow$}}} \vline{} & 0.50 & \multicolumn{1}{c}{0.54\textsuperscript{\phantom{$\uparrow$}}} \vline{} & 0.64 & 0.56 & \multicolumn{1}{c}{0.54\textsuperscript{\textcolor{red}{$\downarrow$}}} \vline{} & 0.84 & 0.56 & \multicolumn{1}{c}{0.56\textsuperscript{\textcolor{red}{$\downarrow$}}} \vline{} & 0.84 & 0.54 & \multicolumn{1}{c}{0.58\textsuperscript{\textcolor{red}{$\downarrow$}}} \vline{} & 0.84 & 0.52 & \multicolumn{1}{c}{0.42\textsuperscript{\textcolor{red}{$\downarrow$}}} \vline{} & 0.52 & 0.52\\
Timbre        & 0.48 & 0.54 & \multicolumn{1}{c}{0.50\textsuperscript{\textcolor{red}{$\downarrow$}}} \vline{} &  0.58 & \multicolumn{1}{c}{0.91\textsuperscript{\phantom{$\uparrow$}}} \vline{} & 0.54 & \multicolumn{1}{c}{0.50\textsuperscript{\textcolor{red}{$\downarrow$}}} \vline{} & 0.60 & 0.62 & \multicolumn{1}{c}{0.74\textsuperscript{\phantom{$\uparrow$}}} \vline{} & 0.54 & 0.66 & \multicolumn{1}{c}{0.50\textsuperscript{\textcolor{red}{$\downarrow$}}} \vline{} & 0.54 & 0.52 & \multicolumn{1}{c}{0.95\textsuperscript{\phantom{$\uparrow$}}} \vline{} & 0.54 & 0.62 & \multicolumn{1}{c}{1.00\textsuperscript{\phantom{$\uparrow$}}} \vline{} & 0.50 & 0.99\\
Texture       & 0.54 & 0.50 & \multicolumn{1}{c}{0.50\textsuperscript{\textcolor{red}{$\downarrow$}}} \vline{} &   0.50 & \multicolumn{1}{c}{0.50\textsuperscript{\phantom{$\uparrow$}}} \vline{} & 0.58 & \multicolumn{1}{c}{0.50\textsuperscript{\textcolor{red}{$\downarrow$}}} \vline{} & 0.62 & 0.56 & \multicolumn{1}{c}{0.51\textsuperscript{\textcolor{red}{$\downarrow$}}} \vline{} & 0.50 & 0.58 & \multicolumn{1}{c}{0.50\textsuperscript{\textcolor{red}{$\downarrow$}}} \vline{} & 0.50 & 0.64 & \multicolumn{1}{c}{0.49\textsuperscript{\textcolor{red}{$\downarrow$}}} \vline{} & 0.50 & 0.52 & \multicolumn{1}{c}{0.69\textsuperscript{\phantom{$\uparrow$}}} \vline{} & 0.54 & 0.70\\
Counting      & 0.56 & 0.50 & \multicolumn{1}{c}{0.50\textsuperscript{\textcolor{red}{$\downarrow$}}} \vline{} &   0.76 & \multicolumn{1}{c}{0.50\textsuperscript{\textcolor{red}{$\downarrow$}}} \vline{} & 0.70 & \multicolumn{1}{c}{0.48\textsuperscript{\textcolor{red}{$\downarrow$}}} \vline{} & 0.82 & 0.74 & \multicolumn{1}{c}{0.60\textsuperscript{\textcolor{red}{$\downarrow$}}} \vline{} & 0.50 & 0.56 & \multicolumn{1}{c}{0.59\textsuperscript{\phantom{$\uparrow$}}} \vline{} & 0.50 & 0.84 & \multicolumn{1}{c}{0.79\textsuperscript{\textcolor{red}{$\downarrow$}}} \vline{} & 0.50 & 0.52 & \multicolumn{1}{c}{0.55\textsuperscript{\phantom{$\uparrow$}}} \vline{} & 0.62 & 0.65\\
\midrule
\textbf{Per-task Average} & 0.56 & 0.63 & \multicolumn{1}{c}{0.50\textsuperscript{\textcolor{red}{$\downarrow$}}} \vline{} &  0.63 & \multicolumn{1}{c}{0.54\textsuperscript{\textcolor{red}{$\downarrow$}}} \vline{} & 0.57 & \multicolumn{1}{c}{0.50\textsuperscript{\textcolor{red}{$\downarrow$}}} \vline{} & 0.60 & 0.62 & \multicolumn{1}{c}{0.58\textsuperscript{\textcolor{red}{$\downarrow$}}} \vline{} & 0.61 & 0.58 & \multicolumn{1}{c}{0.52\textsuperscript{\textcolor{red}{$\downarrow$}}} \vline{} & 0.61 & 0.64 & \multicolumn{1}{c}{0.64\textsuperscript{\phantom{$\uparrow$}}} \vline{} & 0.61 & 0.56 & \multicolumn{1}{c}{0.62\textsuperscript{\phantom{$\uparrow$}}}\vline{} & 0.53 & 0.71\\
\midrule
\multicolumn{22}{c}{\textbf{Accuracy Across Model Groups}} \\
\midrule
\multicolumn{22}{l}{\textbf{(b) Recognition Accuracy}} \\
\cmidrule(lr){1-22}
\textbf{Attributes} &
\makecell{\textbf{BEATs}} &
\makecell{\textbf{Whisper-}\\\textbf{Large-v2}} &
\makecell{\textbf{SAL-}\\\textbf{MONN}} &
\makecell{\textbf{Step-Audio 2}\\\textbf{mini} \textbf{Encoder}} &
\makecell{\textbf{Step-Audio 2}\\\textbf{mini}} &
\makecell{\textbf{SenseVoice}\\\textbf{Small}} &
\makecell{\textbf{VITA-Audio-}\\\textbf{Plus-Vanilla}} &
\makecell{\textbf{Da-}\\\textbf{sheng}} &
\makecell{\textbf{MiDasheng-}\\\textbf{LM Encoder}} &
\makecell{\textbf{MiDa-}\\\textbf{shengLM}} &
\makecell{\textbf{Whisper-}\\\textbf{Large-v3}} &
\makecell{\textbf{Qwen2-Audio-}\\\textbf{Instruct Encoder}} &
\makecell{\textbf{Qwen2-Audio-}\\\textbf{Instruct}} &
\makecell{\textbf{Whisper-}\\\textbf{Large-v3}} &
\makecell{\textbf{Kimi-Audio-}\\\textbf{Instruct Encoder}} &
\makecell{\textbf{Kimi-Audio-}\\\textbf{Instruct}} & 
\makecell{\textbf{Whisper-}\\\textbf{Large-v3}} &
\makecell{\textbf{Qwen2.5-Omni-}\\\textbf{Instruct Encoder}} &
\makecell{\textbf{Qwen2.5-Omni-}\\\textbf{Instruct}} & 
\makecell{\textbf{Qwen3-Omni-}\\\textbf{Instruct Encoder}} &
\makecell{\textbf{Qwen3-Omni-}\\\textbf{Instruct}}\\
\midrule
Pitch         & 0.92 & 0.92 & \multicolumn{1}{c}{0.50\textsuperscript{\textcolor{red}{$\downarrow$}}} \vline{} &  0.92 & \multicolumn{1}{c}{0.57\textsuperscript{\textcolor{red}{$\downarrow$}}} \vline{} & 0.76 & \multicolumn{1}{c}{0.50\textsuperscript{\textcolor{red}{$\downarrow$}}} \vline{} & 0.88 & 0.88 & \multicolumn{1}{c}{0.65\textsuperscript{\textcolor{red}{$\downarrow$}}} \vline{} & 0.86 & 0.96 & \multicolumn{1}{c}{0.52\textsuperscript{\textcolor{red}{$\downarrow$}}} \vline{} & 0.86 & 0.90 & \multicolumn{1}{c}{0.87\textsuperscript{\textcolor{red}{$\downarrow$}}} \vline{} & 0.86 & 0.92 & \multicolumn{1}{c}{0.82\textsuperscript{\textcolor{red}{$\downarrow$}}} \vline{} & 0.94 & 0.90\textsuperscript{\textcolor{red}{$\downarrow$}}\\
Brightness    & 0.96 & 0.92 & \multicolumn{1}{c}{0.50\textsuperscript{\textcolor{red}{$\downarrow$}}} \vline{} &   1.00 & \multicolumn{1}{c}{0.62\textsuperscript{\textcolor{red}{$\downarrow$}}} \vline{} & 0.84 & \multicolumn{1}{c}{0.50\textsuperscript{\textcolor{red}{$\downarrow$}}} \vline{} & 0.88 & 0.98 & \multicolumn{1}{c}{0.75\textsuperscript{\textcolor{red}{$\downarrow$}}} \vline{} & 0.84 & 0.98 & \multicolumn{1}{c}{0.49\textsuperscript{\textcolor{red}{$\downarrow$}}} \vline{} & 0.84 & 0.94 & \multicolumn{1}{c}{0.81\textsuperscript{\textcolor{red}{$\downarrow$}}} \vline{} & 0.84 & 0.86 & \multicolumn{1}{c}{0.84\textsuperscript{\textcolor{red}{$\downarrow$}}} \vline{} & 0.88 & 0.83\textsuperscript{\textcolor{red}{$\downarrow$}}\\
Loudness      & 0.76 & 0.70 & \multicolumn{1}{c}{0.50\textsuperscript{\textcolor{red}{$\downarrow$}}} \vline{} &   0.76 & \multicolumn{1}{c}{0.41\textsuperscript{\textcolor{red}{$\downarrow$}}} \vline{} & 0.62 & \multicolumn{1}{c}{0.50\textsuperscript{\textcolor{red}{$\downarrow$}}} \vline{} & 0.70 & 0.74 & \multicolumn{1}{c}{0.72\textsuperscript{\textcolor{red}{$\downarrow$}}} \vline{} & 0.70 & 0.82 & \multicolumn{1}{c}{0.47\textsuperscript{\textcolor{red}{$\downarrow$}}} \vline{} & 0.70 & 0.80 & \multicolumn{1}{c}{0.70\textsuperscript{\textcolor{red}{$\downarrow$}}} \vline{} & 0.70 & 0.80 & \multicolumn{1}{c}{0.76\textsuperscript{\textcolor{red}{$\downarrow$}}} \vline{} & 0.88 & 0.70\textsuperscript{\textcolor{red}{$\downarrow$}}\\
Velocity      & 0.60 & 0.62 & \multicolumn{1}{c}{0.50\textsuperscript{\textcolor{red}{$\downarrow$}}} \vline{} &   0.64 & \multicolumn{1}{c}{0.53\textsuperscript{\textcolor{red}{$\downarrow$}}} \vline{} & 0.60 & \multicolumn{1}{c}{0.51\textsuperscript{\textcolor{red}{$\downarrow$}}} \vline{} & \cellcolor{lightred}0.58 & 0.66 & \multicolumn{1}{c}{0.46\textsuperscript{\textcolor{red}{$\downarrow$}}} \vline{} & 0.68 & 0.66 & \multicolumn{1}{c}{0.54\textsuperscript{\textcolor{red}{$\downarrow$}}} \vline{} & 0.68 & \cellcolor{lightred}0.58 & \multicolumn{1}{c}{0.50\textsuperscript{\textcolor{red}{$\downarrow$}}} \vline{} & 0.68 & 0.62 & \multicolumn{1}{c}{0.52\textsuperscript{\textcolor{red}{$\downarrow$}}} \vline{} & 0.70 & 0.49\textsuperscript{\textcolor{red}{$\downarrow$}}\\
Duration      & 0.90 & 0.78 & \multicolumn{1}{c}{0.50\textsuperscript{\textcolor{red}{$\downarrow$}}} \vline{} &   0.94 & \multicolumn{1}{c}{0.54\textsuperscript{\textcolor{red}{$\downarrow$}}} \vline{} & 0.90 &\multicolumn{1}{c}{0.50\textsuperscript{\textcolor{red}{$\downarrow$}}} \vline{} & 0.94 & 0.94 & \multicolumn{1}{c}{0.68\textsuperscript{\textcolor{red}{$\downarrow$}}} \vline{} & 0.78 & 1.00 & \multicolumn{1}{c}{0.50\textsuperscript{\textcolor{red}{$\downarrow$}}} \vline{} & 0.78 & 1.00 & \multicolumn{1}{c}{0.75\textsuperscript{\textcolor{red}{$\downarrow$}}} \vline{} & 0.78 & 0.98 & \multicolumn{1}{c}{0.60\textsuperscript{\textcolor{red}{$\downarrow$}}} \vline{} & 0.94 & 0.80\textsuperscript{\textcolor{red}{$\downarrow$}}\\
Tempo         & \cellcolor{lightred}0.50 & \cellcolor{lightred}0.50 & \multicolumn{1}{c}{0.50\textsuperscript{\textcolor{red}{$\downarrow$}}} \vline{}  & 0.68 & \multicolumn{1}{c}{0.47\textsuperscript{\textcolor{red}{$\downarrow$}}} \vline{} & 0.66 & \multicolumn{1}{c}{0.51\textsuperscript{\textcolor{red}{$\downarrow$}}} \vline{} & 0.86 & 0.78 & \multicolumn{1}{c}{0.55\textsuperscript{\textcolor{red}{$\downarrow$}}} \vline{} & \cellcolor{lightred}0.52 & 0.60 & \multicolumn{1}{c}{0.51\textsuperscript{\textcolor{red}{$\downarrow$}}} \vline{} & \cellcolor{lightred}0.52 & 0.64 & \multicolumn{1}{c}{0.55\textsuperscript{\textcolor{red}{$\downarrow$}}} \vline{} & \cellcolor{lightred}0.52 & 0.62 & \multicolumn{1}{c}{0.53\textsuperscript{\textcolor{red}{$\downarrow$}}} \vline{} & \cellcolor{lightred}0.58 & 0.64\textsuperscript{\phantom{$\uparrow$}}\\
Direction     & 0.72 & 0.68 & \multicolumn{1}{c}{0.50\textsuperscript{\textcolor{red}{$\downarrow$}}} \vline{}   & 0.76 & \multicolumn{1}{c}{0.43\textsuperscript{\textcolor{red}{$\downarrow$}}} \vline{} & 0.70 & \multicolumn{1}{c}{0.47\textsuperscript{\textcolor{red}{$\downarrow$}}} \vline{} & 0.74 & 0.74 & \multicolumn{1}{c}{0.44\textsuperscript{\textcolor{red}{$\downarrow$}}} \vline{} & 0.68 & 0.76 & \multicolumn{1}{c}{0.48\textsuperscript{\textcolor{red}{$\downarrow$}}} \vline{} & 0.68 & 0.76 & \multicolumn{1}{c}{0.44\textsuperscript{\textcolor{red}{$\downarrow$}}} \vline{} & 0.68 & 0.74 & \multicolumn{1}{c}{0.49\textsuperscript{\textcolor{red}{$\downarrow$}}} \vline{} & 0.74 & 0.49\textsuperscript{\textcolor{red}{$\downarrow$}}\\
Distance      & \cellcolor{lightred}0.56 & \cellcolor{lightred}0.56 & \multicolumn{1}{c}{0.44\textsuperscript{\textcolor{red}{$\downarrow$}}} \vline{} &   0.60 & \multicolumn{1}{c}{0.46\textsuperscript{\textcolor{red}{$\downarrow$}}} \vline{} & 0.64 & \multicolumn{1}{c}{0.53\textsuperscript{\textcolor{red}{$\downarrow$}}} \vline{} & \cellcolor{lightred}0.58 & \cellcolor{lightred}0.58 & \multicolumn{1}{c}{0.48\textsuperscript{\textcolor{red}{$\downarrow$}}} \vline{} & \cellcolor{lightred}0.56 & 0.62 & \multicolumn{1}{c}{0.52\textsuperscript{\textcolor{red}{$\downarrow$}}} \vline{} & \cellcolor{lightred}0.56 & 0.60 & \multicolumn{1}{c}{0.48\textsuperscript{\textcolor{red}{$\downarrow$}}} \vline{} & \cellcolor{lightred}0.56 & 0.66 & \multicolumn{1}{c}{0.52\textsuperscript{\textcolor{red}{$\downarrow$}}} \vline{} & 0.60 & 0.42\textsuperscript{\textcolor{red}{$\downarrow$}}\\
Reverberation & 1.00 & 1.00 & \multicolumn{1}{c}{0.50\textsuperscript{\textcolor{red}{$\downarrow$}}} \vline{} &   1.00 & \multicolumn{1}{c}{0.52\textsuperscript{\textcolor{red}{$\downarrow$}}} \vline{} & 0.96 & \multicolumn{1}{c}{0.50\textsuperscript{\textcolor{red}{$\downarrow$}}} \vline{} & 1.00 & 1.00 & \multicolumn{1}{c}{0.50\textsuperscript{\textcolor{red}{$\downarrow$}}} \vline{} & 1.00 & 0.98 & \multicolumn{1}{c}{0.50\textsuperscript{\textcolor{red}{$\downarrow$}}} \vline{} & 1.00 & 1.00 & \multicolumn{1}{c}{0.95\textsuperscript{\textcolor{red}{$\downarrow$}}} \vline{} & 1.00 & 0.98 & \multicolumn{1}{c}{0.69\textsuperscript{\textcolor{red}{$\downarrow$}}} \vline{} & 1.00 & 0.83\textsuperscript{\textcolor{red}{$\downarrow$}}\\
Timbre        & \cellcolor{lightred}0.50 & 0.62 & \multicolumn{1}{c}{0.50\textsuperscript{\textcolor{red}{$\downarrow$}}} \vline{} & \cellcolor{lightred}0.52 & \multicolumn{1}{c}{0.98\textsuperscript{\phantom{$\uparrow$}}} \vline{} & \cellcolor{lightred}0.56 & \multicolumn{1}{c}{0.54\textsuperscript{\textcolor{red}{$\downarrow$}}} \vline{} & \cellcolor{lightred}0.54 & \cellcolor{lightred}0.48 & \multicolumn{1}{c}{0.96\textsuperscript{\phantom{$\uparrow$}}} \vline{} & \cellcolor{lightred}0.58 & \cellcolor{lightred}0.52 & \multicolumn{1}{c}{0.67\textsuperscript{\textcolor{red}{$\downarrow$}}} \vline{} & \cellcolor{lightred}0.58 & \cellcolor{lightred}0.56 & \multicolumn{1}{c}{0.99\textsuperscript{\phantom{$\uparrow$}}} \vline{} & \cellcolor{lightred}0.58 & \cellcolor{lightred}0.54 & \multicolumn{1}{c}{0.97\textsuperscript{\phantom{$\uparrow$}}} \vline{} & \cellcolor{lightred}0.58 & 0.98\textsuperscript{\phantom{$\uparrow$}}\\
Texture       & \cellcolor{lightred}0.52 & \cellcolor{lightred}0.50 & \multicolumn{1}{c}{0.50\textsuperscript{\textcolor{red}{$\downarrow$}}} \vline{} &   \cellcolor{lightred}0.56 & \multicolumn{1}{c}{0.65\textsuperscript{\phantom{$\uparrow$}}} \vline{} & \cellcolor{lightred}0.56 & \multicolumn{1}{c}{0.48\textsuperscript{\textcolor{red}{$\downarrow$}}} \vline{} & \cellcolor{lightred}0.52 & 0.66 & \multicolumn{1}{c}{0.69\textsuperscript{\phantom{$\uparrow$}}} \vline{} & \cellcolor{lightred}0.50 & \cellcolor{lightred}0.56 & \multicolumn{1}{c}{0.60\textsuperscript{\phantom{$\uparrow$}}} \vline{} & \cellcolor{lightred}0.50 & \cellcolor{lightred}0.52 & \multicolumn{1}{c}{0.68\textsuperscript{\phantom{$\uparrow$}}} \vline{} & \cellcolor{lightred}0.50 & \cellcolor{lightred}0.54 & \multicolumn{1}{c}{0.80\textsuperscript{\phantom{$\uparrow$}}} \vline{} & \cellcolor{lightred}0.56 & 0.75\textsuperscript{\phantom{$\uparrow$}}\\
Counting      & \cellcolor{lightred}0.54 & \cellcolor{lightred}0.58 & \multicolumn{1}{c}{0.50\textsuperscript{\textcolor{red}{$\downarrow$}}} \vline{} &   \cellcolor{lightred}0.52 & \multicolumn{1}{c}{0.52\textsuperscript{\textcolor{red}{$\downarrow$}}} \vline{} & \cellcolor{lightred}0.52 & \multicolumn{1}{c}{0.40\textsuperscript{\textcolor{red}{$\downarrow$}}} \vline{} & 0.62 & \cellcolor{lightred}0.58 & \multicolumn{1}{c}{0.44\textsuperscript{\textcolor{red}{$\downarrow$}}} \vline{} & \cellcolor{lightred}0.58 & \cellcolor{lightred}0.52 & \multicolumn{1}{c}{0.47\textsuperscript{\textcolor{red}{$\downarrow$}}} \vline{} & \cellcolor{lightred}0.58 & \cellcolor{lightred}0.54 & \multicolumn{1}{c}{0.69\textsuperscript{\phantom{$\uparrow$}}} \vline{} & \cellcolor{lightred}0.58 & \cellcolor{lightred}0.56 & \multicolumn{1}{c}{0.60\textsuperscript{\phantom{$\uparrow$}}} \vline{} & \cellcolor{lightred}0.58 & 0.99\textsuperscript{\phantom{$\uparrow$}}\\
\midrule
\textbf{Per-task Average} & 0.71 & 0.70 & \multicolumn{1}{c}{0.50\textsuperscript{\textcolor{red}{$\downarrow$}}} \vline{} &   0.74 & \multicolumn{1}{c}{0.56\textsuperscript{\textcolor{red}{$\downarrow$}}} \vline{} & 0.69 & \multicolumn{1}{c}{0.50\textsuperscript{\textcolor{red}{$\downarrow$}}} \vline{} & 0.74 & 0.75 & \multicolumn{1}{c}{0.61\textsuperscript{\textcolor{red}{$\downarrow$}}} \vline{} & 0.69 & 0.75 & \multicolumn{1}{c}{0.52\textsuperscript{\textcolor{red}{$\downarrow$}}} \vline{} & 0.69 & 0.74 & \multicolumn{1}{c}{0.70\textsuperscript{\textcolor{red}{$\downarrow$}}} \vline{} & 0.69 & 0.74 & \multicolumn{1}{c}{0.68\textsuperscript{\textcolor{red}{$\downarrow$}}} \vline{} & 0.75 & 0.74\textsuperscript{\textcolor{red}{$\downarrow$}}\\
\bottomrule
\end{tabular}
\end{adjustbox}
\caption{\textbf{Detailed Accuracy Breakdown: Encoders (Probe) vs. E2E Models.} 
We report accuracy across 12 attributes separated by Recognition and Comparison tasks. 
\textit{Pretrained} denotes the off-the-shelf encoder; \textit{Extracted} refers to the encoder state after E2E training (if unfrozen). 
The results confirm a widespread alignment bottleneck that linear probes on encoders consistently achieve robust accuracy ($\geq 0.60$) on signal-level attributes, whereas corresponding E2E models often degrade to near-random performance ($\approx 0.50$). 
Notable exceptions are the Qwen-Omni series, where E2E performance matches or exceeds the probe, indicating a successful preservation of perceptual information during their training.}
\label{tab:comp-rec-across-groups}
\end{table*}

\subsection{The Relational Bottleneck in Encoders}
While the previous analysis highlighted the degradation from encoder to E2E model, we also investigate whether the encoders themselves possess the structural capability to handle relational tasks. Table~\ref{tab:probe_full_acc_with_dimension_allmodels} breaks down the linear probe performance results across the two tasks, Recognition (absolute judgment) and Comparison (relative judgment).

Contrary to human perception, where comparison is cognitively less demanding, our probing results reveal that encoders do not exhibit a ``comparison advantage.'' This suggests that standard audio encoders, primarily trained on global or frame-level classification objectives, lack the inherent mechanisms to explicitly compare distinct temporal segments within a single audio stream, limiting their ability to leverage the contrastive structure of the input.

\begin{table*}[t]
\centering
\setlength{\tabcolsep}{3pt}
\renewcommand{\arraystretch}{1.1}
\begin{adjustbox}{max width=\textwidth}
\begin{tabular}{llccccccccccc}
\toprule
\multicolumn{2}{c}{\textbf{Tasks}} &
\multicolumn{11}{c}{\textbf{Comparison \,|\, Recognition}}\\
\cmidrule(lr){1-2}\cmidrule(lr){3-13}
\textbf{Metric}& \textbf{Accuracy} &
\makecell{\textbf{BEATs}} &
\makecell{\textbf{Whisper-}\\\textbf{Large-v2}} &
\makecell{\textbf{Whisper-}\\\textbf{Large-v3}} &
\makecell{\textbf{SenseVoice}\\\textbf{Small}} &
\makecell{\textbf{Dasheng}} &
\makecell{\textbf{MiDashengLM}\\\textbf{Encoder}} &
\makecell{\textbf{Step-Audio 2}\\\textbf{mini Encoder}} &
\makecell{\textbf{Qwen2-Audio-}\\\textbf{Instruct Encoder}} &
\makecell{\textbf{Kimi-Audio-}\\\textbf{Instruct Encoder}} &
\makecell{\textbf{Qwen2.5-Omni-}\\\textbf{Instruct Encoder}} &
\makecell{\textbf{Qwen3-Omni-}\\\textbf{Instruct Encoder}} \\
\midrule
\multirow{4}{*}{\makecell[l]{Spectral\& \\Amplitude}}
  & Pitch
    & 0.66\,|\,0.92
    & 0.76\,|\,0.92
    & 0.74\,|\,0.86
    & 0.66\,|\,0.76
    & 0.56\,|\,0.88
    & 0.82\,|\,0.88
    & 0.72\,|\,0.92
    & 0.60\,|\,0.96
    & 0.74\,|\,0.90
    & 0.66\,|\,0.92
    & 0.58\,|\,0.94 \\
  & Brightness
    & 0.70\,|\,0.96
    & 0.92\,|\,0.92
    & 0.84\,|\,0.84
    & 0.58\,|\,0.84
    & 0.62\,|\,0.88
    & 0.76\,|\,0.98
    & 0.70\,|\,1.00
    & 0.60\,|\,0.98
    & 0.74\,|\,0.94 
    & 0.60\,|\,0.86
    & 0.52\,|\,0.88 \\
  & Loudness
    & 0.58\,|\,0.76
    & 0.58\,|\,0.70
    & 0.58\,|\,0.70
    & 0.62\,|\,0.62
    & 0.62\,|\,0.70
    & 0.68\,|\,0.74
    & 0.66\,|\,0.76
    & 0.66\,|\,0.82
    & 0.70\,|\,0.80 
    & 0.62\,|\,0.80
    & 0.60\,|\,0.88 \\
  & Velocity
    & 0.54\,|\,0.60
    & 0.56\,|\,0.62
    & 0.56\,|\,0.68
    & 0.54\,|\,0.60
    & 0.56\,|\,0.58
    & 0.52\,|\,0.66
    & 0.60\,|\,0.64
    & 0.56\,|\,0.66
    & 0.58\,|\,0.58 
    & 0.56\,|\,0.62
    & 0.50\,|\,0.70 \\
\addlinespace[2pt]
\multirow{2}{*}{Temporal}
  & Duration
    & 0.52\,|\,0.90
    & 0.56\,|\,0.78
    & 0.50\,|\,0.78
    & 0.50\,|\,0.90
    & 0.58\,|\,0.94
    & 0.56\,|\,0.94
    & 0.58\,|\,0.94
    & 0.56\,|\,1.00
    & 0.62\,|\,1.00 
    & 0.54\,|\,0.98
    & 0.50\,|\,0.94 \\
  & Tempo
    & 0.48\,|\,0.50
    & 0.58\,|\,0.50
    & 0.62\,|\,0.52
    & 0.54\,|\,0.66
    & 0.56\,|\,0.86
    & 0.56\,|\,0.78
    & 0.60\,|\,0.68
    & 0.60\,|\,0.60
    & 0.54\,|\,0.64 
    & 0.52\,|\,0.62
    & 0.52\,|\,0.58 \\
\addlinespace[2pt]
\multirow{3}{*}{\makecell[l]{Spatial\&\\ Environment}}
  & Direction
    & 0.52\,|\,0.72
    & 0.68\,|\,0.68
    & 0.54\,|\,0.68
    & 0.48\,|\,0.70
    & 0.48\,|\,0.74
    & 0.46\,|\,0.74
    & 0.54\,|\,0.76
    & 0.52\,|\,0.76
    & 0.54\,|\,0.76 
    & 0.52\,|\,0.74
    & 0.44\,|\,0.74 \\
  & Distance
    & 0.58\,|\,0.56
    & 0.62\,|\,0.56
    & 0.56\,|\,0.56
    & 0.56\,|\,0.64
    & 0.54\,|\,0.58
    & 0.60\,|\,0.58
    & 0.58\,|\,0.60
    & 0.52\,|\,0.62
    & 0.64\,|\,0.60 
    & 0.52\,|\,0.66
    & 0.52\,|\,0.60 \\
  & Reverberation
    & 0.54\,|\,1.00
    & 0.80\,|\,1.00
    & 0.84\,|\,1.00
    & 0.50\,|\,0.96
    & 0.64\,|\,1.00
    & 0.56\,|\,1.00
    & 0.68\,|\,1.00
    & 0.56\,|\,0.98
    & 0.54\,|\,1.00 
    & 0.52\,|\,0.98
    & 0.52\,|\,1.00 \\
\addlinespace[2pt]
\multirow{2}{*}{Timbre}
  & Timbre
    & 0.48\,|\,0.50
    & 0.54\,|\,0.62
    & 0.54\,|\,0.58
    & 0.54\,|\,0.56
    & 0.60\,|\,0.54
    & 0.62\,|\,0.48
    & 0.58\,|\,0.52
    & 0.66\,|\,0.52
    & 0.52\,|\,0.56 
    & 0.62\,|\,0.54 
    & 0.50\,|\,0.58 \\
  & Texture
    & 0.54\,|\,0.52
    & 0.50\,|\,0.50
    & 0.50\,|\,0.50
    & 0.58\,|\,0.56
    & 0.62\,|\,0.52
    & 0.56\,|\,0.66
    & 0.50\,|\,0.56
    & 0.58\,|\,0.56
    & 0.64\,|\,0.52 
    & 0.52\,|\,0.54
    & 0.54\,|\,0.56 \\
\addlinespace[2pt]
\multirow{1}{*}{Scene Level}
  & Counting
    & 0.56\,|\,0.54
    & 0.50\,|\,0.58
    & 0.50\,|\,0.58
    & 0.70\,|\,0.52
    & 0.82\,|\,0.62
    & 0.74\,|\,0.58
    & 0.76\,|\,0.52
    & 0.56\,|\,0.52
    & 0.84\,|\,0.54 
    & 0.52\,|\,0.56
    & 0.62\,|\,0.58 \\
\midrule
\multicolumn{2}{c}{\textbf{Per-task Average}}
    & 0.56\,|\,0.71
    & 0.63\,|\,0.70
    & 0.61\,|\,0.69
    & 0.57\,|\,0.69
    & 0.60\,|\,0.74
    & 0.62\,|\,0.75
    & 0.63\,|\,0.74
    & 0.58\,|\,0.75
    & 0.64\,|\,0.74 
    & 0.56\,|\,0.74
    & 0.53\,|\,0.75 \\
\multicolumn{2}{c}{\textbf{Overall Average}}
    & 0.63
    & 0.67
    & 0.65
    & 0.63
    & 0.67
    & 0.69
    & 0.68
    & 0.67
    & 0.69 	
    & 0.65
    & 0.64 \\
\bottomrule
\end{tabular}
\end{adjustbox}
\vspace{4pt}
 \caption{\textbf{Probe Accuracy across Task Dimensions: Evidence of a Relational Bottleneck.} This table reports the mean accuracy of linear probes trained on frozen pretrained encoders and their unfrozen encoders, aggregated by task type (Recognition vs. Comparison) across all 12 attributes. Crucially, comparison performance consistently remains at or below recognition performance, mirroring the behavior of full E2E models. This indicates a \textit{shared relational bottleneck} that without architectural mechanisms to explicitly model contrastive relationships between segments, even raw perceptual representations fail to leverage the comparative structure of the task, treating comparison as merely a more complex form of recognition.}  
\label{tab:probe_full_acc_with_dimension_allmodels}
\end{table*}

\subsection{Attribute-wise Patterns in Encoders}

Beyond identifying bottlenecks, our probing setup allows us to analyze the plasticity of audio encoders during the E2E training process. Table~\ref{tab:model_group_accuracies} compares the probing accuracy of \textit{Pretrained Encoders} versus \textit{Extracted Encoders}. We observe that E2E training with an unfrozen audio encoder can yield small gains. 

To deeper inspection in Table~\ref{tab:probe_full_acc_with_dimension_allmodels}, we find a non-uniform adaptation capability across different physical attributes. 
Comparing pre-trained encoders and extracted encoders, we can tell that unfreezing the encoder helps the Tempo and Distance attributes more noticeably, yet brings limited gains for Timbre, Texture,and Counting.

\section{Implications}
Concretely, our findings suggest two distinct pathways for future development:
(i) \textbf{For perceptual recognition.} Prioritize \emph{alignment strategies}. Since encoders already encode rich signal cues as shown by our probes, future work should focus on preventing representational degradation during LLM integration, potentially via adaptable projectors or the partial unfreezing strategies observed in Omni-models.
(ii) \textbf{For relational reasoning.} Innovate in \emph{encoder pretraining}. Current encoders lack the mechanisms to explicitly compare inputs. Addressing this requires incorporating architectural inductive biases, such as native cross-segment attention or contrastive objectives, to ensure relational information is captured before reaching the LLM.

\begin{table}[t]
\centering
\footnotesize
\setlength{\tabcolsep}{6pt}
\renewcommand{\arraystretch}{1.12}
\begin{adjustbox}{max width=\columnwidth}
\begin{tabular}{lccc}
\toprule
\makecell[l]{\textbf{Model}\\ \textbf{Groups}} &
\makecell{\textbf{Comparison}\\ \textbf{Accuracy}} &
\makecell{\textbf{Recognition}\\ \textbf{Accuracy}} &
\makecell{\textbf{Overall}\\ \textbf{Accuracy}} \\
\midrule
BEATs                   & 0.56\textsuperscript{\phantom{$\uparrow$}} & \textbf{0.71}\textsuperscript{\phantom{$\uparrow$}} & 0.63\textsuperscript{\phantom{$\uparrow$}} \\
Whisper-Large-v2        & \textbf{0.63}\textsuperscript{\phantom{$\uparrow$}} & 0.70\textsuperscript{\phantom{$\uparrow$}} & \textbf{0.67}\textsuperscript{\phantom{$\uparrow$}} \\
Salmonn                 & 0.50\textsuperscript{\textcolor{red}{$\downarrow$}} & 0.50\textsuperscript{\textcolor{red}{$\downarrow$}} & 0.50\textsuperscript{\textcolor{red}{$\downarrow$}} \\
\midrule
Whisper-Large-v3        & 0.61\textsuperscript{\phantom{$\uparrow$}} & 0.69\textsuperscript{\phantom{$\uparrow$}} & 0.65\textsuperscript{\phantom{$\uparrow$}} \\
Step-Audio 2 mini Encoder     & \textbf{0.63}\textsuperscript{\phantom{$\uparrow$}} & \textbf{0.74}\textsuperscript{\phantom{$\uparrow$}} & \textbf{0.68}\textsuperscript{\phantom{$\uparrow$}} \\
Step-Audio 2 mini & 0.54\textsuperscript{\textcolor{red}{$\downarrow$}} & 0.56\textsuperscript{\textcolor{red}{$\downarrow$}} & 0.55\textsuperscript{\textcolor{red}{$\downarrow$}} \\
\midrule
SenseVoiceSmall         & \textbf{0.57}\textsuperscript{\phantom{$\uparrow$}} & \textbf{0.69}\textsuperscript{\phantom{$\uparrow$}} & \textbf{0.63}\textsuperscript{\phantom{$\uparrow$}} \\
VITA-Audio-Plus-Vanilla              & 0.50\textsuperscript{\textcolor{red}{$\downarrow$}} & 0.50\textsuperscript{\textcolor{red}{$\downarrow$}} & 0.50\textsuperscript{\textcolor{red}{$\downarrow$}} \\
\midrule
Dasheng                 & 0.60\textsuperscript{\phantom{$\uparrow$}} & 0.74\textsuperscript{\phantom{$\uparrow$}} & 0.67\textsuperscript{\phantom{$\uparrow$}} \\
MiDashengLM Encoder     & \textbf{0.62}\textsuperscript{\phantom{$\uparrow$}} & \textbf{0.75}\textsuperscript{\phantom{$\uparrow$}} & \textbf{0.69}\textsuperscript{\phantom{$\uparrow$}} \\
MiDashengLM             & 0.58\textsuperscript{\textcolor{red}{$\downarrow$}} & 0.61\textsuperscript{\textcolor{red}{$\downarrow$}} & 0.60\textsuperscript{\textcolor{red}{$\downarrow$}} \\
\midrule
Whisper-Large-v3        & \textbf{0.61}\textsuperscript{\phantom{$\uparrow$}} & 0.69\textsuperscript{\phantom{$\uparrow$}} & 0.65\textsuperscript{\phantom{$\uparrow$}} \\
Qwen2-Audio-Instruct Encoder     & 0.58\textsuperscript{\phantom{$\uparrow$}} & \textbf{0.75}\textsuperscript{\phantom{$\uparrow$}} & \textbf{0.67}\textsuperscript{\phantom{$\uparrow$}} \\
Qwen2-Audio-Instruct           & 0.52\textsuperscript{\textcolor{red}{$\downarrow$}} & 0.52\textsuperscript{\textcolor{red}{$\downarrow$}} & 0.52\textsuperscript{\textcolor{red}{$\downarrow$}} \\
\midrule
Whisper-Large-v3        & 0.61\textsuperscript{\phantom{$\downarrow$}} & 0.69\textsuperscript{\phantom{$\downarrow$}} & 0.65\textsuperscript{\phantom{$\downarrow$}} \\
Kimi-Audio-Instruct Encoder     & \textbf{0.64}\textsuperscript{\phantom{$\uparrow$}} & \textbf{0.74}\textsuperscript{\phantom{$\uparrow$}} & \textbf{0.69}\textsuperscript{\phantom{$\uparrow$}}\\ 
Kimi-Audio-Instruct     & \textbf{0.64}\textsuperscript{\phantom{$\uparrow$}} & 0.70\textsuperscript{\textcolor{red}{$\downarrow$}} & 0.67\textsuperscript{\textcolor{red}{$\downarrow$}} \\
\midrule
Whisper-Large-v3 & 0.61\textsuperscript{\phantom{$\uparrow$}} & 0.69\textsuperscript{\phantom{$\uparrow$}} & \textbf{0.65}\textsuperscript{\phantom{$\uparrow$}} \\
Qwen2.5-Omni-Instruct Encoder    & 0.56\textsuperscript{\phantom{$\downarrow$}} & \textbf{0.74}\textsuperscript{\phantom{$\uparrow$}} & \textbf{0.65}\textsuperscript{\phantom{$\uparrow$}} \\
Qwen2.5-Omni-Instruct            & \textbf{0.62}\textsuperscript{\phantom{$\uparrow$}} & 0.68\textsuperscript{\textcolor{red}{$\downarrow$}} & \textbf{0.65}\textsuperscript{\phantom{$\uparrow$}} \\
\midrule
Qwen3-Omni-Instruct Encoder     & 0.53\textsuperscript{\phantom{$\downarrow$}} & \textbf{0.75}\textsuperscript{\phantom{$\uparrow$}} & 0.64\textsuperscript{\phantom{$\downarrow$}} \\
Qwen3-Omni-Instruct     & \textbf{0.71}\textsuperscript{\phantom{$\uparrow$}} & 0.74\textsuperscript{\textcolor{red}{$\downarrow$}} & \textbf{0.72}\textsuperscript{\phantom{$\uparrow$}} \\
\bottomrule
\end{tabular}
\end{adjustbox}
\caption{This table contrasts the performance of off-the-shelf Pretrained Encoders against Extracted Encoders that were updated during training and E2E models. We can find that E2E training with an unfrozen audio encoder can yield small gains}  
\label{tab:model_group_accuracies}
\end{table}

\begin{table*}[t]
\centering
\footnotesize
\setlength{\tabcolsep}{4pt}
\begin{adjustbox}{max width=0.95\textwidth}
\begin{tabular}{lccccccccccccc}
\toprule \toprule
\multirow{2.5}{*}{\textbf{Models}} & \multicolumn{4}{c}{\textbf{Spectral \& Amplitude}} &  \multicolumn{2}{c}{\textbf{Temporal}} & \multicolumn{3}{c}{\textbf{Spatial\& Environment}} &  \multicolumn{2}{c}{\textbf{Timbre}} & {\textbf{Scene Level}}& \multirow{2.5}{*}{\textbf{Avg.}}  \\ 
\cmidrule(lr){2-5} \cmidrule(lr){6-7} \cmidrule(lr){8-10} \cmidrule(lr){11-12}  \cmidrule(lr){13-13} 
& Pitch & Brightness & Loudness & Velocity & Duration & Tempo & Direction & Distance & Reverberation & Texture & Timbre & Counting & \\
\midrule
Random Guess & 0.50 & 0.50 & 0.50 & 0.50 & 0.50 & 0.50 & 0.50 & 0.50 & 0.50 & 0.50 & 0.50 & 0.50 & 0.50\\
\cdashline{1-14}
\noalign{\vskip 0.5mm}
Human &1.00 &0.93 &0.97 &0.83 &1.00 &0.97 &0.83 &0.87 &1.00 &1.00 &1.00 &1.00 &0.95 \\
\midrule \midrule
\multicolumn{14}{c}{\textbf{Large Audio Language Models (LALMs)}} \\ 
\midrule \midrule
BAT & \cellcolor{green1!89}0.05 & \cellcolor{green1!98}0.00 & \cellcolor{green1!95}0.04 & \cellcolor{green1!87}0.08 & \cellcolor{green1!95}0.03 & \cellcolor{green1!100}0.00 & \cellcolor{green1!99}0.00 & \cellcolor{green1!100}0.00 & \cellcolor{green1!100}0.00 & \cellcolor{green1!40}0.27 & \cellcolor{green1!75}0.09 & \cellcolor{green1!98}0.00 & \cellcolor{green1!90}0.05\\

MU-LLaMA & \cellcolor{green1!40}0.34 & \cellcolor{green1!86}0.08 & \cellcolor{green1!99}0.00 & \cellcolor{green1!38}0.32 & \cellcolor{green1!92}0.04 & \cellcolor{green1!74}0.12 & \cellcolor{green1!78}0.10 & \cellcolor{green1!73}0.16 & \cellcolor{green1!50}0.24 & \cellcolor{green1!100}0.00 & \cellcolor{green1!100}0.00 & \cellcolor{green1!85}0.06 & \cellcolor{green1!76}0.12\\

LLaMA-Omni & \cellcolor{green1!28}0.38 & 0.50 & \cellcolor{green1!9}0.51 & \cellcolor{green1!1}0.50 & \cellcolor{green1!4}0.46 & \cellcolor{green1!1}0.51 & 0.50 & 0.50 & \cellcolor{green1!4}0.49 & 0.50 & 0.50 & \cellcolor{green1!14}0.44 & \cellcolor{green1!5}0.48\\

Audio Flamingo 2 & 0.50 & \cellcolor{green1!12}0.47 & 0.53 & 0.50 & 0.49 & 0.52 & \cellcolor{green1!14}0.45 & \cellcolor{green1!15}0.39 & \cellcolor{green1!3}0.49 & \cellcolor{green1!4}0.48 & \cellcolor{green1!3}0.50 & 0.50 & \cellcolor{green1!4}0.49\\

GLM-4-Voice & 0.50 & 0.51 & \cellcolor{green1!2}0.47 & \cellcolor{green1!2}0.49 & \cellcolor{green1!7}0.47 & 0.54 & 0.44 & 0.50 & 0.49 & 0.50 & 0.50 & \cellcolor{green1!10}0.46 & \cellcolor{green1!2}0.49\\

VITA-Audio-Plus-Vanilla & 0.52 & 0.51 & 0.50 & 0.45 & 0.51 & 0.49 & 0.45 &  0.54 & 0.54 & 0.50 & 0.50 & 0.48 & 0.50\\

Audio Flamingo 3 & 0.50 & 0.50 & 0.50 & 0.50 & 0.44 & 0.52 & 0.51 & 0.52 & 0.50 & 0.50 & 0.50 & 0.50 & 0.50 \\

Voxtral-Mini & 0.53 & 0.45 & 0.53 & 0.51 & \cellcolor{green1!1}0.49 & 0.52 & 0.51 & 0.47 & 0.50 & 0.52 & 0.50 & 0.52 & 0.50\\

Baichuan-Audio-Instruct & 0.48 & 0.52 & 0.49 & 0.48 & 0.53 & 0.51 & 0.45 & 0.47 & 0.50 & 0.54 & 0.50 & 0.49 & 0.50\\

LLaMA-Omni2 & 0.49 & 0.47 & 0.42 & 0.47 & 0.48 & 0.49 & 0.56 & 0.56 & 0.55 & 0.54 & 0.50 & 0.50 & 0.50\\

SALMONN & 0.50 & 0.50 & 0.53 & 0.50 & 0.49 & 0.52 & 0.51 & 0.47 & 0.50 & 0.50 & 0.50 & 0.50 & 0.50\\

R1-AQA & 0.50 & 0.49 & 0.50 & 0.50 & 0.46 & 0.51 & 0.50 & 0.48 & 0.52 & 0.51 & 0.48 & 0.49 & 0.50\\

Qwen2-Audio-Instruct & 0.43 & 0.49 & 0.49 & 0.56 & 0.52 & 0.55 & \underline{0.52} & 0.54 & 0.56 & 0.50 & 0.50 &  0.59 & 0.52\\

Step-Audio 2 mini& 0.47 & 0.55 & 0.57 & 0.48 & 0.47 & 0.52 & 0.51 & 0.50 & 0.48 & 0.91 & 0.50 & 0.50 & 0.54\\

MiDashengLM & 0.56 & 0.64 & 0.61 & \underline{0.62} &0.58 & 0.55 & 0.48 & 0.54 & 0.54 & 0.74 & 0.51 & 0.60 & 0.58\\

MiMo-Audio-Instruct &0.61 	&0.68 	&0.57 &	0.54 &	0.51 	&0.48 	&0.50 &	0.54 	&\underline{0.63} &	0.85 &	0.56 &	0.50 &	0.58 \\

Kimi-Audio-Instruct & \underline{0.79} & 0.81 & \cellcolor{green1!1}0.66 & 0.53 & 0.64 & 0.58 & 0.43 & 0.39 & 0.58 & 0.95 & 0.49 & \underline{0.79} & 0.64\\

\cdashline{1-14}
\noalign{\vskip 0.5mm}
GPT-4o-Audio &0.71& 	0.68 &	0.55 &0.55 &	0.44 &	0.49 &	0.49 &	0.56 &	0.50 &	0.88 &	0.43 &	0.66 &	0.58\\
\midrule \midrule
\multicolumn{14}{c}{\textbf{Large Audio Reasoning Models (LARMs)}}  \\ 
\midrule \midrule
Mellow & \cellcolor{green1!84}0.07 & \cellcolor{green1!80}0.08 & \cellcolor{green1!75}0.14 & \cellcolor{green1!60}0.20 & \cellcolor{green1!57}0.18 & \cellcolor{green1!45}0.28 & \cellcolor{green1!68}0.16 & \cellcolor{green1!58}0.20 & \cellcolor{green1!49}0.28 & \cellcolor{green1!67}0.16 & \cellcolor{green1!92}0.03 & \cellcolor{green1!62}0.15 & \cellcolor{green1!66}0.16\\

GAMA & \cellcolor{green1!28}0.38 & \cellcolor{green1!16}0.42 & \cellcolor{green1!23}0.38 & \cellcolor{green1!7}0.45 & \cellcolor{green1!50}0.24 & \cellcolor{green1!5}0.49 & \cellcolor{green1!41}0.29 & \cellcolor{green1!30}0.36 & \cellcolor{green1!36}0.26 & 0.48 & \cellcolor{green1!19}0.41 & \cellcolor{green1!41}0.33 & \cellcolor{green1!25}0.37\\

Audio Flamingo 2 Sound-CoT & 0.52 & 0.50 & 0.55 & 0.49 & 0.50 & 0.52 & 0.51 & 0.47 & 0.49 & 0.50 & 0.51 & 0.49 & 0.50\\

Audio Flamingo 3 (think mode) & 0.50 & 0.50 & \cellcolor{green1!3}0.51 & 0.50 & \cellcolor{green1!4}0.49 & \cellcolor{green1!1}0.52 & \cellcolor{green1!6}0.46 & \cellcolor{green1!5}0.52 & 0.50 & 0.50 & 0.50 & \cellcolor{green1!1}0.44 & \cellcolor{green1!2}0.50\\

R1-AQA (think mode) & 0.51 & 0.51 & 0.48 & \cellcolor{green1!2}0.50 & 0.49 & 0.54 & 0.51 & \cellcolor{green1!1}0.47 & 0.52 & \cellcolor{green1!2}0.60 & 0.56 & \cellcolor{green1!5}0.50 & \cellcolor{green1!1}0.52\\

Step-Audio 2 mini Think&\cellcolor{green1!4}0.48 &\cellcolor{green1!6}0.44 &\cellcolor{green1!4}0.51 &\cellcolor{green1!2}0.53 &\cellcolor{green1!4}0.53 &\cellcolor{green1!3}0.49 &\cellcolor{green1!3}0.51 &\cellcolor{green1!5}0.46 &\cellcolor{green1!3}0.50 &\cellcolor{green1!4}0.70 &\cellcolor{green1!3}0.67 &\cellcolor{green1!7}0.46 &\cellcolor{green1!4}0.52 \\

Audio-Reasoner & 0.65 & 0.62 & 0.55 & 0.57 & 0.56 & \cellcolor{green1!1}0.49 & \cellcolor{green1!4}0.49 & \cellcolor{green1!6}0.49 & \cellcolor{green1!5}0.48 & \cellcolor{green1!2}0.65 & \cellcolor{green1!1}0.54 & \cellcolor{green1!4}0.64 & \cellcolor{green1!2}0.56\\

Step-Audio-R1 &0.70 &0.65 &0.54 &0.50 &0.50 &0.55 &0.44 &\textbf{0.67} &0.57 &0.96 &0.58 &\textbf{0.80} &0.62 \\

MiMo-Audio (think mode)	&\underline{0.79} &	0.74 &	0.64 &	0.46 &	\underline{0.67} &	0.54 &	0.50 	&0.57 &	 \cellcolor{green1!1}0.61 &	0.69 &	0.55 	&0.74 &	0.63 \\

\midrule \midrule
\multicolumn{14}{c}{\textbf{Omni Language Models (OLMs)}}\\ 
\midrule \midrule
OpenOmni & 0.43 & 0.48 & 0.51 & 0.50 & 0.51 & 0.52 & \underline{0.52} & 0.49 & 0.52 & 0.50 & 0.54 & 0.50 & 0.50\\

Baichuan-Omni-1.5 & 0.56 & 0.47 & 0.54 & 0.52 & 0.48 & 0.51 & 0.51 & 0.44 & 0.51 & 0.57 & 0.46 & 0.52 & 0.51\\

VITA-1.5 & 0.53 & 0.52 & 0.54 & 0.52 & 0.52 & 0.51 & 0.51 & 0.50 & 0.49 & 0.48 & 0.44 & 0.51 & 0.51\\

HumanOmni & 0.50 & 0.52 & 0.54 & 0.51 & 0.47 & 0.48 & \underline{0.52} & 0.48 & 0.49 & 0.80 & 0.50 & 0.50 & 0.53\\

Ming-Lite-Omni-1.5&	0.56 &	0.53 	&0.55 &	0.50 &	0.61 &	0.60 &	0.50 &	0.48 &	0.50 &	0.92 &	0.56 &	0.69 &	0.58 \\

Ola & 0.72 & 0.73 & 0.68 & 0.49 & 0.55 & 0.55 & \textbf{0.54} & 0.48 & 0.47 & 0.97 & 0.47 & 0.58 & 0.60\\

Qwen2.5-Omni & 0.60 & \underline{0.89} & \underline{0.73} & 0.54 & 0.54 & 0.50 & 0.50 & 0.52 & 0.42 & \textbf{1.00} & \underline{0.69} & 0.55 & 0.62\\

Qwen3-Omni-Instruct&	\textbf{0.83} &	\textbf{0.92} &	\textbf{0.80} &	\textbf{0.64} &	\textbf{0.73} &	\textbf{0.66} &	\underline{0.52}& 	\underline{0.60} &	0.52 &	\underline{0.99} &	\textbf{0.70} &	0.65 &	\textbf{0.71} \\

\cdashline{1-14}
\noalign{\vskip 0.5mm}
Gemini-2.5-Flash &0.77 &	0.82 &	0.68 &	0.60 &	0.54 &	\underline{0.64} &	0.50 	&0.54 &	\textbf{0.75} &	0.91 &	0.61 &	0.78 	&\underline{0.68} \\
\bottomrule \bottomrule
\end{tabular}
\end{adjustbox}
\vspace{2pt}
\begin{tikzpicture}[baseline]
  \fill[left color=green1!0, right color=green1!100]
       (0,0) rectangle (\dimexpr0.7\linewidth\relax, 1.8ex);
  \node[anchor=west,font=\small\bfseries] at (-2cm,0.8ex) {\textit{\%abstention}};
\end{tikzpicture}
\\[-3ex]
\caption{Comparison task accuracy across audio attributes.}
\label{tab:comparison_full}
\end{table*}

\begin{table*}[t]
\centering
\footnotesize
\setlength{\tabcolsep}{4pt}
\begin{adjustbox}{max width=0.95\textwidth}
\begin{tabular}{lccccccccccccc}
\toprule \toprule
\multirow{2.5}{*}{\textbf{Models}} & \multicolumn{4}{c}{\textbf{Spectral \& Amplitude}} &  \multicolumn{2}{c}{\textbf{Temporal}} & \multicolumn{3}{c}{\textbf{Spatial\& Environment}} &  \multicolumn{2}{c}{\textbf{Timbre}} & {\textbf{Scene Level}}& \multirow{2.5}{*}{\textbf{Avg.}}  \\ 
\cmidrule(lr){2-5} \cmidrule(lr){6-7} \cmidrule(lr){8-10} \cmidrule(lr){11-12}  \cmidrule(lr){13-13} 
& Pitch & Brightness & Loudness & Velocity & Duration & Tempo & Direction & Distance & Reverberation & Texture & Timbre & Counting & \\
\midrule
Random Guess & 0.50 & 0.50 & 0.50 & 0.50 & 0.50 & 0.50 & 0.50 & 0.50 & 0.50 & 0.50 & 0.50 & 0.50 & 0.50\\
\cdashline{1-14}
\noalign{\vskip 0.5mm}
Human & 0.87 &0.93 &0.77 &0.83 &0.83 &0.77 &0.83 &0.73 &1.00 &1.00 &0.93 &1.00 &0.88 
\\
\midrule \midrule
\multicolumn{14}{c}{\textbf{Large Audio Language Models (LALMs)}} \\ 
\midrule \midrule
BAT & \cellcolor{green1!100}0.00 & \cellcolor{green1!100}0.00 & \cellcolor{green1!100}0.00 & \cellcolor{green1!100}0.00 & \cellcolor{green1!100}0.00 & \cellcolor{green1!100}0.00 & \cellcolor{green1!100}0.00 & \cellcolor{green1!96}0.03 & \cellcolor{green1!83}0.11 & \cellcolor{green1!91}0.06 & \cellcolor{green1!97}0.02 & \cellcolor{green1!100}0.00 & \cellcolor{green1!97}0.02\\

MU-LLaMA & \cellcolor{green1!100}0.00 & \cellcolor{green1!95}0.03 & \cellcolor{green1!100}0.00 & \cellcolor{green1!87}0.06 & \cellcolor{green1!100}0.00 & \cellcolor{green1!80}0.11 & \cellcolor{green1!98}0.00 & \cellcolor{green1!99}0.01 & \cellcolor{green1!78}0.10 & \cellcolor{green1!14}0.39 & \cellcolor{green1!5}0.48 & \cellcolor{green1!99}0.01 & \cellcolor{green1!80}0.10\\

LLaMA-Omni & \cellcolor{green1!19}0.43 & \cellcolor{green1!1}0.49 & \cellcolor{green1!11}0.45 & \cellcolor{green1!10}0.47 & \cellcolor{green1!10}0.46 & \cellcolor{green1!2}0.48 & 0.49 & \cellcolor{green1!2}0.40 & \cellcolor{green1!12}0.40 & 0.54 & 0.50 & 0.51 & \cellcolor{green1!6}0.47\\

GLM-4-Voice & 0.50 & 0.50 & 0.51 & \textbf{0.55} & 0.48 & 0.50 & 0.50 & \cellcolor{green1!1}0.44 & 0.46 & 0.49 & 0.50 & \cellcolor{green1!28}0.38 & \cellcolor{green1!2}0.48\\

Audio Flamingo 3 & 0.46 & 0.50 & 0.50 & 0.50 & 0.50 & 0.50 & 0.49 & 0.48 & 0.50 & 0.50 & 0.50 & 0.41 & 0.49\\

Voxtral-Mini & 0.52 & 0.43 & 0.49 & \cellcolor{green1!7}0.46 & 0.51 & 0.51 & 0.50 & 0.46 & 0.49 & 0.55 & 0.49 & 0.50 & \cellcolor{green1!1}0.49\\

LLaMA-Omni2 & 0.54 & 0.50 & 0.50 & \cellcolor{green1!7}0.44 & 0.50 & 0.44 & 0.50 & 0.52 & 0.50 & 0.50 & 0.47 & 0.46 & \cellcolor{green1!1}0.49\\

VITA-Audio-Plus-Vanilla & 0.50 & 0.50 & 0.50 & 0.51 & 0.50 & 0.51 & 0.47 & 0.53 & 0.50 & 0.54 & 0.48 & 0.40 & 0.50\\

SALMONN & 0.50 & 0.50 & 0.50 & 0.50 & 0.50 & 0.50 & 0.50 & 0.44 & 0.50 & 0.50 & 0.50 & 0.50 & 0.50\\

Audio Flamingo 2 & 0.50 & 0.54 & 0.50 & 0.50 & 0.48 & 0.50 & 0.50 & 0.44 & 0.50 & \cellcolor{green1!1}0.60 & 0.56 & 0.50 & 0.51\\

Qwen2-Audio-Instruct & 0.52 & \cellcolor{green1!8}0.49 & \cellcolor{green1!8}0.47 & \cellcolor{green1!1}\underline{0.54} & 0.50 & 0.51 & 0.48 & 0.52 & 0.50 & 0.67 & 0.60 & 0.47 & \cellcolor{green1!1}0.52\\

Baichuan-Audio-Instruct & 0.49 & 0.50 & 0.53 & 0.50 & 0.50 & 0.49 & 0.42 & 0.39 & 0.55 & 0.81 & 0.54 & 0.49 & 0.52\\

Step-Audio 2 mini& 0.57 & 0.62 & 0.41 & 0.53 & 0.54 & 0.47 & 0.43 & 0.46 & 0.52 & \underline{0.98} & 0.65 & 0.52 & 0.56\\

R1-AQA & 0.64 & 0.72 & 0.57 & 0.52 & 0.50 & 0.47 & 0.49 & 0.51 & 0.54 & \textbf{0.99} & \underline{0.79} & 0.46 & 0.60\\

MiMo-Audio-Instruct&	0.81 &	0.72 &	0.58 	&0.51 	&0.56 &	0.50 &	0.50 	&0.48 	&0.59 	&0.93 &	0.58 &	0.49 &	0.60 \\

MiDashengLM & 0.65 & 0.75 & \underline{0.72} & 0.46 & 0.68 & 0.55 & 0.44 & 0.48 & 0.50 & 0.96 & 0.69 & 0.44 & 0.61\\

Kimi-Audio-Instruct & \underline{0.87} & 0.81 & 0.70 & 0.50 & 0.75 & 0.55 & \cellcolor{green1!6}0.44 & \cellcolor{green1!5}0.48 & \textbf{0.95} & \textbf{0.99} & 0.68 & 0.69 & \cellcolor{green1!1}\underline{0.70}\\

\cdashline{1-14}
\noalign{\vskip 0.5mm}
GPT-4o-Audio& 0.71 &	0.78 &	\cellcolor{green1!3}0.49 &	\cellcolor{green1!2}0.52 	&\cellcolor{green1!1}0.54 &	0.51 &	\cellcolor{green1!1}\underline{0.51} &	\cellcolor{green1!1}\textbf{0.54} &	\cellcolor{green1!4}0.50 &	0.92 	&0.49 &	0.68 	&\cellcolor{green1!1}0.60 \\

\midrule \midrule
\multicolumn{14}{c}{\textbf{Large Audio Reasoning Models (LARMs)}}  \\ 
\midrule \midrule
Mellow & \cellcolor{green1!76}0.11 & \cellcolor{green1!68}0.19 & \cellcolor{green1!90}0.05 & \cellcolor{green1!84}0.08 & \cellcolor{green1!45}0.24 & \cellcolor{green1!57}0.20 & \cellcolor{green1!77}0.12 & \cellcolor{green1!63}0.15 & \cellcolor{green1!47}0.28 & \cellcolor{green1!34}0.32 & \cellcolor{green1!33}0.39 & \cellcolor{green1!23}0.38 & \cellcolor{green1!58}0.21\\

GAMA & \cellcolor{green1!55}0.21 & \cellcolor{green1!52}0.20 & \cellcolor{green1!61}0.21 & \cellcolor{green1!67}0.16 & \cellcolor{green1!23}0.30 & \cellcolor{green1!47}0.28 & \cellcolor{green1!39}0.31 & \cellcolor{green1!51}0.19 & \cellcolor{green1!30}0.39 & \cellcolor{green1!1}0.76 & \cellcolor{green1!4}0.46 & \cellcolor{green1!27}0.30 & \cellcolor{green1!38}0.31\\

Audio Flamingo 2 Sound-CoT & \cellcolor{green1!52}0.26 & \cellcolor{green1!18}0.53 & \cellcolor{green1!37}0.34 & \cellcolor{green1!52}0.22 & 0.52 & \cellcolor{green1!6}0.48 & 0.50 & \cellcolor{green1!2}0.45 & 0.50 & \cellcolor{green1!5}0.66 & \cellcolor{green1!17}0.47 & \cellcolor{green1!3}0.49 & \cellcolor{green1!16}0.45\\

Audio Flamingo 3 (think mode) & 0.52 & \cellcolor{green1!5}0.47 & 0.50 & 0.50 & 0.48 & 0.44 & \underline{0.51} & \cellcolor{green1!3}0.44 & 0.50 & 0.50 & 0.51 & 0.45 & \cellcolor{green1!1}0.49\\

R1-AQA (think mode) & 0.65 & \cellcolor{green1!28}0.45 & 0.45 & \cellcolor{green1!1}0.51 & 0.50 & 0.48 & \cellcolor{green1!1}\textbf{0.52} & \cellcolor{green1!5}0.47 & 0.51 & \cellcolor{green1!4}0.94 & \cellcolor{green1!2}0.66 & \cellcolor{green1!54}0.24 & \cellcolor{green1!8}0.53\\

Audio-Reasoner & \cellcolor{green1!6}0.69 & \cellcolor{green1!2}\textbf{0.84} & \cellcolor{green1!1}0.57 & \cellcolor{green1!2}0.51 & \cellcolor{green1!5}0.53 & \cellcolor{green1!1}0.48 & \cellcolor{green1!4}0.46 & \cellcolor{green1!2}0.43 & 0.52 & 0.78 & \cellcolor{green1!2}0.53 & \cellcolor{green1!54}0.17 & \cellcolor{green1!7}0.54\\

Step-Audio 2 mini Think&\cellcolor{green1!5}0.52 &\cellcolor{green1!3}0.62 &\cellcolor{green1!6}0.52 &\cellcolor{green1!1}0.41 &\cellcolor{green1!5}0.53 &\cellcolor{green1!2}0.54 &\cellcolor{green1!8}0.49 &\cellcolor{green1!7}0.44 &\cellcolor{green1!1}0.52 &\cellcolor{green1!3}0.92 &\cellcolor{green1!2}0.46 &\cellcolor{green1!16}0.54 &\cellcolor{green1!5}0.54 \\

Step-Audio-R1 &0.57 &0.68 &0.54 &0.50 &0.50 &\underline{0.58} &\underline{0.51} &\underline{0.53} &\underline{0.83} &0.95 &0.62 &\underline{0.73} &0.63 \\

MiMo-Audio (think mode)& 	0.83 & 	0.76 	& \cellcolor{green1!1}0.56 & 	\cellcolor{green1!3}0.49 & 	0.71 	& 0.51 & 	0.49 & 	0.48 & 	0.73 	& 0.91 & 	0.68 & 	\cellcolor{green1!2}0.71 	& \cellcolor{green1!1}0.66 \\

\midrule \midrule

\multicolumn{14}{c}{\textbf{Omni Language Models (OLMs)}}\\ 
\midrule \midrule
VITA-1.5 & 0.49 & 0.49 & 0.50 & 0.45 & 0.50 & 0.51 & 0.46 & 0.41 & 0.61 & 0.47 & 0.43 & 0.41 & 0.48\\

OpenOmni & 0.50 & 0.48 & 0.50 & 0.51 & 0.53 & 0.48 & 0.49 & 0.44 & 0.53 & 0.51 & 0.53 & 0.51 & 0.50\\

Baichuan-Omni-1.5 & 0.58 & 0.47 & 0.51 & 0.51 & 0.48 & 0.49 & 0.49 & 0.39 & 0.46 & 0.84 & 0.56 & 0.59 & 0.53\\

Ming-Lite-Omni-1.5	&0.57 &	0.50 &	0.50 &	0.50 	&0.44 &	0.50 &	\underline{0.51} &	0.45 &	0.56 &	0.95 &	0.57 &	0.61 &	0.56 \\

HumanOmni & 0.59 & 0.70 & 0.50 & 0.46 & 0.53 & 0.52 & \textbf{0.52} & 0.36 & 0.49 & 0.91 & 0.64 & 0.68 & 0.58\\

Ola & 0.75 & 0.72 & 0.60 & 0.46 & 0.57 & 0.48 & 0.43 & 0.45 & 0.68 & 0.97 & 0.59 & 0.58 & 0.61\\

Qwen2.5-Omni & 0.82 & \textbf{0.84} & \textbf{0.76} & 0.52 & 0.60 & 0.53 & 0.49 & 0.52 & 0.69 & 0.97 & \textbf{0.80} & 0.60 & 0.68\\

Qwen3-Omni-Instruct	&\textbf{0.90} &	\underline{0.83} &	0.70 	&0.49 &	\underline{0.80} &	\textbf{0.64} &	0.49 &	0.42 &	\underline{0.83} 	&\underline{0.98} 	&0.75 &	\textbf{0.99} &	\textbf{0.74} \\

\cdashline{1-14}
\noalign{\vskip 0.5mm}
Gemini-2.5-Flash& 0.77 &	0.75 &	\cellcolor{green1!2}0.57 &	\cellcolor{green1!5}\underline{0.54} &	\textbf{0.81} &	0.55 &	0.47 &	0.46 &	0.68 	&0.97 	&0.63& 	0.63 	&\cellcolor{green1!1}0.65 
\\
\bottomrule \bottomrule
\end{tabular}
\end{adjustbox}
\vspace{2pt}
\begin{tikzpicture}[baseline]
  \fill[left color=green1!0, right color=green1!100]
       (0,0) rectangle (\dimexpr0.7\linewidth\relax, 1.8ex);
  \node[anchor=west,font=\small\bfseries] at (-2cm,0.8ex) {\textit{\%abstention}};
\end{tikzpicture}
\\[-3ex]
\caption{Recognition task accuracy across audio attributes.}
\label{tab:recognition_full}
\end{table*}

\end{document}